\documentclass[%
 aip,
 jmp,%
 amsmath,amssymb,
preprint,
superscriptaddress,
]{revtex4-2}

\usepackage{graphicx}
\usepackage{dcolumn}
\usepackage{bm}
\usepackage{nicematrix}
\usepackage{siunitx}
\DeclareSIUnit\hartree{\text {\ensuremath {E}}_{\mathrm {h}}}
\DeclareSIUnit\bohr{\text {\ensuremath {a}}_{0}}
\usepackage{multirow}
\usepackage{upgreek}
\usepackage[version=4]{mhchem}
\usepackage[modules={scheme}]{chemmacros}
\usepackage{bigdelim} 

\begin{document}

\title{Heavier chalcogenofenchones for fundamental gas-phase studies of molecular chirality}% Force line breaks with \\

\author{Manjinder Kour}
\affiliation{Fachbereich Chemie, Philipps Universit\"at Marburg, Hans-Meerwein Str.4, 35032 Marburg, Germany}
\author{Denis Kargin}
\affiliation{Institut f\"ur Chemie und CINSaT, Universit\"at Kassel, Heinrich-Plett-Str. 40, 34132 Kassel, Germany}
\author{Eileen D\"oring}
\affiliation{Institut f\"ur Physik und CINSaT, Universit\"at Kassel, Heinrich-Plett-Str. 40, 34132 Kassel, Germany}
\author{Sudheendran Vasudevan}
\affiliation{Institut f\"ur Physik und CINSaT, Universit\"at Kassel, Heinrich-Plett-Str. 40, 34132 Kassel, Germany}
\author{Martin Maurer}
\affiliation{Institut f\"ur Chemie und CINSaT, Universit\"at Kassel, Heinrich-Plett-Str. 40, 34132 Kassel, Germany}
\author{Pascal Stahl}
\affiliation{Institut f\"ur Physik und CINSaT, Universit\"at Kassel, Heinrich-Plett-Str. 40, 34132 Kassel, Germany}
\author{Igor Vidanovi\'{c}}
\affiliation{Institut f\"ur Chemie und CINSaT, Universit\"at Kassel, Heinrich-Plett-Str. 40, 34132 Kassel, Germany}
\author{Clemens Bruhn}
\affiliation{Institut f\"ur Chemie und CINSaT, Universit\"at Kassel, Heinrich-Plett-Str. 40, 34132 Kassel, Germany}
\author{Wenhao Sun}
\affiliation{Deutsches Elektronen-Synchrotron DESY,
Notkestr. 85, 22607 Hamburg, Germany}
\author{Steffen M. Giesen}
\affiliation{Fachbereich Chemie, Philipps Universit\"at Marburg, Hans-Meerwein Str.4, 35032 Marburg, Germany}
\author{Thomas Baumert}
\affiliation{Institut f\"ur Physik und CINSaT, Universit\"at Kassel, Heinrich-Plett-Str. 40, 34132 Kassel, Germany}
\author{Robert Berger*}
\affiliation{Fachbereich Chemie, Philipps Universit\"at Marburg, Hans-Meerwein Str.4, 35032 Marburg, Germany}
\author{Hendrike Braun}
\affiliation{Institut f\"ur Physik und CINSaT, Universit\"at Kassel, Heinrich-Plett-Str. 40, 34132 Kassel, Germany}
\author{Guido W. Fuchs}
\affiliation{Institut f\"ur Physik und CINSaT, Universit\"at Kassel, Heinrich-Plett-Str. 40, 34132 Kassel, Germany}
\author{Thomas F. Giesen}
\affiliation{Institut f\"ur Physik und CINSaT, Universit\"at Kassel, Heinrich-Plett-Str. 40, 34132 Kassel, Germany}
\author{Rudolf Pietschnig}
\affiliation{Institut f\"ur Chemie und CINSaT, Universit\"at Kassel, Heinrich-Plett-Str. 40, 34132 Kassel, Germany}
\author{Melanie Schnell}
\affiliation{Deutsches Elektronen-Synchrotron DESY,
Notkestr. 85, 22607 Hamburg, Germany}
\affiliation{Institut f{\"u}r Physikalische Chemie,
Christian-Albrechts-Universit{\"a}t zu Kiel,
Max-Eyth-Str. 1, 24118 Kiel, Germany}
\author{Arne Senftleben}
\affiliation{Institut f\"ur Physik und CINSaT, Universit\"at Kassel, Heinrich-Plett-Str. 40, 34132 Kassel, Germany}

\date{\today}

\begin{abstract}
Monoterpene ketones are frequently studied compounds that enjoy great popularity both in chemistry and in physics due to comparatively high volatility, stability, conformational rigidity and commercial availability. Herein, we explore the heavier chalcogenoketone derivatives of fenchone as promising benchmark systems—synthetically accessible in enantiomerically pure form—for systematic studies of nuclear charge (\textit{Z}) dependent properties in chiral compounds. Synthesis, structural characterization, thorough gas-phase rotational and vibrational spectroscopy as well as accompanying quantum chemical studies on the density-functional-theory level reported in this work foreshadow subsequent applications of this compound class for fundamental investigations of molecular chirality under well-defined conditions.
\end{abstract}


\maketitle

%--------------------------------------------------------------------

\section{Introduction}

Terpenoids form a large class of natural products with relevance in diverse fields of research and application such as i) drug development, with the anti-malaria drug Artimisinin as well as Paclitaxel, which is administered in cancer treatments, featuring as prominent examples \cite{Augustin:2020,Kamran:2022}, ii) flavor and fragrance industries, with linalool, coumarine and fenchone serving as few of the numerous representatives \cite{Sharmeen:2021}, and iii) environmental protection and climate dynamics, with terpenoids being emitted by plants to the atmosphere, where they undergo numerous reactions with radicals to produce secondary organic aerosols \cite{Waring:2011}.

Being natural products or derivatives of them, terpenoids add richly to the chiral pool of precursors used in the laboratory syntheses of enantiomerically pure substances. At the same token, access to the non-natural enantiomers of these terpenoids requires then specific chemical protocols. In this context, fenchone \textbf{1-O} and its constitutional isomer camphor \textbf{2} are both early examples in the history of natural product chemistry for which total synthesis was developed to provide separate enantiomers and racemic mixtures. 

Due to their rigidity, high volatility, and commercial availability both in racemic and in enantiomerically pure form, these two terpenoids served as important cornerstones in research on physical properties of enantiomers: Camphor was, for example, among the first molecular systems \cite{nafie:1976} for which vibrational circular dichroism could be detected \cite{holzwarth:1974,nafie:1976}. 
Bromocamphor, camphor and fenchone were selected for the first experimental observations \cite{bowering:2001,garcia:2003,hergenhahn:2004,lischke:2004,powis:2008} of the theoretically predicted photoelectron circular dichroism (PECD) \cite{ritchie:1976} in single-photon excitation and subsequently also in multiphoton excitation \cite{Lux:2012}.
Early synchrotron radiation studies on circular dichroism in one-photon transitions between bound states were also performed with camphor and fenchone \cite{gedanken:1986,pulm:1997}.
And finally, also one of the pioneering high-resolution searches for resonance frequency splittings between rovibrational transitions of enantiomers induced by parity-violating electroweak interactions were performed with camphor \cite{arimondo:1977} to provide an initial upper bound of $\Delta \nu/\nu$ of about $10^{-8}$. However, this bound was more than 10 orders of magnitude larger than the effect sizes expected for systems composed of light elements only (see e.g. Refs.~\onlinecite{berger:2004a,quack:2008,schwerdtfeger:2010,berger:2019} for a review and Ref.~\onlinecite{berger:2001} for corresponding estimates for fluorooxirane) and numerically estimated specifically for camphor in Ref.~\onlinecite{schwerdtfeger:2004}. 

In particular, when searching for larger parity-violating effects in chiral molecules, one could in principle benefit from the established  \cite{zeldovich:1977,harris:1978,hegstrom:1980,gorshkov:1982} steep scaling with nuclear charge $Z$ by choosing heavy-element derivatives of champhor or fenchone. One possible route is to introduce further substituents as realized with bromocamphor mentioned before, the other is to replace oxygen with one of the heavier chalcogens S, Se, Te or even Po. The latter route has the additional charm to systematically alter the chromophoric system for infrared (IR) and optical studies of chirality. This is the key motivation to establish chalcogenofenchones \textbf{1-X} (with X from the group of chalcogens) as a versatile and tunable research platform for systematic gas-phase studies on molecular chirality. Thus, we have determined and analyzed an extensive set of experimental and theoretical data on this compound class that we report in two papers:

In the present paper, we report on the synthesis and spectroscopic characterization of thiofenchone (\textbf{1-S}) and selenofenchone (\textbf{1-Se}) in solution, before we introduce results from a gas-phase mircowave (MW) study to determine structural parameters of these compounds. Subsequently, we present experimental gas-phase IR overview spectra and compare them to accompanying theoretical spectra, obtained from electronic structure theory for the series of fenchone homologues down to polonofenchone. We close with demonstrating the potential of the chalcogenofenchone family for high-resolution IR rovibrational spectroscopy, for which one motivation stems from the search for molecular parity violation in chiral molecules.

In an accompanying paper,\cite{vasudevan:2025} we report on single-photon absorption and multi-photon photoelectron spectra in the ultraviolet and visible range of the electromagnetic spectrum to determine excite state energies of the chalcogenofenchones in the gas phase and present detailed studies of photoelectron circular dichroism after multiphoton excitation.

\section{Experimental and Computational Details}

\subsection{Synthesis of \textbf{1-S} and \textbf{1-Se}}

Heavier chalcogenofenchones \textbf{1-S}, \textbf{1-Se} are accessible via formal chalcogen exchange reactions from
fenchone for which the integrity of the stereocenters is ensured. \textbf{1-S} has been prepared using \textit{Lawesson's}
reagent (2,4-Bis(4-methoxyphenyl)-1,3,2,4-dithiadiphosphetane-2,4-disulfide, \ce{(CH3OC6H4PS2)2}) in 90~\% isolated yield. Preparation of selenocarbonyl compounds can be achieved by a variety
of different selenium transfer reagents including \ce{H2Se}, \ce{NaSeH},\cite{jigami:1998} hydrazones with diselenium dichloride\cite{okazaki:1983,guziec:1984} or disilenium dibromide\cite{guziec:1984},
conversion of aldehydes to selenoaldehydes with \ce{(Me3Si)2Se} and catalytic amounts of \textit{n}-BuLi,\cite{segi:1988} conversion of ketones to selenoketones with \ce{(Me2Al)2Se},\cite{li:1999} \ce{(Me3Si)2Se}/\ce{BF3\cdot Et2O}\cite{takikawa:1994} or Woollins’ reagent\cite{woolins:2005}.
Owing to the steric hindrance of the carbonyl carbon atom in fenchone these led to incomplete
conversion to \textbf{1-Se} together with inseparable impurities of byproducts.

\begin{scheme}
%\begin{figure}
    \centering
    \includegraphics[width=\linewidth]{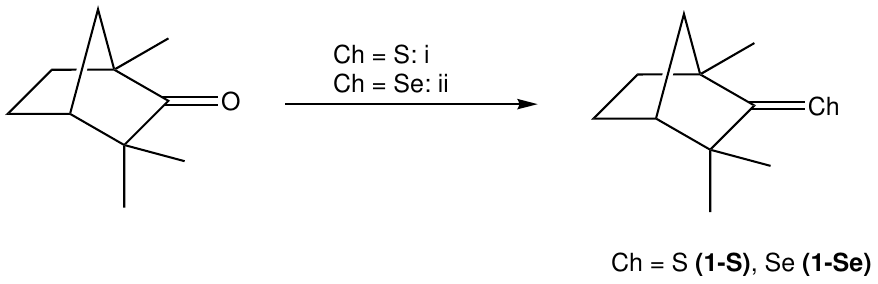}
    \caption{Preparation of heavier chalcogenofenchones \textbf{1-S} and \textbf{1-Se} (using i: (CH\textsubscript{3}OC\textsubscript{6}H\textsubscript{4}PS\textsubscript{2})\textsubscript{2}, ii: (C\textsubscript{8}H\textsubscript{14}B)\textsubscript{2}Se).\label{scheme:synthesis}}
%\end{figure}
\end{scheme}

Better results were obtained with bis(1,5-bicyclooctanediylboryl)monoselenide (\ce{(C8H14B)2Se}) as selenation reagent of
choice in this work. We vary the procedure reported by Takikawa, based on Köster’s method for
preparation of bis(1,5-bicyclooctanediylboryl)monoselenide.\cite{koster:1992,shimada:1997} The monoselenide was
prepared \textit{in situ} in mesitylene without the need for an autoclave and used without further purification.\cite{shimada:1997}
Furthermore, we modified the work-up procedure, increasing the yields to 45~\%. For the work-up, it was
mandatory to separate selenoketones from byproducts. Instead of aqueous work-up with quenching the
reaction with 10~\% aqueous NaOH-solution as in the reported literature, we performed separation of the volatile reagents and products in vacuum either by distillation in stationary vacuum or condensation into a cold trap at reduced pressure. Without distillation, column chromatography of the crude reaction mixture resulted in decomposition and release of red selenium. To separate the mesitylene, employed in
the preparation of bis(1,5-bicyclooctanediylboryl)monoselenide, a Vigreux column setup had to be
used, owing to the volatility of the selenofenchone. Remaining traces of fenchone and mesitylene were
separated via column chromatography. It should be noted that the purified selenofenchone is
moderately air and moisture stable and could be handled under atmospheric conditions without
immediate extrusion of elemental, red selenium. All the synthesized compounds were confirmed
by IR, \textsuperscript{1}H-NMR, \textsuperscript{13}C-NMR spectroscopy and mass spectrometry. The purity of the samples was confirmed by CHN-elemental analysis. In the Appendix, we provide more details on the corresponding instrumentation and spectral characterization.

\subsection{Experimental Methods}
Gas-phase IR absorption spectra of fenchone (\textbf{1-O}), thiofenchone (\textbf{1-S}) and selenofenchone (\textbf{1-Se}) were recorded in the range from \qtyrange[range-phrase=--,range-units=single]{800}{4000}{\per\centi\meter} using an infrared Fourier-Transform (IR-FT) spectrometer (Bruker VERTEX 80) at \SI{0.25}{\per\centi\meter} spectral resolution. A commercially available fenchone sample was used, while thiofenchone and selenofenchone were synthesized as described in Section 3.1. In order to avoid pressure broadening of the spectral lines, the sample pressure was maintained below \SI{100}{Pa}. A sensitive liquid nitrogen-cooled MCT-detector was utilized to record averaged spectra of the samples and of the empty sample cell for background subtraction. However, residual spectral features of gaseous H$_2$O and CO$_2$ persist in the recorded spectra of the samples.  
To demonstrate the feasibility of rotationally resolved vibrational spectra of chalcogenofenchones, a small fraction of the fenchone (1-O) spectrum was recorded around \SI{1023}{\per\centi\meter} with a high-resolution quantum cascade laser (QCL) spectrometer with a resolution of \SI{3e-4}{\per\centi\meter}. In order to reduce the spectral line density of fenchone, the sample was adiabatically cooled in a supersonic jet of helium buffer gas to approximately \SI{30}{K}. Figure \ref{fig:fenchone-highres} shows a small portion of the recorded fenchone spectrum, which consists of dense rovibrational lines of full-width at half maximum (FWHM) of \SI{38}{\mega\hertz} (\SI{0.001}{\per\centi\meter}).

Gas-phase MW spectra of \textbf{1-S} and \textbf{1-Se} were recorded in the range from \qtyrange[range-phrase=--,range-units=single]{2}{8}{\giga\hertz} using the broadband chirped-pulse Fourier transform microwave (FTMW) spectrometer COMPACT. A detailed description of the spectrometer is available elsewhere.\cite{schmitz:2012} To generate sufficient vapor pressure, the samples of \textbf{1-S} and \textbf{1-Se} were heated to \SI{65}{\celsius} and \SI{85}{\celsius}, respectively. Neon was used as a carrier gas to introduce the vapor into the spectrometer via a supersonic expansion at a stagnation pressure of \SI{3}{\bar}, resulting in a rotational temperature ($T_\mathrm{rot}$) of approximately \SI{1}{\kelvin}. For each molecule, one million free induction decays (FIDs) were acquired in the time domain and fast Fourier transformed to obtain the frequency-domain spectrum. The rotational transitions exhibit a FWHM linewidth of approximately \SI{60}{\kilo\hertz}, with a frequency accuracy of \SI{10}{\kilo\hertz}.

\subsection{Computational Methods}

The density functional theory (DFT) calculations were performed using  the GAUSSIAN16 program package.\cite{g16} Becke’s three-parameter
hybrid exchange functional (B3) in combination with the correlation functional of Lee, Yang and Parr
(LYP) was employed in the B3LYP Kohn--Sham DFT calculations. The aug-cc-pVTZ basis set was chosen for all atoms lighter than Te
in all the cases. Relativistic small-core pseudo-potentials were used on Te and Po in combination with the corresponding aug-cc-pVTZ-PP basis sets. The integral evaluation
made use of the grid denoted as “UltraFine”. Criteria for the self-consistent field (SCF) energy convergence were chosen to be \SI{e-8}{\hartree}
and the maximal change of a density matrix element was required to remain on the order of \SI{e-6}{} or below.
Threshold values for maximum gradient components and displacements during the energy minimization of the
structures were chosen to be \SI{4.5e-4}{\hartree\per\bohr} and \SI{1.8e-3}{\bohr} respectively. 
The root-mean-square (RMS) gradient components and displacements were converged to \SI{3e-4}{\hartree\per\bohr} and \SI{1.2e-3}{\bohr}, respectively. Harmonic vibrational wavenumbers including
IR intensities were calculated for the fully optimized structures. Only
real harmonic vibrational wavenumbers were obtained for all structures, confirming the localization of
local minima on the potential energy surfaces. An overall scaling factor of 0.968 has been applied in this work when harmonic vibrational wavenumbers are reported to facilitate direct comparison with experimentally observed fundamentals. This factor corresponds to the rounded value of 0.9676 recommended in Ref.~\onlinecite{sinha:2004} for the use with the present combination of functional and basis set (B3LYP/aug-cc-pVTZ). 

Calculations of anharmonic vibrational wavenumbers were performed on the same level of theory at the equilibrium structures. The
second order vibrational perturbation theory (VPT2) model was applied to anharmonic force fields (full cubic force field together with the semidiagonal part of the quartic force field). It
was used at the default settings whereas the resonances obtained due to the lack of overlap between
variational and deperturbed states were removed by setting relevant modes frozen. For all the frozen
modes, the IR absorption intensity calculated for the corresponding harmonic modes was later used also for the
anharmonic cases, which is indicated in the tables by values given in parentheses. These modes are highlighted in cyan in corresponding plots. IR absorption spectra computed in the double harmonic approximation are displayed
using an overall scaling factor and a Lorentzian lineshape function with half-width at half-maximum (HWHM) of \SI{1.5}{\per\centi\meter}.
Infrared absorption spectra computed within VPT2 to account for anharmonicities are plotted unscaled with the similar lineshape function.

Predictions of electroweak parity-violating effects were performed within a quasi-relativistic (two-component) zeroth-order regular approximation approach to electroweak quantum chemistry \cite{lenthe:1993,Berger:2005a,Berger:2005b}, using a modified version \cite{wullen:2010} of the program package Turbomole \cite{ahlrichs:1989}. Equilibrium structures of the molecules were taken from the DFT calculations described above. We employed an established even tempered basis set for the chalcogens and carbon atoms\cite{Bruck:2023}, while using an aug-cc-pVTZ basis for hydrogen, which allows comparison to earlier calculations with similar approaches \cite{laerdahl:1999,Berger:2005b,Bruck:2023}. A finite nuclear model with Gaussian nuclear density distribution was selected. The value of Fermi's constant was set to $G_\mathrm{F}=\SI{2.22249e-14}{\hartree\bohr^3}$ in this numerical study and the value of the Weinberg parameter was $\sin^2(\theta_\mathrm{w}) \approx 0.2319$.

\section{Results and Discussion}

\noindent The compounds fenchone (\textbf{1-O}), thiofenchone (\textbf{1-S}), selenofenchone (\textbf{1-Se}), tellurofenchone (\textbf{1-Te}) and polonofenchone (\textbf{1-Po}) 
have been investigated thoroughly in this work. Compounds \textbf{1-O}, \textbf{1-S} and \textbf{1-Se} are studied herein both experimentally and computationally, but compounds \textbf{1-Te}, \textbf{1-Po} are only included in the theoretical studies and have to the best of our knowledge not been synthesized so far. The compounds \textbf{1-X} comprise a series of geometrically related molecules with one C-X unit. DFT calculations were performed to study structures and vibrational properties; these properties are not reported so far in a comparably comprehensive manner for this class of compounds. 

\subsection{Structural determination}

Spectral characterization of \textbf{1-S} and \textbf{1-Se} in the liquid or solution phase was already available in the literature to a certain degree and our results for these compounds synthesized in the course of this work are in agreement with those data. The UV/vis absorption spectra for \textbf{1-S} and \textbf{1-Se} in solution are featuring values of $\epsilon$(\SI{488}{\nano\meter}, n-pentane)~=~\SI{10}{\liter\per\mol\per\centi\meter} for \textbf{1-S} [$\epsilon$(\SI{487}{\nano\meter}, benzene)~=~\SI{9}{\liter\per\mol\per\centi\meter},\cite{ramnath:1983} $\epsilon$(\SI{487}{\nano\meter}, cyclohexane)~=~\SI{12}{\liter\per\mol\per\centi\meter}\,\cite{andersen:1982}]
and $\epsilon$(\SI{625}{\nano\meter}, n-hexane)~=~\SI{39}{\liter\per\mol\per\centi\meter} for \textbf{1-Se} [$\epsilon$(\SI{616}{\nano\meter}, cyclohexane)~=~\SI{42}{\liter\per\mol\per\centi\meter},\cite{back:1976} $\epsilon$(\SI{626}{\nano\meter}, cyclohexane)~=~\SI{35}{\liter\per\mol\per\centi\meter}\, \cite{andersen:1982}]. The chiral nature of the enantiomers of \textbf{1-S} and \textbf{1-Se} was corroborated by CD spectroscopy in hexane solution. Details of the assignment and discussion of the Cotton-effects is available in the literature.\cite{wijekoon:1983} 
The assignment of the NMR data required some additional effort owing to minor deviations from published values. For the \textsuperscript{1}H-NMR spectral analysis, assignment of the diastereotopic methylene protons was refined using 1D and 2D NOE-experiments including the respective proton couplings (see Appendix). During assignment of the \textsuperscript{13}C-NMR chemical shifts, we stumbled on a discrepancy in case of selenofenchone. In previous publications \cite{wijekoon:1983,cullen:1982} the three most shielded \textsuperscript{13}C-NMR chemical shifts were assigned to the methyl groups similar to the assignment in fenchone (see Table~\ref{tab:nmr-data}). However, according to 2D-HSQC-NMR measurements, the chemical shifts of C5, C8 and C9 for selenofenchone have to be re-assigned. Based on these 2D-measurements, the chemical shift of the methylene carbon atom C5 (\SI{25.2}{ppm}) in fact resonates in between the shifts of the methyl carbon atoms C10 (\SI{21.0}{ppm}) and C9 (\SI{26.7}{ppm}), while the resonance of methyl carbon atom C8 is observed at \SI{27.8}{ppm} at lowest field. This assignment fits nicely the trend observed in thiofenchone, where the resonance of the methylene carbon atom is found at values between the methyl carbon atoms as well. 

\begin{table}[h]
  \caption{\textsuperscript{13}C-NMR (\ce{CDCl3}) shifts of fenchone, thiofenchone and selenofenchone (in ppm).\label{tab:nmr-data}}  
  \begin{tabular}{lrrrc}
  \hline
   {Position} & {\textbf{1-O}} & {\textbf{1-S}} & {\textbf{1-Se}} & Numbering Scheme \\\hline
   C1 & 53.8 & 66.3 & 72.9 & \multirow{10}{7cm}{\includegraphics[width=6.5cm]{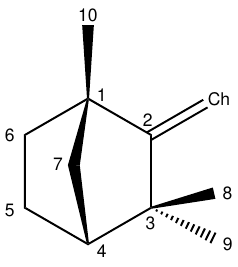}}\\
   C2 &222.6 &280.5 &293.4 \\
   C3 & 43.1 & 57.7 & 63.7 \\
   C4 & 45.3 & 46.9 & 47.5 \\
   C5 & 24.9 & 25.0 & 25.2 \\
   C6 & 31.8 & 35.4 & 33.1 \\
   C7 & 41.5 & 43.7 & 42.8 \\
   C8 & 23.3 & 28.6 & 27.8 \\
   C9 & 21.6 & 26.4 & 26.7 \\
  C10 & 14.6 & 19.1 & 21.0 \\\hline
  \end{tabular}  
\end{table}

\begin{figure}
\centering
\includegraphics[width=0.50\linewidth]{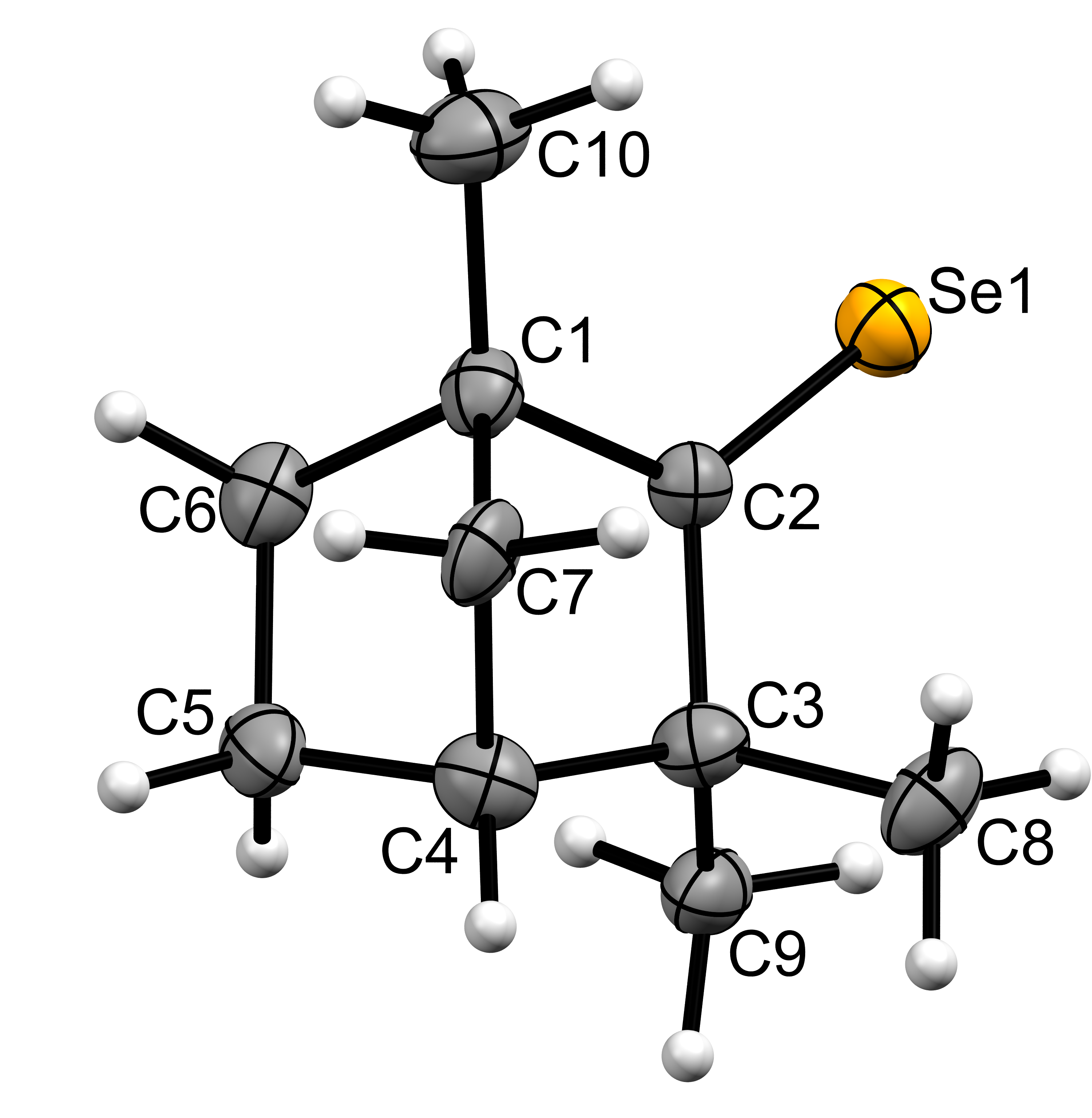}
\caption{Molecular structure of one of the three independent molecules of (1$S$,$4$R)-configured \textbf{1-Se} in the solid state with probability ellipsoids drawn at 30\,\% probability level.\label{fig:Se-cryst}}
\end{figure}

The crystal structure of \textbf{1-O} was reported for instance in Ref.~\onlinecite{bond:2001} (see also Table~\ref{tab:structures}). For \textbf{1-Se}, we are not aware of a previously reported crystal structure. In the present work, the compound was found to crystallize in the non-centrosymmetric space group P2$_1$2$_1$2$_1$ with three independent molecules in the asymmetric unit. The absolute configuration was established by a compound containing a chiral reference molecule (starting material (1$S$,4$R$)-\textbf{1-O}) of known absolute configuration. In addition, the Flack parameter with a value of 0 within the standard deviation (0.02(6)) corroborates the stereochemical assignment. Due to poor crystal quality ($R$(int) = 0.1163), the connectivity of the atoms can be deduced, (see Figure~\ref{fig:Se-cryst}) but only the selenium-carbon bond length will be discussed. The structure shows no peculiarities with the selenium-carbon bond length in the range between 1.755(17)--1.808(17)~\AA, which are comparable to literature values (1.773~\AA) \cite{brooks:1991} as well as values obtained in the present work from DFT calculations.

\begin{figure}[!htb]
\centering
\includegraphics[width=\linewidth]{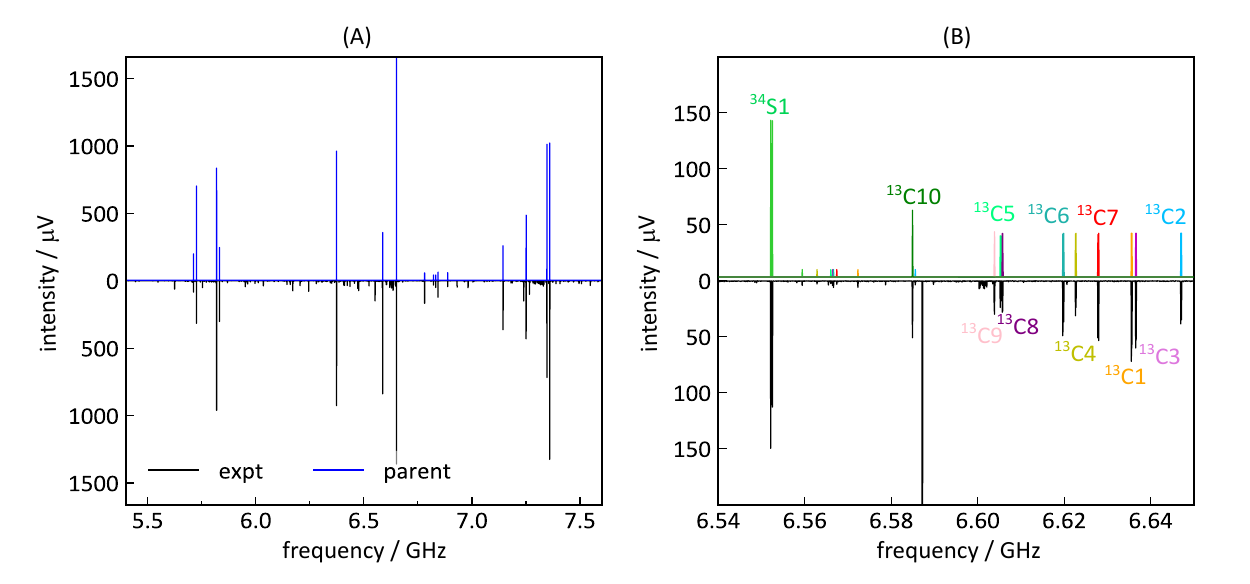}
\caption{(A) Portion of the microwave spectrum of \textbf{1-S}, obtained with an average of 1$\times 10^{6}$ FID acquisitions. The lower trace (black) shows the experimental spectrum, while the upper trace (blue) presents the simulated spectrum of the parent species, calculated using the spectroscopically fitted parameters at a rotational temperature of 1 K. (B) Zoomed-in view highlighting the transition doublets of $J'_{{K_a}'{K_c}'}-J''_{{K_a}''{K_c}''} =$ 4$_{1,4}$--3$_{1,3}$ and 4$_{0,4}$--3$_{0,3}$, arising from one $^{34}$S and ten $^{13}$C singly substituted isotopologues of \textbf{1-S} in natural abundance. \label{fig:1-S_MW_spectra}}
\end{figure}

\begin{figure}[!htb]
\centering
\includegraphics[width=\linewidth]{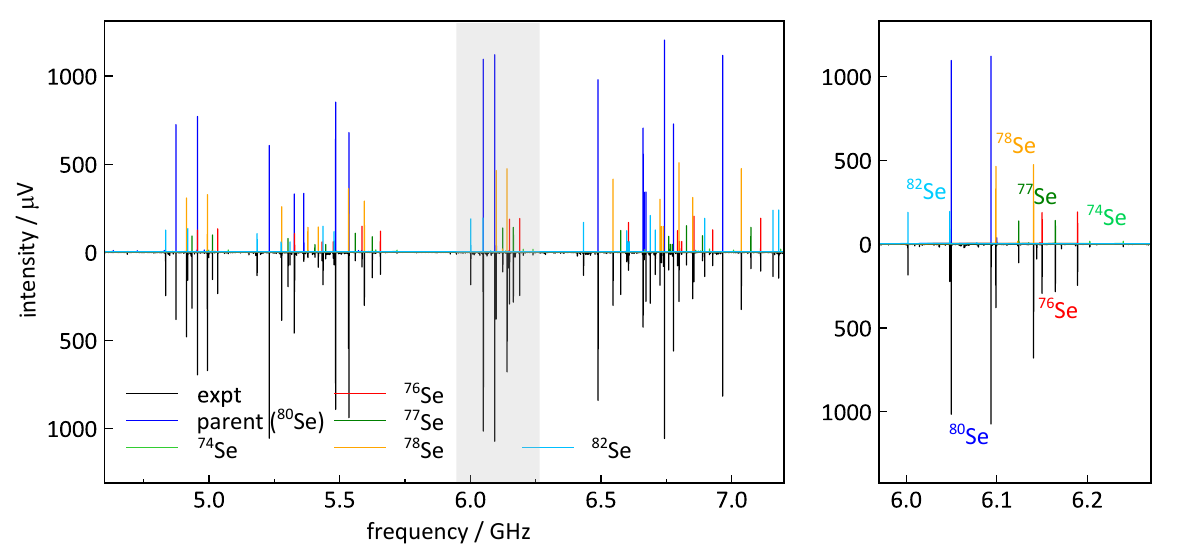}
\caption{Portion of the microwave spectrum of \textbf{1-Se}, obtained with an average of 1$\times 10^{6}$ FID acquisitions. The lower trace (black) shows the experimental spectrum, while the upper trace presents the simulated spectra of the six singly substituted Se isotopologues ($^{74}$Se, $^{76}$Se, $^{77}$Se, $^{78}$Se, $^{80}$Se and $^{82}$Se) of \textbf{1-Se}, calculated using the spectroscopically fitted parameters and their natural abundances at a rotational temperature of 1 K. The right panel highlights the transition doublets of $J'_{{K_a}'{K_c}'}-J''_{{K_a}''{K_c}''} =$ 5$_{1,5}$--4$_{1,4}$ and 5$_{0,5}$--4$_{0,4}$.}
\label{fig:1-Se_MW_spectra}
\end{figure}

Besides X-ray diffraction studies to obtain the crystal structure of \textbf{1-Se}, also ground state microwave spectroscopic studies of \textbf{1-S} and \textbf{1-Se} in the gas phase were conducted in this work. In the recorded rotational spectra, the signal-to-noise ratios were sufficient to detect the \ce{^13C} and \ce{^34S} monosubstituted isotopologues of \textbf{1-S}, and the \ce{^13C}, \ce{^74Se}, \ce{^76Se}, \ce{^77Se}, \ce{^78Se}, and \ce{^82Se} monosubstituted isotopologues of \textbf{1-Se} at their natural abundances, in addition to the normal species that is composed of the most abundant isotope of each element. Figures~\ref{fig:1-S_MW_spectra} and \ref{fig:1-Se_MW_spectra} show portions of the corresponding MW spectra of \textbf{1-S} and \textbf{1-Se}, respectively. The rotational transitions observed  were assigned and fitted with Watson’s $A$-reduced Hamiltonian in the $III^\mathrm{r}$ representation for \textbf{1-S} and the $I^\mathrm{r}$ representation for \textbf{1-Se}, employing Pickett’s SPFIT program.\cite{pickett:1991,kisiel} Based on the fitted rotational constants (see Tables~\ref{tab:1-S_RC} and \ref{tab:1-Se_RC} in the appendix), the effective $r_0$ structure (see Tables~\ref{tab:1-S,Se_r}--\ref{tab:1-S,Se_tau} in the appendix for details and Table~\ref{tab:structures} below for a selection of parameters) of the heavy-atom backbone was well determined through the STRFIT routine, which performs a nonlinear least-squares fit of internal coordinates to the experimental moments of inertia.\cite{kisiel,kisiel:2003} The three isotopologues with \ce{^74Se}, \ce{^76Se} and \ce{^77Se} were excluded from this fit, whereas all hydrogen-related structural parameters were kept frozen at their input values during this fitting procedure.

Input structural parameters for this analysis of the microwave spectra as well as structural parameters for the subsequent theoretical vibrational study were obtained from quantum chemical energy minimizations of the molecular arrangements. Selected geometrical parameters for \textbf{1-O} to \textbf{1-Po} in the gas phase are presented in Table~\ref{tab:structures}. Bond lengths ($r$ in Å) and angles ($\alpha$ in degrees) around the C=X unit of the chalcogenofenchones determined i) experimentally in the condensed phase via X-ray diffraction in Ref.~\onlinecite{bond:2001} and in the gas phase via microwave spectroscopy in Ref.~\onlinecite{loru:2016} for \textbf{1-O}, ii) experimentally by gas-phase MW spectroscopy for \textbf{1-S} and \textbf{1-Se} in the present work as well as iii) theoretically for the series of chalcogenofenchones \textbf{1-O}--\textbf{1-Po} in the present work are compared. 

\begin{table}
\caption{Experimental (X-ray diffraction for \textbf{1-O} from Ref.~\onlinecite{bond:2001},  microwave (MW, $r_0$) for \textbf{1-O} from Ref.~\onlinecite{loru:2016} and MW ($r_0$) for \textbf{1-S}, \textbf{1-Se} from this work), as well as calculated (DFT, $r_\mathrm{e}$) geometric parameters of fenchone and its heavier homologues. Bond lengths $r$ in {\AA} and bond angles $\alpha$ in degrees.\label{tab:structures}}
\begin{tabular}{ll%
                 S[table-format=1.4(2),round-mode=figures,round-precision=4]%
                 S[table-format=1.4(2),round-mode=figures,round-precision=4]%
                 S[table-format=1.4(2),round-mode=figures,round-precision=4]%
                 S[table-format=3.2(2),round-mode=figures,round-precision=4]%
                 S[table-format=3.2(2),round-mode=figures,round-precision=4]%
                 S[table-format=3.2(2),round-mode=figures,round-precision=4]}
\hline
Mol.           & {Method} 
                        & {$r$(C1-C2)} 
                                       & {$r$(C2-C3)} 
                                                      & {$r$(C2-X)} 
                                                                  & {$\alpha$(C1-C2-X)} 
                                                                              & {$\alpha$(X-C2-C3)} 
                                                                                          & {$\alpha$(C1-C2-C3)} \\
\hline
\textbf{1-O}   & {X-ray\footnotemark} 
                        & 1.517(5)     & 1.526(5)     & 1.215(4)  & 126.6(3)  & 125.8(3)  & 107.6(3)  \\
               & {MW\footnotemark}
                        & 1.526(29)    & 1.535(31)    & 1.214(5)  &           & 125.8(31) & 107.5(12) \\
               & {DFT}  & 1.528933     & 1.545207     & 1.206635  & 126.7084  & 125.6991  & 107.5898  \\
\textbf{1-S}   & {MW}   & 1.5059(43) & 1.5226(30) & 1.6341(30) 
                                                                  & 126.60(16) & 125.52(20) & 107.87(22)\\
               & {DFT}  & 1.517866     & 1.537909     & 1.623510  & 126.9622  & 125.9604  & 107.0740  \\
\textbf{1-Se}  & {MW}   & 1.505(15)& 1.524(13)& 1.773(21)
                                                                  & 126.61(73) & 125.55(71) & 107.8(13)\\
               & {DFT}  & 1.512433     & 1.532989     & 1.777177  & 126.8086  & 125.8375  & 107.3495  \\
\textbf{1-Te}  & {DFT}  & 1.508282     & 1.527270     & 1.967111  & 127.0372  & 125.8209  & 107.1125  \\
\textbf{1-Po}  & {DFT}  & 1.507493     & 1.528807     & 2.101277  & 126.7925  & 125.8064  & 107.3966  \\\hline
\end{tabular}
\footnotetext[1]{From Ref.~\onlinecite{bond:2001}.}
\footnotetext[2]{From Ref.~\onlinecite{loru:2016}.}
\end{table}

The calculated overall geometric parameters for fenchone corroborate well with the corresponding experimental values: All reported parameters for \textbf{1-O} show little or no deviation, with the maximum absolute deviation between bond lengths being \SI{0.019}{\angstrom}, whereas bond angles agree within the stated standard deviations. The reported experimental \cite{bond:2001,loru:2016} and calculated C=O internuclear distances match well (deviation of about \SI{0.008}{\angstrom}), which is very useful for spectroscopic studies. 

The deviation between experimental and computed C=X bond length in \textbf{1-S} is of similar magnitude (\SI{0.01}{\angstrom}) as for \textbf{1-O}, whereas the C=Se bond length in \textbf{1-Se} from the MW study and the DFT calculations agrees within the experimental standard deviation. 
Furthermore, inspection of computed structures in the series from \textbf{1-O} to \textbf{1-Po} indicates that the C-X internuclear distance grows by $\approx$~\SI{0.89}{\angstrom} from \textbf{1-O} to \textbf{1-Po}, consistent with the increase of the atomic radii within the series of chalcogens. This consecutive enlargement can drive interesting features in the vibrations which are discussed in the next subsection. 
Both measured and computed C-X bond lengths for the chalcogenofenchones can be compared to bond lengths expected by virtue of Pyykk{\"o}'s compilation of covalent radii for the various elements in single, double and triple bonded situations \cite{pyykko:2015}. From this data set, one would anticipate C-X double bond lengths (X=O, S, Se, Te and Po) of \qtylist[list-units=bracket]{1.24;1.61;1.74;1.95;2.02}{\angstrom}, respectively. Overall, these appear to be reasonably well reproduced with absolute deviations being smallest for \textbf{1-S} and \textbf{1-Te} (below \SI{0.02}{\angstrom}) followed by \textbf{1-O} and \textbf{1-Se} (below \SI{0.04}{\angstrom}). Only for the parent fenchone, the measured and computed C-X bond length is shorter than the sum of the corresponding covalent radii for double bonds, in the other cases the bond length is (slightly) longer. A comparatively large deviation, however, is found for \textbf{1-Po}, which is with \SI{2.101}{\angstrom} computed to be about halfway between the sum of covalent radii for a C-Po double bond and a C-Po single bond. 

The two C-C bonds adjacent to the C=X moiety show only a marginal length decrease of $\approx$~\SI{-0.02}{\angstrom} in the chalcogenofenchone series \textbf{1-O} to \textbf{1-Po}, confirming the rigidity of the bicyclic ring system. Also the reported bond angles just change by one degree or even less with respect to \textbf{1-O} when substituting oxygen by heavy-elemental chalcogens (see also Table~\ref{tab:1-S,Se_alpha} in the appendix), with the important bond angles around the C=X bond deviating negligibly (less than \SI{0.9}{\degree}) when comparing experimental values and quantum chemically computed ones. 

Structural inspection indicates that main differences in vibrational features are to be expected for modes in which motions of the C=X moiety participate. Most vibrations involving the C-H bond framework and the skeleton, on the other hand, should not get affected much upon moving from \textbf{1-O} to \textbf{1-Po} and should thus cause roughly similar features.

\subsection{Infrared spectral features}

Due to the presence of 27 atoms for each of the compounds, the number of fundamental vibrational transitions is 75. Given the molecular structure of the chalcogenofenchones, we can expect among these fundamentals---besides vibrations involving the whole skeleton---in particular the following characteristic transitions\cite{herzberg:1991,harris:1989,hesse:2008,williams:2019} in this compound family : i) strong methyl (\ce{CH3}) asymmetric C-H stretching vibrations around \SI{2960}{\per\centi\meter}, ii) strong methylene (\ce{CH2}) asymmetric C-H stretching modes near \SI{2925}{\per\centi\meter}, iii) a weak methine (\ce{CH}) C-H stretching transition at about \SI{2885}{\per\centi\meter}, iv) strong \ce{CH3} symmetric C-H stretching modes at about \SI{2870}{\per\centi\meter}, v) strong \ce{CH2} symmetric C-H stretching fundamentals around \SI{2855}{\per\centi\meter}, vi) several methyl asymmetric C-H bending modes and methylene scissoring modes of medium strength in the approximate range \qtyrange[range-phrase=--,range-units=single]{1480}{1430}{\per\centi\meter}, vii) a characteristic doublet caused by the coupled \ce{CH3} umbrella modes from the geminal dimethyl substituents at C3 near \SI{1380}{\per\centi\meter}, viii) a signal of medium strength from the \ce{CH3} umbrella mode of the methyl group attached to C1 in the range \qtyrange[range-phrase=--,range-units=single]{1390}{1370}{\per\centi\meter}, ix) various methylene wagging and twisting fundamentals in the approximate range \qtyrange[range-phrase=--,range-units=single]{1350}{1150}{\per\centi\meter} and finally x) great variations due to alterations in the C=X moiety, which impact in particular on the valence C=X stretching fundamental that is expected in \textbf{1-O} to yield a strong signal above \SI{1700}{\per\centi\meter}. 

As evident from \textbf{1-O}, owing to the good concordances in
geometrical parameters as compared to the corresponding experimental ones, an overall good correlation in the
vibrational spectra could be made in this work. Experimental (\textbf{1-O}--\textbf{1-Se}) and predicted (\textbf{1-O}--\textbf{1-Po}) gas-phase IR
spectra are presented in Figures \ref{fig:Ovib}--\ref{fig:Povib}. 
The positions of experimental and theoretical bands align well
with each other. Discussion of some important assignments is presented below.

\subsubsection{Assignments by comparing IR spectra computed in the double harmonic approximation to experimental gas-phase spectra}

It is noteworthy that we scaled the harmonic vibrational wavenumbers in all the reported cases by a common factor of 0.968. The calculated spectra are shown with Lorentzian line shapes convoluting the underlying computed signals that are additionally displayed in stick representation. In the recorded spectral range
(\qtyrange[range-phrase=--,range-units=single]{4000}{800}{\per\centi\meter}) we observe for the reported cases \textbf{1-O}, \textbf{1-S} and \textbf{1-Se} in comparison to the measured spectra that (i) the overall profile seems to resemble the experimental one well and (ii) the intensity pattern is also fairly well reproduced, which shows in general
further improvement when anharmonic effects are accounted for, with more details given below. The selected
vibrational modes were characterized in all cases by visual inspection of the atomic displacements along normal modes and for the fenchone case also by comparison with assignments reported in the literature \cite{devlin:1996,devlin:1997,longhi:2006}.

\paragraph{Fenchone (\textbf{1-O})}

\begin{figure}[htbp]
\centering
  \includegraphics[width=1.05\textwidth]{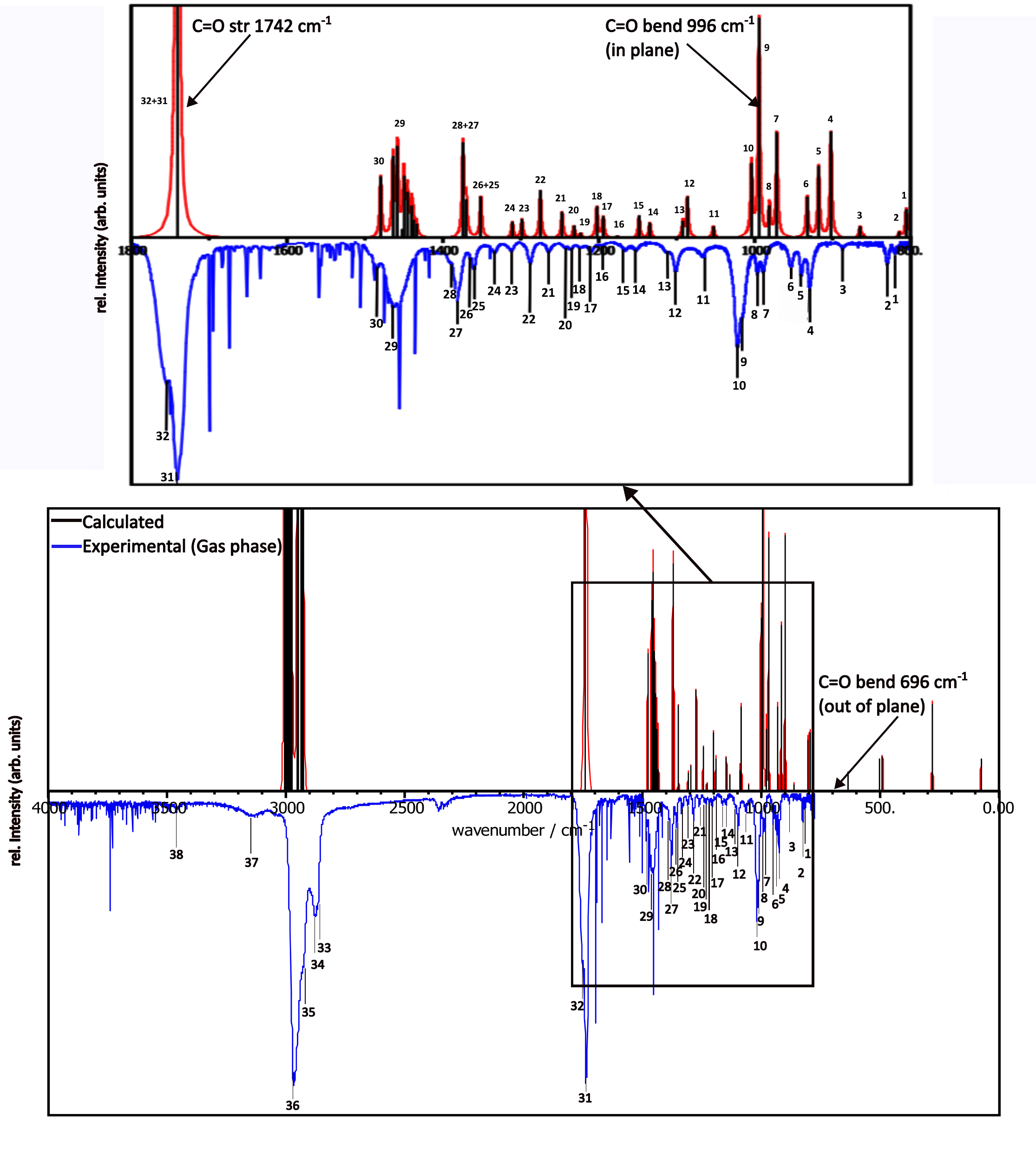}
  \caption{Combined theoretical (red) and low-resolution experimental gas-phase vibrational overview spectra (blue) of \textbf{1-O}. The 
  theoretical spectrum was computed on the DFT level within the double-harmonic approximation and is displayed using an overall scaling factor of the harmonic vibrational wavenumbers of 0.968 and a Lorentzian lineshape function with HWHM of \SI{1.5}{\per\centi\meter}. The underlying theoretical stick spectrum is indicated in black. \label{fig:Ovib}}
\end{figure}

The experimentally obtained IR spectrum and the corresponding calculated spectrum of \textbf{1-O} are shown in
Figure~\ref{fig:Ovib}. A peak listing is provided in Table~\ref{tab:Ofundamentals}. The intense band of the carbonyl stretching fudamental computed at \SI{1741}{\per\centi\meter} is identified clearly (peak number 32 at \SI{1742}{\per\centi\meter}, first overtone: peak 38 at \SI{3462}{\per\centi\meter}). Another strong band corresponding to in-plane C=O bending is predicted at \SI{995}{\per\centi\meter} and tentatively assigned to peak number 9 in the experimental spectrum at \SI{1017}{\per\centi\meter}, but also peak number 10 at \SI{1023}{\per\centi\meter} could alternatively be attributed to this transition as per its higher intensity. As evident from Figure~\ref{fig:Ovib}, the predicted bands appear close to the observed ones. Another
band at \SI{696}{\per\centi\meter} with marginal intensity is predicted for the C=O out-of-plane vibration. Fenchone
has nine nearly degenerate \ce{CH3} and \ce{CH2} asymmetric stretching modes (C8, C9, C10 and C5, C6, C7), three \ce{CH3} symmetric stretching modes (C8, C9, C10), three \ce{CH2} symmetric stretching modes (C5, C6, C7) and one \ce{CH} stretching mode (C4) which are predicted at wavenumbers 3012--2979, 2934--2930, 2955--2942 and \SI{2976}{\per\centi\meter}, respectively.
Antisymmetric and symmetric deformation, wagging and rocking modes of the \ce{CH3} and \ce{CH2} groups
along with ring torsions are predicted between \qtylist[list-units=bracket]{1483;1006}{\per\centi\meter}. The doublet arising from the coupled \ce{CH3} umbrella modes of the geminal dimethyl unit is predicted to resonate at \qtylist[list-units=bracket]{1376;1353}{\per\centi\meter} and to envelope the umbrella mode of the remaining \ce{CH3} group at \SI{1372}{\per\centi\meter} according to the DFT calculations, whereas the corresponding experimentally observed transitions arise at only slightly higher wavenumbers (peaks 28-25). The fourth signal observed in this region is possibly caused by a combination band (see section 3.3.2 below). 
It is worth mentioning that for fenchone, the calculated harmonic wavenumbers (after scaling with a common factor for all fenchone derivatives) correlate from fairly well to extremely well with the experimental values. For instance the C=O stretching wavenumber deviates by less than \SI{1}{\per\centi\meter}.
Moreover, the resonance wavenumbers obtained here from our experimental gas-phase IR study of \textbf{1-O} agree for most bands in the range between \qtylist[list-units=bracket]{1400;800}{\per\centi\meter} to within \SI{2}{\per\centi\meter} with previous IR spectra of Refs.~\onlinecite{devlin:1996,devlin:1997} recorded for a dilute solution of \textbf{1-O} dissolved in tetracloromethane (\ce{CCl4}) and carbondisulfide (\ce{CS2}). This indicates that solvent effect in the chosen non-polar media appear to play only a minor role in this spectral region. Larger deviations are observed only for peaks number 9, 26 and 28, for the latter of which no correspondence is reported in the mentioned earlier works. For a solution phase investigation and analysis of \textbf{1-O} in \ce{CCl4}, which extended also to the C-H stretching fundamental and selected overtone regions in a combined IR absorption and vibrational circular dichroism study, we mention here also Refs.~\onlinecite{longhi:2004,longhi:2006}.

\begin{table}[htp]
\footnotesize
\caption{Calculated harmonic (scaled by 0.968) and anharmonic wavenumbers $\Tilde{\nu}$ (\SI{}{\per\centi\meter}), integrated absorption coefficients $A$ (\SI{}{\kilo\meter\per\mol}), experimental transition wavenumbers, intensities (in arbitrary units, a.u.) and assignments of selected bands (peak number (\#) in Figure~\ref{fig:Ovib}) of \textbf{1-O}.\label{tab:Ofundamentals}}
\begin{tabular}{cccccccc}
\hline
Mode & \multicolumn{2}{c}{Harmonic}  & \multicolumn{2}{c}{Anharmonic} & \multicolumn{3}{c}{Experimental} \\ 
\ & $\Tilde{\nu}$   & $A$ & $\Tilde{\nu}$   & $A$ & \# & $\Tilde{\nu}$   & $I_\mathrm{exp}$ \\
\ &  (cm$^{-1}$)  & (\SI{}{\kilo\meter\per\mol}) & \ (cm$^{-1}$) & (\SI{}{\kilo\meter\per\mol}) &   &  (cm$^{-1}$) & (a.u.)  \\
\hline
2$\nu_{17}$&        &       &  3522.3 &  3.0  & 	\hspace{0.3cm} 38   & 3462.44 & 0.024 \\
&	 & 	&	& &	\hspace{0.3cm} 37	& 3143.4 & 0.033 \\
$\nu_{ 1}$ & 3011.5 &  20.0 &  2913.0 &  12.4 & \rdelim\}{16}{0.3cm}[] \multirow{3}{*}{\,\,\,\,\,}& &  \\
$\nu_{ 2}$ & 3008.6 &  35.6 &  2941.2 &  29.8 & & &\\
$\nu_{ 3}$ & 3002.6 &  36.4 &  3021.9 &  24.5 & & &\\
$\nu_{ 4}$ & 2996.5 &  31.1 &  2923.7 &   7.8 & & &\\
$\nu_{ 5}$ & 2994.8 &  17.2 &  2967.1 &  41.3 & & &\\
$\nu_{ 6}$ & 2993.1 &  16.6 &  2925.5 &  16.2 & & &\\
$\nu_{ 7}$ & 2989.3 &  16.9 &  2909.6 &   6.0 & \hspace{0.3cm} 36 & 2967.88 & 0.436\\
$\nu_{ 8}$ & 2987.5 &  13.4 &  3049.0 &  32.7 & \hspace{0.3cm} 35 & 2934.82 & 0.272\\
$\nu_{ 9}$ & 2979.3 &  31.8 &  2983.9 & 185.4 & \hspace{0.3cm} 34 & 2881.58 & 0.186\\
$\nu_{10}$ & 2976.2 &  25.7 &  2950.0 &  71.4 & \hspace{0.3cm} 33 & 2870.58 & 0.176\\
$\nu_{11}$ & 2955.1 &  36.4 &  2927.2 &  23.4 & & &\\
$\nu_{12}$ & 2948.1 &  27.9 &  2916.3 &  46.7 & & &\\
$\nu_{13}$ & 2942.5 &  30.1 &  2893.4 &  34.0 & & &\\
$\nu_{14}$ & 2934.7 &  32.2 &  2899.1 &  18.4 & & &\\
$\nu_{15}$ & 2930.7 &  15.7 &  2875.9 &   8.3 & & &\\
$\nu_{16}$ & 2930.1 &  27.0 &  2938.6 &   4.0 & & &\\
\multirow{2}{*}{$\nu_{17}$} & \multirow{2}{*}{1741.9} & \multirow{2}{*}{209.8} &  \multirow{2}{*}{1771.8} &  \multirow{2}{*}{43.6} & \rdelim\}{2}{0.3cm}[] 31 & 1756.46 & 0.258 \\
 			  & 	   & 	   &		 &			&	\hspace{0.3cm} 32	& 1741.83 & 0.441\\
$\nu_{18}$ & 1483.2 &   6.5 &  1490.0 &   4.7 & \hspace{0.3cm}30 & 1485.78 & 0.051\\
$\nu_{19}$ & 1466.9 &   9.0 &  1515.4 &  (9.0)& \rdelim\}{4}{0.3cm}[] \multirow{4}{*}{29} & \multirow{4}{*}{1463.56} &\multirow{4}{*}{0.121} \\
$\nu_{20}$ & 1461.3 &  10.0 &  1509.6 & (10.0)& & &\\
$\nu_{21}$ & 1454.5 &   0.7 &  1462.2 &   0.3 & & &\\
$\nu_{22}$ & 1453.1 &   7.5 &  1460.0 &   0.9 &  &  &\\

\hline
\end{tabular}
\end{table}

\begin{table}[htp]
\footnotesize
\begin{tabular}{cccccccc}
\hline
Mode & \multicolumn{2}{c}{Harmonic}  & \multicolumn{2}{c}{Anharmonic} & \multicolumn{3}{c}{Experimental} \\ 
\ & $\Tilde{\nu}$   & $A$ & $\Tilde{\nu}$   & $A$ & \# & $\Tilde{\nu}$   & $I_\mathrm{exp}$ \\
\ &  (cm$^{-1}$)  & (\SI{}{\kilo\meter\per\mol}) & \ (cm$^{-1}$) & (\SI{}{\kilo\meter\per\mol}) &   &  (cm$^{-1}$) & (a.u.)  \\
\hline

$\nu_{23}$ & 1448.0 &   5.4 &  1456.1 &   3.0 & \rdelim\}{4}{0.3cm}[] \multirow{4}{*}{29} & \multirow{4}{*}{1463.56} &\multirow{4}{*}{0.121}\\
$\nu_{24}$ & 1442.5 &   3.2 &  1490.2 &  (3.2)& & &\\
$\nu_{25}$ & 1441.5 &   2.2 & 1451.9 &   1.2 & & &\\
$\nu_{26}$ & 1436.0 &   1.6 & 1445.8 &   0.0 & & &\\
$\nu_{27}$ & 1376.1 &  10.5 & 1394.4 &   8.7 & \hspace{0.3cm}28 & 1391.07 & 0.038 \\
$\nu_{28}$ & 1372.2 &   4.3 & 1384.6 &   2.7 & \hspace{0.3cm}27 & 1382.76 & 0.111 \\
\multirow{2}{*}{$\nu_{29}$} & \multirow{2}{*}{1353.0} &   \multirow{2}{*}{4.6} & \multirow{2}{*}{1368.9} &   \multirow{2}{*}{1.0} & \rdelim\}{2}{0.3cm}[] 26  & 1365.37 & 0.033 \\
 			  & 	   & 	   &		 &			&	\hspace{0.3cm} 25	& 1361.01 & 0.051\\
$\nu_{30}$ & 1312.1 &   1.8 & 1326.9 &   0.3 & \hspace{0.3cm}24 & 1334.78 & 0.020 \\
$\nu_{31}$ & 1299.9 &   2.0 & 1317.2 &   0.6 & \hspace{0.3cm}23 & 1313.1  & 0.016 \\
$\nu_{32}$ & 1276.4 &   5.2 & 1296.8 &   0.6 & \hspace{0.3cm}22 & 1289.21 & 0.039\\
$\nu_{33}$ & 1248.3 &   2.8 & 1257.6 &   2.1 & \hspace{0.3cm}21 & 1265.35 & 0.018 \\
$\nu_{34}$ & 1233.2 &   1.3 & 1251.0 &   0.8 & \rdelim\}{5}{0.4cm}[] \multirow{5}{*}{ 18,} & 1244.16 & 0.010 \\
$\nu_{35}$ & 1224.4 &   0.5 & 1240.8 &   0.2 & \hspace{0.3cm}  20, 19, & 1237.2  & 0.015 \\
$\nu_{36}$ & 1203.1 &   3.5 & 1213.0 &   0.0 &   & 1224.01 & 0.011 \\
$\nu_{37}$ & 1195.4 &   2.3 & 1206.3 &   1.4 & \hspace{0.3cm} 17, 16 & 1210.92 & 0.018 \\
$\nu_{38}$ & 1177.6 &   0.3 & 1188.0 &   0.0 &   & 1196.57 & 0.005 \\
$\nu_{39}$ & 1149.5 &   2.6 & 1156.1 &   0.4 & \hspace{0.3cm}15 & 1166.98& 0.015 \\
$\nu_{40}$ & 1136.0 &   1.7 & 1145.7 &   0.8 & \hspace{0.3cm}14 & 1153.64& 0.015 \\
$\nu_{41}$ & 1093.0 &   1.9 & 1106.1 &   2.9 & \hspace{0.3cm} 13 & 1114   & 0.021 \\
$\nu_{42}$ & 1088.0 &   4.5 & 1104.2 &   2.8 & \hspace{0.3cm}12 & 1102.09& 0.054 \\
$\nu_{43}$ & 1055.0 &   1.4 & 1069.2 &   0.2 & \hspace{0.3cm}11 & 1068.27& 0.026 \\
$\nu_{44}$ & 1006.1 &   7.6 & 1019.7 &   5.5 & \hspace{0.3cm}10 & 1023.17& 0.196 \\
$\nu_{45}$ &  996.0 &  23.3 & 1008.0 &  19.0 & \hspace{0.3cm}9  & 1017.39 &0.143  \\
$\nu_{46}$ &  983.3 &   4.0 &  990.1 &   2.3 & \hspace{0.3cm}8  & 997.39 & 0.058 \\
$\nu_{47}$ &  972.6 &  13.4 &  981.6 &  16.6 & \hspace{0.3cm}7  & 989.96 & 0.058 \\
$\nu_{48}$ &  933.0 &   4.8 &  948.3 &   2.2 & \hspace{0.3cm}6  & 954.4  & 0.046 \\
\hline
\end{tabular}
\end{table}

\begin{table}[htp]
\footnotesize
\begin{tabular}{cccccccc}
\hline
Mode & \multicolumn{2}{c}{Harmonic}  & \multicolumn{2}{c}{Anharmonic} & \multicolumn{3}{c}{Experimental} \\ 
\ & $\Tilde{\nu}$   & $A$ & $\Tilde{\nu}$   & $A$ & \# & $\Tilde{\nu}$   & $I_\mathrm{exp}$ \\
\ &  (cm$^{-1}$)  & (\SI{}{\kilo\meter\per\mol}) & \ (cm$^{-1}$) & (\SI{}{\kilo\meter\per\mol}) &   &  (cm$^{-1}$) & (a.u.)  \\
\hline
$\nu_{49}$ &  929.5 &   0.4 &  936.2 &   3.0 & & &  \\
$\nu_{50}$ &  918.6 &   7.7 &  929.6 &  11.9 & 5 & 940.97 & 0.063 \\
$\nu_{51}$ &  902.6 &  12.2 &  906.2 &   3.4 & 4 & 930.1 & 0.087 \\
$\nu_{52}$ &  864.8 &   1.2 &  874.0 &   0.6 & 3 & 887.79 & 0.016 \\
$\nu_{53}$ &  860.6 &   0.3 &  869.2 &   0.7 & & &  \\
$\nu_{54}$ &  815.4 &   0.6 &  830.4 &   1.2 & & &  \\
$\nu_{55}$ &  806.4 &   3.4 &  814.8 &   3.1 & 2 & 830.71 & 0.043 \\
$\nu_{56}$ &  800.0 &   3.5 &  811.0 &   4.5 & 1 & 820.87 &  0.022\\
$\nu_{57}$ &  759.2 &   0.4 &  769.4 &   0.3 & & &\\
$\nu_{58}$ &  695.7 &   0.7 &  706.6 &   0.6 & & &\\
$\nu_{59}$ &  637.9 &   1.8 &  646.7 &   1.8 & & &\\
$\nu_{60}$ &  576.2 &   0.6 &  588.3 &   0.5 & & &\\
$\nu_{61}$ &  505.5 &   2.4 &  520.5 &   1.3 & & &\\
$\nu_{62}$ &  493.6 &   2.4 &  508.3 &   2.1 & & &\\
$\nu_{63}$ &  432.9 &   0.5 &  446.6 &   0.2 & & &\\
$\nu_{64}$ &  410.2 &   0.7 &  420.2 &   0.4 & & &\\
$\nu_{65}$ &  356.9 &   0.4 &  360.1 &   0.5 & & &\\
$\nu_{66}$ &  315.8 &   0.1 &  324.8 &   0.1 & & &\\
$\nu_{67}$ &  289.3 &   0.9 &  295.6 &   0.2 & & &\\
$\nu_{68}$ &  283.5 &   4.7 &  300.5 &   5.1 & & &\\
$\nu_{69}$ &  251.9 &   0.3 &  261.3 &   0.5 & & &\\
$\nu_{70}$ &  221.1 &   0.8 &  160.2 &   0.2 & & &\\
$\nu_{71}$ &  203.1 &   0.7 &  222.7 &   1.5 & & &\\
$\nu_{72}$ &  193.4 &   0.7 &  175.1 &   0.2 & & &\\
$\nu_{73}$ &  184.1 &   0.0 &  283.8 &   1.3 & & & \\
$\nu_{74}$ &  176.4 &   1.0 &  104.1 &   0.1 & & &\\
$\nu_{75}$ &   83.3 &   2.4 &   96.2 &   2.4 & & &\\
\hline
\end{tabular}
\end{table}

%------------------------------------------------------------------------------------------
\clearpage
%------------------------------------------------------------------------------------------

\paragraph{Thiofenchone (\textbf{1-S})} 

\begin{figure}[htbp]
\centering
  \includegraphics[width=0.97\textwidth]{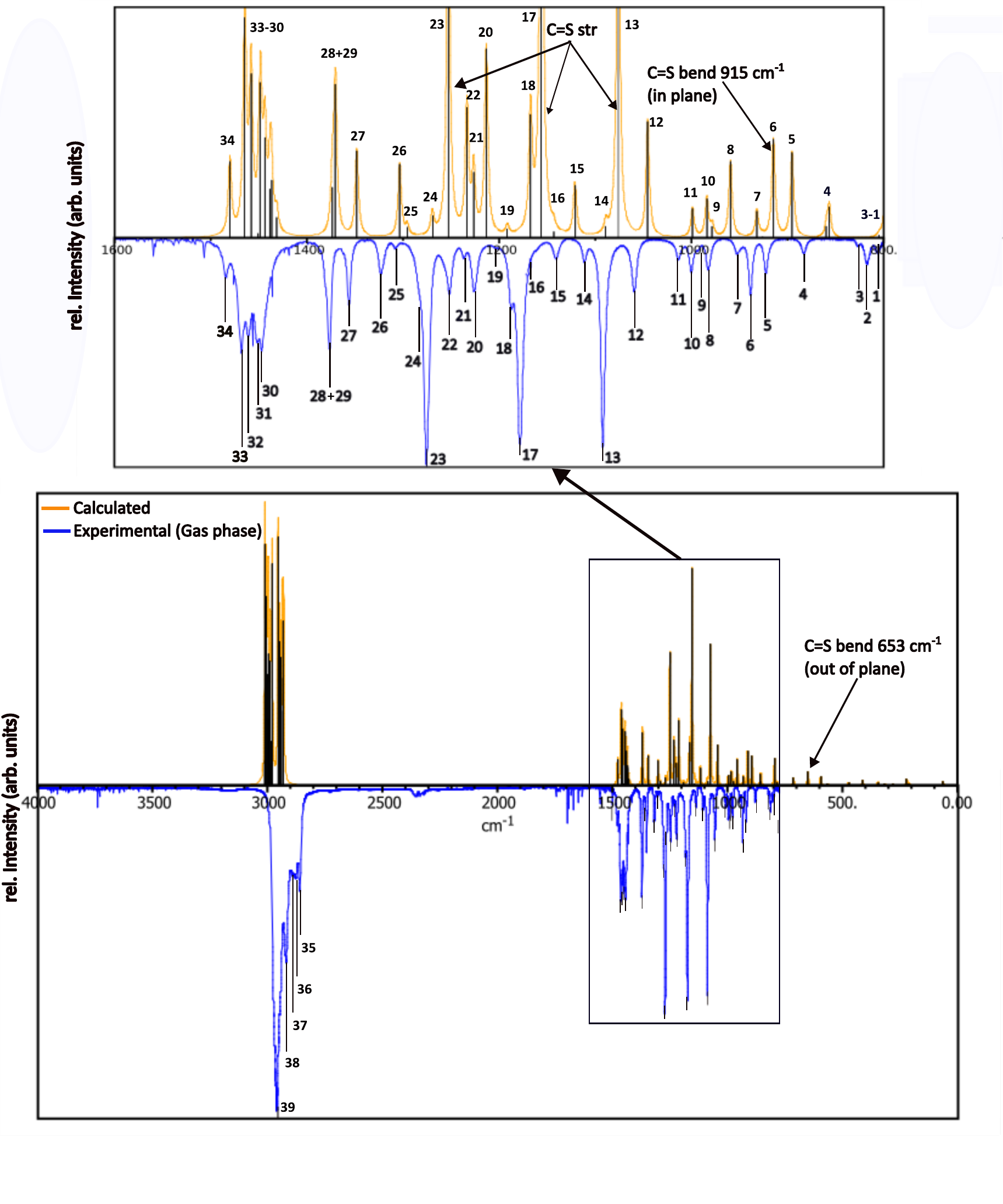}
  \caption{Combined theoretical (orange) and low-resolution experimental gas-phase vibrational overview spectra (blue) of \textbf{1-S}. Theoretical spectrum was computed on the DFT level within the double-harmonic approximation and is displayed using an overall scaling factor of the harmonic vibrational wavenumbers of 0.968 and a Lorentzian lineshape function with HWHM of \SI{1.5}{\per\centi\meter}. The underlying theoretical stick spectrum is indicated in black. For some fundamentals the approximate character, as obtained by visual inspection of the corresponding normal modes, is indicated.\label{fig:Svib}}
\end{figure}

Figure~\ref{fig:Svib} compares the theoretically predicted IR spectrum and the corresponding experimental gas-phase IR
spectrum of \textbf{1-S}, whereas Table~\ref{tab:Sfundamentals} lists the predicted and measured bands in detail. Like for \textbf{1-O}, bands observed in the calculated spectrum are in accordance with the measured spectrum. A most intense band in the low wavenumbers region at \SI{1158}{\per\centi\meter} is observed mainly for C=S stretching, which underwent a large downshift to the lower wavenumber region compared to carbonyl stretching in \textbf{1-O}. It is found by visual inspection that two more bands
at \qtylist[list-units=brackets]{1254;1078}{\per\centi\meter} involve a substantial vibration in C=S bond along with the other bonding
vibrations. Hence, three bands computed at \qtylist[list-units=bracket]{1254;1158;1078}{\per\centi\meter} and observed in the gas-phase IR spectrum at \qtylist[list-units=bracket]{1276;1179;1086}{\per\centi\meter} (peaks 23, 17 and 13) may be tentatively designated as
the “-C=S I, II and III bands”. Another characteristic in-plane and out-of-plane C=S bending is predicted at \qtylist[list-units=bracket]{915;653}{\per\centi\meter}, respectively. The corresponding in-plane bending mode is observed experimentally at \SI{953}{\per\centi\meter} (peak 6). In the literature,
bands at \qtylist[list-units=bracket]{1470;1450;1275;1180;1090}{\per\centi\meter} are listed for \textbf{1-S} dissolved in tetrachloromethane, but without a specific assignment.\cite{wijekoon:1983} The last three of these correspond to the "-C=S I, II and III bands" mentioned above, whereas the first two find their correspondence in the signal group of peaks 34-30 observed in our present work that can be assigned to the host of methyl asymmetric bending modes and methylene scissoring modes. These transitions feature also prominently in the solution-phase IR spectrum measured in the present work under attenuated total reflection and resonate at \qtylist[list-units=bracket]{1467;1447;1275;1177;1091}{\per\centi\meter} (see Figure~\ref{fig:S-IR} in the Appendix).
The umbrella mode doublet expected for the geminal dimethyl substituents at C4 and the umbrella mode for the methyl group at C1 correspond to the peak group 29--27 (theo: \qtylist[list-units=bracket]{1375;1372;1350}{\per\centi\meter}; exp: \qtylist[list-units=bracket]{1380;1377;1357}{\per\centi\meter}). The signal at lower wavenumber corresponds to the symmetric umbrella motion of the geminal dimethyl substituents, whereas the asymmetric combination of the umbrella motions is perturbed by the umbrella motion of the bridge head methyl group, giving rise to the two higher wavenumber signals.  
Various stretching modes of three \ce{CH3} (C8, C9, C10), three \ce{CH2} (C5, C6, C7) are predicted theoretically between \qtyrange[range-phrase=--,range-units=single]{3016}{2932}{\per\centi\meter}
and \qtyrange[range-phrase=--,range-units=single]{3006}{2944}{\per\centi\meter}, respectively, whereas the aliphatic C(4)-H group is assigned to the relatively intense band predicted at
\SI{2980}{\per\centi\meter}.

\begin{table}[htp]
\footnotesize
\caption{Calculated harmonic (scaled by 0.968) and anharmonic wavenumbers $\Tilde{\nu}$ (cm\textsuperscript{-1}), integrated absorption coefficients $A$ (\SI{}{\kilo\meter\per\mol}), experimental transition wavenumbers, intensities (in arbitrary units, a.u.) and assignments of selected bands (peak number (\#) in Figure~\ref{fig:Svib}) of \textbf{1-S}.\label{tab:Sfundamentals}}

\begin{tabular}{cccccccc}
\hline
Mode & \multicolumn{2}{c}{Harmonic}  & \multicolumn{2}{c}{Anharmonic} & \multicolumn{3}{c}{Experimental} \\ 
\ & $\Tilde{\nu}$   & $A$ & $\Tilde{\nu}$   & $A$ & \# & $\Tilde{\nu}$   & $I_\mathrm{exp}$ \\
\ &  (cm$^{-1}$)  & (\SI{}{\kilo\meter\per\mol}) & \ (cm$^{-1}$) & (\SI{}{\kilo\meter\per\mol}) &   &  (cm$^{-1}$) & (a.u.)  \\
\hline
$\nu_{1}$ & 3016.0 &  7.9 & 2960.5 & 26.4 & \rdelim\}{16}{0.3cm}[] \multirow{3}{*}{\,\,\,\,\,\,\,\,\,\,\,\,\,\,}  &  & \\
$\nu_{2}$ & 3014.4 & 40.0 & 2957.5 & 70.9 &  &  & \\
$\nu_{3}$ & 3006.6 & 33.0 & 3012.9 & 35.3 &  &  & \\
$\nu_{4}$ & 3001.4 & 25.0 & 2944.0 & 29.9 &  &  & \\
$\nu_{5}$ & 2999.8 & 12.5 & 2964.3 & 15.5 &  &  & \\
$\nu_{6}$ & 2997.9 & 21.8 & 2940.6 &  3.5 & \hspace{0.3cm} 39 &  2964.46 & 1.791 \\
$\nu_{7}$ & 2994.6 &  5.3 & 2906.3 &  2.1 & \hspace{0.3cm} 38 &  2924.65 & 1.003 \\
$\nu_{8}$ & 2990.5 & 20.4 & 3022.8 & 26.8 & \hspace{0.3cm} 37 &  2894.66 & 0.510 \\
$\nu_{9}$ & 2986.1 &  9.4 & 2973.8 & 69.7 & \hspace{0.3cm} 36 &  2882.55 & 0.527 \\
$\nu_{10}$& 2980.3 & 44.3 & 2953.9 &  8.7 & \hspace{0.3cm} 35 &  2865    & 0.593 \\
$\nu_{11}$& 2956.9 & 41.6 & 2933.5 &  2.3 &  &  & \\
$\nu_{12}$& 2950.4 & 27.1 & 2904.4 & 16.6 &  &  & \\
$\nu_{13}$& 2944.8 & 23.9 & 2915.1 & 79.2 &  &  & \\
$\nu_{14}$& 2937.0 & 32.1 & 2911.5 & 22.0 &  &  & \\
$\nu_{15}$& 2932.9 & 13.0 & 2870.1 & 13.3 &  &  & \\
$\nu_{16}$& 2932.2 & 20.7 & 2921.7 &  4.8 &  &  & \\
$\nu_{17}$& 1482.9 &  4.7 & 1531.9 & (4.7)& \hspace{0.3cm} 34 &  1484.48 & 0.222 \\
$\nu_{18}$& 1466.9 & 14.0 & 1476.4 &  5.5 & \rdelim\}{8}{0.3cm}[] \multirow{7}{*}{ \,\,\,\,\,\,\,\,\,\,\,\,\,\,}&  & \\
$\nu_{19}$& 1461.5 & 10.6 & 1470.2 &  2.6 &  &   & \\
$\nu_{20}$& 1452.8 &  1.4 & 1500.8 & (1.4)& \hspace{0.4cm} 33 &  1468.17 & 0.658 \\
$\nu_{21}$& 1450.5 &  9.2 & 1458.4 &  2.6 & \hspace{0.4cm} 32 & 1461.42 & 0.563\\
$\nu_{22}$& 1446.0 &  6.8 & 1455.0 &  1.3 & \hspace{0.4cm} 31 & 1452.16 & 0.596 \\
$\nu_{23}$& 1441.5 &  2.1 & 1445.3 &  0.9 & \hspace{0.4cm} 30 & 1447.52 & 0.648  \\
$\nu_{24}$& 1439.6 &  4.6 & 1443.3 &  2.1 &  & &\\
$\nu_{25}$& 1433.3 &  1.4 & 1435.0 &  0.8 &  &  &  \\
 \hline
\end{tabular}
\end{table}

\begin{table}[htp]
\footnotesize
\begin{tabular}{cccccccc}
\hline
Mode & \multicolumn{2}{c}{Harmonic}  & \multicolumn{2}{c}{Anharmonic} & \multicolumn{3}{c}{Experimental} \\ 
\ & $\Tilde{\nu}$   & $A$ & $\Tilde{\nu}$   & $A$ & \# & $\Tilde{\nu}$   & $I_\mathrm{exp}$ \\
\ &  (cm$^{-1}$)  & (\SI{}{\kilo\meter\per\mol}) & \ (cm$^{-1}$) & (\SI{}{\kilo\meter\per\mol}) &   &  (cm$^{-1}$) & (a.u.)  \\
\hline
$\nu_{26}$& 1375.2 &  2.1 & 1386.4 &  1.6 & \rdelim\}{2}{0.3cm}[] 29 & 1380.04 & 0.316 \\
$\nu_{27}$& 1372.8 & 11.0 & 1383.0 &  9.7 & \hspace{0.4cm} 28 &  1376.84 & 0.633 \\
$\nu_{28}$& 1350.2 &  5.8 & 1360.8 &  4.8 & \hspace{0.3cm}27 & 1356.65 & 0.380 \\
$\nu_{29}$& 1305.0 &  4.7 & 1310.1 &  3.5 & \hspace{0.3cm}26 & 1323.11 & 0.202 \\
$\nu_{30}$& 1297.0 &  0.7 & 1303.4 &  0.4 & \hspace{0.3cm}25 & 1308.12 & 0.059 \\
$\nu_{31}$& 1271.0 &  1.5 & 1275.4 &  0.4 & \hspace{0.3cm}24 & 1283.91 & 0.299 \\
$\nu_{32}$& 1254.0 & 25.7 & 1261.0 & 12.7 & \hspace{0.3cm}23 & 1275.82 & 1.301 \\
$\nu_{33}$& 1235.0 &  8.2 & 1245.5 &  3.9 & \hspace{0.3cm}22 & 1252.15 & 0.317 \\
$\nu_{34}$& 1228.0 &  3.9 & 1224.5 &  0.9 & \hspace{0.3cm}21 & 1236.9  & 0.111 \\
$\nu_{35}$& 1215.3 & 12.8 & 1219.8 &  0.2 & \hspace{0.3cm}20 & 1226.2  & 0.301 \\
$\nu_{36}$& 1192.1 &  0.6 & 1197.9 &  0.5 & \hspace{0.3cm}19 & 1203.19 & 0.055 \\
$\nu_{37}$& 1168.9 &  8.7 & 1175.8 &  2.4 & \hspace{0.3cm}18 & 1187.64 & 0.403 \\
$\nu_{38}$& 1157.6 & 40.4 & 1166.1 &  0.4 & \hspace{0.3cm}17 & 1178.57 & 1.230 \\
$\nu_{39}$& 1144.9 &  0.5 & 1152.0 &  0.5 & \hspace{0.3cm}16 & 1164.14 & 0.076 \\
$\nu_{40}$& 1122.6 &  3.5 & 1127.6 &  1.3 & \hspace{0.3cm}15 & 1141.07 & 0.116 \\
$\nu_{41}$& 1090.7 &  0.5 & 1098.8 &  0.4 & \hspace{0.3cm}14 & 1111.48 & 0.136 \\
$\nu_{42}$& 1078.2 & 25.8 & 1086.3 & 14.5 & \hspace{0.3cm}13 & 1092.41 & 1.196 \\
$\nu_{43}$& 1047.4 &  7.6 & 1054.9 &  4.0 & \hspace{0.3cm}12 & 1059.89 & 0.306 \\
$\nu_{44}$& 1000.9 &  1.8 & 1007.4 &  1.7 & \hspace{0.3cm}11 & 1013.81 & 0.122 \\
$\nu_{45}$&  986.2 &  2.6 &  997.3 &  3.1 & \hspace{0.3cm}10 & 1000.6  & 0.192 \\
$\nu_{46}$&  980.4 &  0.8 &  988.6 &  0.8 & \hspace{0.3cm}9 & 994.18  & 0.072 \\
$\nu_{47}$&  960.0 &  5.0 &  970.2 &  4.1 & \hspace{0.3cm}8 & 982.85  & 0.176 \\
$\nu_{48}$&  932.2 &  1.8 &  941.6 &  0.5 & \hspace{0.3cm}7 & 952.66  & 0.088 \\
$\nu_{49}$& 926.9 &  0.1 &  936.4 &  0.3 &  & & \\
$\nu_{50}$&  915.4 &  6.5 &  923.1 &  7.6 & \hspace{0.3cm}6 & 938.97  & 0.323 \\
$\nu_{51}$&  895.5 &  5.5 &  902.4 &  4.8 & \hspace{0.3cm}5 & 922.89  & 0.200\\ \hline

\end{tabular}
\end{table}

\begin{table}[htp]
\footnotesize
\begin{tabular}{cccccccc}
\hline
Mode & \multicolumn{2}{c}{Harmonic}  & \multicolumn{2}{c}{Anharmonic} & \multicolumn{3}{c}{Experimental} \\ 
\ & $\Tilde{\nu}$   & $A$ & $\Tilde{\nu}$   & $A$ & \# & $\Tilde{\nu}$   & $I_\mathrm{exp}$ \\
\ &  (cm$^{-1}$)  & (\SI{}{\kilo\meter\per\mol}) & \ (cm$^{-1}$) & (\SI{}{\kilo\meter\per\mol}) &   &  (cm$^{-1}$) & (a.u.)  \\
\hline
$\nu_{52}$&  860.5 &  1.0 &  867.5 &  0.4 & \rdelim\}{2}{0.3cm}[] \multirow{2}{*}{4}  & \multirow{2}{*}{883.61}  & \multirow{2}{*}{0.089} \\
$\nu_{53}$&  857.0 &  1.8 &  864.8 &  1.5 &  & &\\
$\nu_{54}$&  806.1 &  0.3 &  823.7 &  0.6 & \rdelim\}{3}{0.3cm}[] 3& 826.89  & 0.045 \\
$\nu_{55}$&  799.2 &  4.8 &  809.2 &  1.3 & \hspace{0.4cm}2 & 817.89  & 0.142 \\
$\nu_{56}$&  787.2 &  0.6 &  784.1 &  1.1 & \hspace{0.4cm}1 & 806.62  & 0.074 \\
$\nu_{57}$&  717.1 &  1.6 &  725.5 &  1.4 &  & & \\
$\nu_{58}$&  653.5 &  2.5 &  662.2 &  1.7 &  & & \\ 
$\nu_{59}$&  597.6 &  1.6 &  607.2 &  1.6 &  & & \\
$\nu_{60}$&  567.9 &  0.4 &  579.9 &  0.1 &  & & \\
$\nu_{61}$&  477.8 &  0.6 &  488.5 &  0.1 &  & & \\
$\nu_{62}$&  462.9 &  0.2 &  468.5 &  0.1 &  & & \\
$\nu_{63}$&  416.4 &  1.0 &  423.0 &  0.7 &  & & \\
$\nu_{64}$&  402.7 &  0.2 &  410.3 &  0.2 &  & & \\
$\nu_{65}$&  350.9 &  0.6 &  353.3 &  0.7 &  & & \\
$\nu_{66}$&  321.3 &  0.3 &  320.5 &  0.1 &  & & \\
$\nu_{67}$&  286.6 &  0.3 &  284.3 &  0.2 &  & & \\
$\nu_{68}$&  264.9 &  0.1 &  260.5 &  0.2 &  & & \\
$\nu_{69}$&  239.7 &  0.1 &  278.5 &  0.1 &  & & \\
$\nu_{70}$&  224.9 &  1.1 &  239.5 &  1.1 &  & & \\
$\nu_{71}$&  216.3 &  0.5 &  220.5 &  0.2 &  & & \\
$\nu_{72}$&  210.3 &  0.1 &  190.5 &  0.3 &  & & \\
$\nu_{73}$&  198.6 &  0.0 &  136.4 &  0.1 &  & & \\
$\nu_{74}$&  174.4 &  0.0 &  124.5 &  0.0 &  & & \\
$\nu_{75}$&   71.2 &  0.7 &   54.4 &  0.7 &  & & \\
\hline
\end{tabular}
\end{table}

%------------------------------------------------------------------------------------------
\clearpage
%------------------------------------------------------------------------------------------
\paragraph{Selenofenchone (\textbf{1-Se})}

\begin{figure}[htbp]
\centering
  \includegraphics[width=1.03\textwidth]{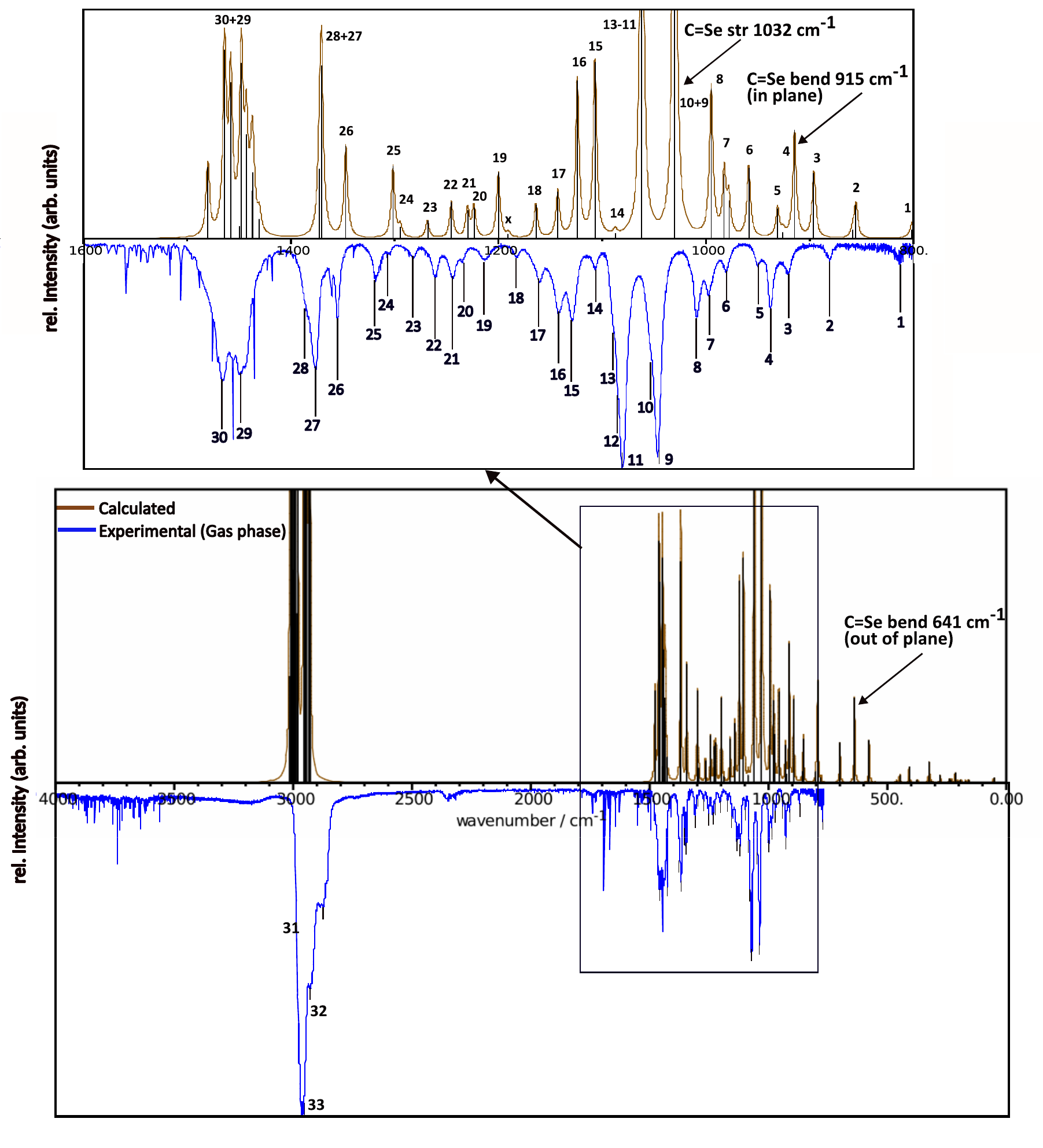}
  \caption{Combined theoretical (brown) and low-resolution experimental gas-phase vibrational overview spectra (black) of \textbf{1-Se}. Theoretical spectrum was computed on the DFT level within the double-harmonic approximation and is displayed using an overall scaling factor of the harmonic vibrational wavenumbers of 0.968 and a Lorentzian lineshape function with HWHM of \SI{1.5}{\per\centi\meter}. The underlying theoretical stick spectrum is indicated in black. For some fundamentals the approximate character, as obtained by visual inspection of the corresponding normal modes, is indicated.\label{fig:Sevib}}
\end{figure}

\textbf{1-Se}, one of the very first isolable saturated selenoketones and first synthesized in 1975, \cite{back:1975,back:1976} is the least explored case
among the chosen molecules \textbf{1-O}, \textbf{1-S} and \textbf{1-Se}, especially in the vibrational spectroscopy. As \textbf{1-Se} is mostly employed as reagent, only few findings  have been reported in the past.\cite{fung:1980,andersen:1982,wijekoon:1983,guziec:1984} Following the accuracy trend as of \textbf{1-O}
and \textbf{1-S}, a good agreement is observed in the predicted IR spectrum and the corresponding experimental spectrum of \textbf{1-Se} (Figure~\ref{fig:Sevib}) where bands observed in the calculated spectrum align overall well with the measured spectrum (see Table~\ref{tab:Sefundamentals} for a detailed list of bands). 
For example, the deviation between the predicted and experimental wavenumbers for the major vibrational modes are \SI{0.1}{\per\centi\meter} for C=O stretching in \textbf{1-O}, \SI{21}{\per\centi\meter} for C=S stretching in \textbf{1-S} and \SI{15}{\per\centi\meter} for C=Se stretching in \textbf{1-Se}, demonstrating the overall reliability of the predictions. 

A most intense band in the low wavenumbers region predicted at \SI{1032}{\per\centi\meter}, observed at \SI{1047}{\per\centi\meter} (peak 9), is
mainly a C=Se stretching mode, which experienced a lower downshift in comparison to the previous two
consecutive cases. Besides, two bands obtained in the quantum chemical calculations at \qtylist[list-units=bracket]{1064;996}{\per\centi\meter}, experimentally observed at \qtylist[list-units=bracket]{1081;1009}{\per\centi\meter} (peaks 11, 8), are also benefiting pronouncedly from C=Se stretching contributions, so that we could also refer to the three of them as the “-C=Se I, II and III bands”. The vibrational bands predicted at \SI{915}{\per\centi\meter} (found experimentally at \SI{938}{\per\centi\meter}, peak 4) and \SI{641}{\per\centi\meter} could be assigned to
C=Se in-plane and out-of-plane bending, respectively. 

Our DFT calculations locate the umbrella mode doublet of the geminal dimethyl substitutents on C4 at about \SI{1373}{\per\centi\meter} and \SI{1350}{\per\centi\meter}. As opposed to \textbf{1-S}, according to the DFT calculations in the double harmonic approximation for \textbf{1-Se} the band with \emph{lower} wavenumber corresponds to a \emph{symmetric} umbrella motion of the two methyl groups, whereas the accompanying higher wavenumber signal of the asymmetric combination of the umbrella modes is split into two bands due to coupling with the umbrella mode of the bridge head methyl group. These three bands resonate in the experimental gas-phase IR spectrum at \qtylist[list-units=bracket]{1380;1377;1157}{\per\centi\meter} (peaks 29--27).

With very small changes in higher region as
compared to \textbf{1-S}, stretching modes of \ce{CH3} (C8, C9, C10), \ce{CH2} (C5, C6, C7) and C(4)H are
quantum chemically obtained between \qtyrange[range-phrase=--,range-units=single]{3016}{2932}{\per\centi\meter}, \qtyrange[range-phrase=--,range-units=single]{3007}{2945}{\per\centi\meter} and \SI{2980}{\per\centi\meter}, respectively. 

In previous reports, experimental IR bands for \textbf{1-Se} in a tetrachloromethane solution were listed at \qtylist[list-units=bracket]{1470;1450;1380;1360;1080;1050}{\per\centi\meter} in Ref.~\onlinecite{back:1976} and at \qtylist[list-units=bracket]{1465;1455;1075;1045}{\per\centi\meter} in Ref.~\onlinecite{wijekoon:1983},
but given therein without a specific assignment. In our solution-phase IR spectra measured via attenuated total reflection, these bands are located at \qtylist[list-units=bracket]{1465;1444;1376;1355;1079;1046}{\per\centi\meter} (see Figure~\ref{fig:Se-IR} in the Appendix). Based on our analysis, these bands can be assigned to \ce{CH3}, and \ce{CH2} deformation modes ($\approx$~\SI{1465}{\per\centi\meter} and $\approx$~\SI{1445}{\per\centi\meter}), doublet of the geminal methyl groups ($\approx$~\SI{1375} and $\approx$~\SI{1355}) and C=Se stretching modes (I: $\approx$~\SI{1080}{\per\centi\meter}; II: $\approx$~\SI{1045}{\per\centi\meter}), consistent with the theoretical spectrum and visual inspection of normal modes. 

\begin{table}[htp]
\footnotesize
\caption{Calculated harmonic (scaled by 0.968) and anharmonic wavenumbers $\Tilde{\nu}$ (cm\textsuperscript{-1}), integrated absorption coefficients $A$ (\SI{}{\kilo\meter\per\mol}), experimental transition wavenumbers, intensities (in arbitrary units, a.u.) and assignments of selected bands (peak number (\#) in Figure~\ref{fig:Sevib}) of \textbf{1-Se}.\label{tab:Sefundamentals}}
\begin{tabular}{cccccccc}
\hline
Mode & \multicolumn{2}{c}{Harmonic}  & \multicolumn{2}{c}{Anharmonic} & \multicolumn{3}{c}{Experimental} \\ 
\ & $\Tilde{\nu}$   & $A$ & $\Tilde{\nu}$   & $A$ & \# & $\Tilde{\nu}$   & $I_\mathrm{exp}$ \\
\ &  (cm$^{-1}$)  & (\SI{}{\kilo\meter\per\mol}) & \ (cm$^{-1}$) & (\SI{}{\kilo\meter\per\mol}) &   &  (cm$^{-1}$) & (a.u.)  \\
\hline
$\nu_{1}$ & 3017.6 &  7.4 &  3117.4 &  (7.4)& \rdelim\}{16}{0.3cm}[] \multirow{3}{*}{\,\,\,\,\,}& &  \\
$\nu_{2}$ & 3015.9 & 37.9 &  3115.6 & (37.9)& & &  \\
$\nu_{3}$ & 3007.8 & 31.9 &  2973.2 &  82.0 & & &  \\
$\nu_{4}$ & 3002.8 & 23.0 &  2968.6 &  42.8 & & &  \\
$\nu_{5}$ & 3001.0 & 11.2 &  2981.7 & 110.1 & & &  \\
$\nu_{6}$ & 2999.0 & 21.8 &  3028.3 &  11.3 & &  &  \\
$\nu_{7}$ & 2995.8 &  3.2 &  3045.4 &  24.4 & \hspace{0.3cm} 33 & 2966.37 & 0.452  \\
$\nu_{8}$ & 2991.0 & 22.6 &  2927.3 &  24.6 &\hspace{0.3cm} 32 & 2934.97 & 0.287  \\
$\nu_{9}$ & 2987.9 &  6.3 &  2999.5 &  46.4 & \hspace{0.3cm} 31 & 2886.43 & 0.166  \\
$\nu_{10}$& 2981.3 & 46.9 &  2961.6 &  86.5 & & &  \\
$\nu_{11}$& 2957.5 & 48.0 &  2915.0 &  26.3 & &  &  \\
$\nu_{12}$& 2951.2 & 26.7 &  2929.5 &  52.8 & & &  \\
$\nu_{13}$& 2944.9 & 22.5 &  2931.3 &  76.2 & & &  \\
$\nu_{14}$& 2936.7 & 33.7 &  2924.2 &  24.5 & & &  \\
$\nu_{15}$& 2932.6 & 12.1 &  2942.7 &  95.4 & & &  \\
$\nu_{16}$& 2932.1 & 20.5 &  2951.4 & 364.0 & &  &  \\
$\nu_{17}$& 1482.7 &  4.5 &  1531.7 & (4.5)& \rdelim\}{3}{0.3cm}[] \multirow{3}{*}{30} & \multirow{3}{*}{1466.58} & \multirow{3}{*}{0.144} \\
$\nu_{18}$& 1466.3 & 11.9 &  1507.6 &  2.3 &  & &  \\
$\nu_{19}$& 1461.1 & 10.3 &  1509.4 &(10.3)&  & &  \\ 
$\nu_{20}$& 1452.1 &  2.0 &  1469.8 &  0.0 & \rdelim\}{6}{0.3cm}[] \multirow{6}{*}{29} & \multirow{6}{*}{1449.29} & \multirow{6}{*}{0.136} \\
$\nu_{21}$& 1449.8 & 10.5 &  1467.0 &  3.7 &  & &  \\
$\nu_{22}$& 1445.1 &  7.0 &  1446.8 &  0.8 &  & &  \\
$\nu_{23}$& 1440.9 &  2.0 &  1494.0 &  4.1 & & &  \\ 
$\nu_{24}$& 1438.5 &  5.2 &  1448.6 &  3.2 &  & & \\
$\nu_{25}$& 1432.4 &  1.4 &  1450.2 &  0.9 &  &  &\\\hline
\end{tabular}
\end{table}

\begin{table}[htp]
\footnotesize
\begin{tabular}{cccccccc}
\hline
Mode & \multicolumn{2}{c}{Harmonic}  & \multicolumn{2}{c}{Anharmonic} & \multicolumn{3}{c}{Experimental} \\ 
\ & $\Tilde{\nu}$   & $A$ & $\Tilde{\nu}$   & $A$ & \# & $\Tilde{\nu}$   & $I_\mathrm{exp}$ \\
\ &  (cm$^{-1}$)  & (\SI{}{\kilo\meter\per\mol}) & \ (cm$^{-1}$) & (\SI{}{\kilo\meter\per\mol}) &   &  (cm$^{-1}$) & (a.u.)  \\
\hline

$\nu_{26}$& 1373.8 &  2.5 &  1415.8 &  6.6 & \hspace{0.3cm} 28 & 1384.69 & 0.077  \\
$\nu_{27}$& 1372.5 & 13.2 &  1383.0 &  4.5 & \hspace{0.3cm} 27 & 1377.55 & 0.130  \\
$\nu_{28}$& 1348.9 &  6.0 &  1384.2 &  3.8 & \hspace{0.3cm} 26 & 1355.93 & 0.082  \\
$\nu_{29}$& 1302.7 &  4.7 &  1312.6 &  2.2 & \hspace{0.3cm} 25 & 1318.85 & 0.035  \\
$\nu_{30}$& 1295.7 &  0.7 &  1310.3 &  0.7 & \hspace{0.3cm} 24 & 1307.48 & 0.011  \\
$\nu_{31}$& 1269.8 &  1.2 &  1286.6 &  0.2 &\hspace{0.3cm}  23 & 1283.11 & 0.015  \\
$\nu_{32}$& 1245.5 &  2.4 &  1262.8 &  1.1 & \hspace{0.3cm} 22 & 1261.83 & 0.034  \\
$\nu_{33}$& 1230.9 &  1.8 &  1248.9 &  2.4 & \hspace{0.3cm} 21 & 1245.04 & 0.036  \\
$\nu_{34}$& 1225.3 &  1.8 &  1241.0 &  5.1 & \hspace{0.3cm} 20 & 1236.17 & 0.019  \\
$\nu_{35}$& 1200.8 &  4.3 &  1212.5 &  0.3 &\hspace{0.3cm}  19 & 1213.96 & 0.017  \\
$\nu_{36}$& 1190.4 &  0.3 &  1203.6 &  0.3 & & &  \\
$\nu_{37}$& 1164.9 &  2.3 &  1178.8 &  3.7 &\hspace{0.3cm}  18 & 1183.25 & 0.014  \\
$\nu_{38}$& 1143.6 &  2.6 &  1154.9 &  0.2 & \hspace{0.3cm} 17 & 1161.42 & 0.039  \\
$\nu_{39}$& 1124.9 & 11.1 &  1137.8 &  7.2 & \hspace{0.3cm} 16 & 1142.45 & 0.074  \\
$\nu_{40}$& 1107.4 & 11.6 &  1120.0 &  5.0 &\hspace{0.3cm}  15 & 1130.04 & 0.079  \\
$\nu_{41}$& 1089.0 & 0.2 &  1103.2 &  1.6  & \hspace{0.3cm} 14 & 1107.53 & 0.028  \\
\multirow{3}{*}{$\nu_{42}$}& \multirow{3}{*}{1063.9} & \multirow{3}{*}{34.8} &  \multirow{3}{*}{1080.6} &  \multirow{3}{*}{32.5} & \rdelim\}{3}{0.3cm}[] 13 & 1090.01 & 0.090 \\
			 &		  & 	&			&			&	\hspace{0.3cm} 12	& 1085.4	& 0.158\\
			 &		  & 	&			&			&	\hspace{0.3cm} 11	& 1081.02	& 0.232\\
\multirow{2}{*}{$\nu_{43}$}& \multirow{2}{*}{1032.2} & \multirow{2}{*}{42.4} &  \multirow{2}{*}{1053.0} &  \multirow{2}{*}{3.7} & \rdelim\}{2}{0.3cm}[] 10 & 1053.15 & 0.118  \\
 			&		  & 	&			&	&	\hspace{0.3cm} 9	& 1047.39 & 0.223\\ 
$\nu_{44}$& 996.4 & 9.5 &  1033.2 &  4.3 & \hspace{0.3cm} 8 & 1009.44 & 0.078  \\
$\nu_{45}$& 984.3 & 4.6 &  1001.5 &  1.9 & \rdelim\}{2}{0.3cm}[] \multirow{2}{*}{7}  & \multirow{2}{*}{998.48}  & \multirow{2}{*}{0.053} \\
$\nu_{46}$& 979.8 & 2.4 &  994.0 &  3.6 &  & &  \\
$\nu_{47}$& 959.1 & 4.8 &  978.2 &  5.1 & \hspace{0.3cm} 6 & 981.56  & 0.030  \\
$\nu_{48}$& 931.0 & 1.9 &  941.6 &  2.7 & \rdelim\}{2}{0.3cm}[] \multirow{2}{*}{5} & \multirow{2}{*}{950.64}  & \multirow{2}{*}{0.023}  \\ 
$\nu_{49}$& 925.6 & 0.5 &  952.9 &  1.3 & & &  \\\hline
\end{tabular}
\end{table}

\begin{table}[htp]
\footnotesize

\begin{tabular}{cccccccc}
\hline
Mode & \multicolumn{2}{c}{Harmonic}  & \multicolumn{2}{c}{Anharmonic} & \multicolumn{3}{c}{Experimental} \\ 
\ & $\Tilde{\nu}$   & $A$ & $\Tilde{\nu}$   & $A$ & \# & $\Tilde{\nu}$   & $I_\mathrm{exp}$ \\
\ &  (cm$^{-1}$)  & (\SI{}{\kilo\meter\per\mol}) & \ (cm$^{-1}$) & (\SI{}{\kilo\meter\per\mol}) &   &  (cm$^{-1}$) & (a.u.)  \\
\hline
$\nu_{50}$& 914.6 & 7.2 &  924.0 &  3.3 & \hspace{0.3cm} 4 & 938.37  & 0.067  \\ 
$\nu_{51}$& 895.7 & 4.3 &  902.2 &  5.0 & \hspace{0.3cm}3  & 921.9   & 0.032 \\
$\nu_{52}$& 858.8 & 0.9 &  869.6 &  0.4 & \rdelim\}{2}{0.3cm}[] \multirow{2}{*}{2}  & \multirow{2}{*}{881.89}  & \multirow{2}{*}{0.017} \\ 
$\nu_{53}$& 854.9 & 1.9 &  863.4 &  3.8 &  &  &\\
$\nu_{54}$& 802.5 & 1.3 &  822.6 &  0.2 & \hspace{0.3cm}1 &  815.33  & 0.013\\
$\nu_{55}$& 796.7 & 4.4 &  802.3 &  3.3 &  &  &\\
$\nu_{56}$& 781.6 & 0.3 &  793.4 &  2.0 &  &  &\\
$\nu_{57}$& 702.5 & 2.1 &  716.3 &  2.0 &  &  &\\
$\nu_{58}$& 640.9 & 4.3 &  655.0 &  3.6 &  &  &\\
$\nu_{59}$& 579.4 & 2.2 &  591.9 &  2.4 &  &  &\\
$\nu_{60}$& 555.5 & 0.0 &  572.0 &  0.1 &  &  & \\
$\nu_{61}$& 468.7 & 0.1 &  486.8 &  0.1 &  &  & \\
$\nu_{62}$& 452.8 & 0.4 &  476.0 &  0.2 &  &  &\\ 
$\nu_{63}$& 412.5 & 0.8 &  427.8 &  0.6 &  &  &\\
$\nu_{64}$& 377.8 & 0.1 &  390.4 &  0.2 &  &  &\\
$\nu_{65}$& 328.8 & 1.1 &  340.2 &  0.8 &  &  &\\
$\nu_{66}$& 319.5 & 0.1 &  355.7 &  0.4 &  &  &\\
$\nu_{67}$& 282.3 & 0.4 &  307.2 &  0.7 &  &  &\\
$\nu_{68}$& 244.3 & 0.2 &  264.3 &  0.6 &  &  &\\
$\nu_{69}$& 236.5 & 0.2 &  236.2 &  0.0 &  &  &\\
$\nu_{70}$& 220.7 & 0.5 &  324.4 &  0.2 &  &  &\\
$\nu_{71}$& 211.3 & 0.1 &  284.1 &  0.2 &  &  &\\
$\nu_{72}$& 199.1 & 0.1 &  153.0 &  0.1 &  &  &\\
$\nu_{73}$& 175.4 & 0.1 &  199.3 &  0.2 &  &  &\\
$\nu_{74}$& 165.6 & 0.2 &  170.8 &  0.1 &  &  &\\
$\nu_{75}$&  62.0 & 0.2 &   77.3 &  0.2 &  &  &\\ \hline

\end{tabular}
\end{table}

%------------------------------------------------------------------------------------------
\clearpage
%------------------------------------------------------------------------------------------

\paragraph{Tellurofenchone (\textbf{1-Te}) and Polonofenchone (\textbf{1-Po})}

\begin{figure}[htbp]
\centering
  \includegraphics[width=1.05\textwidth]{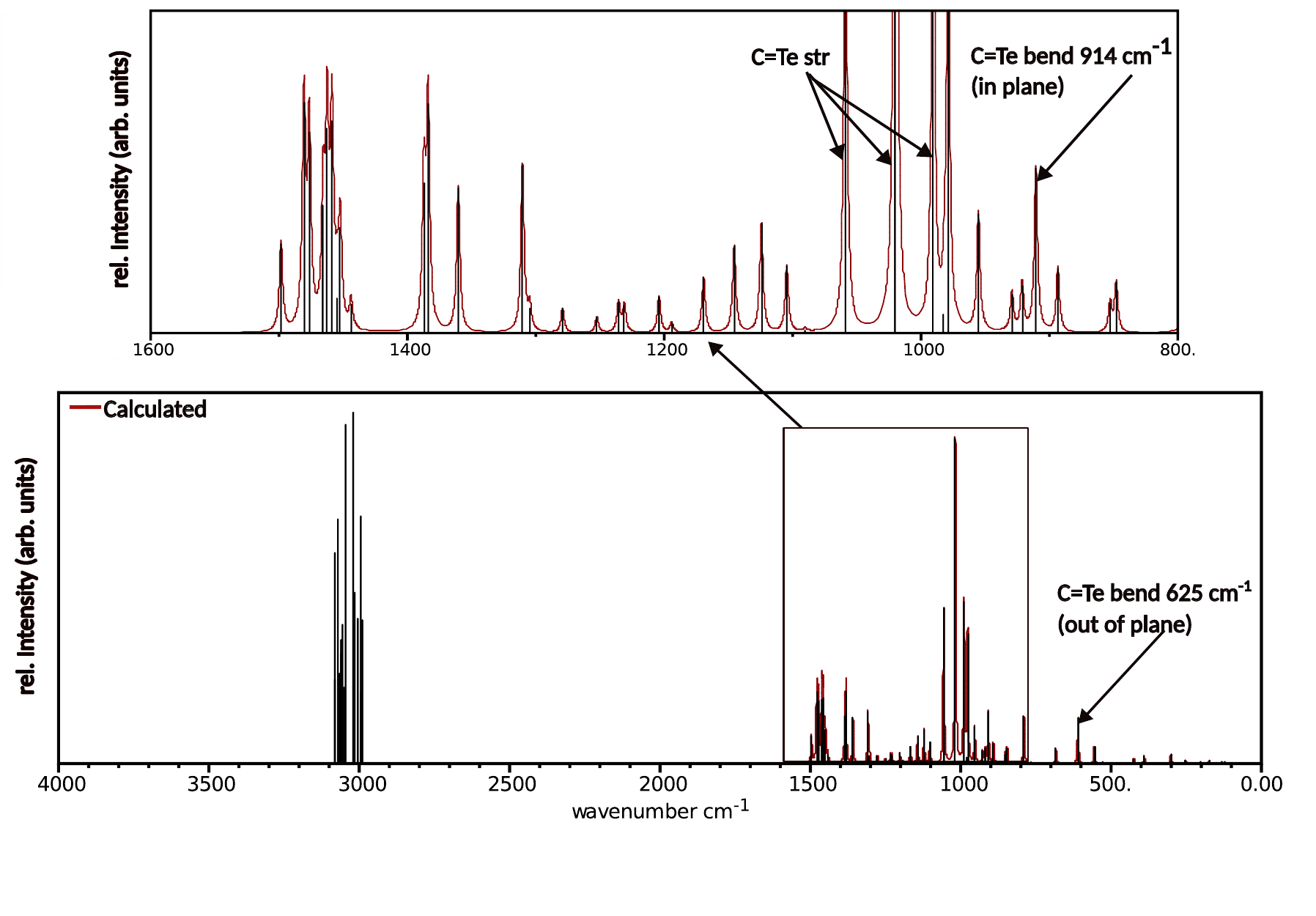}
  \caption{Theoretical gas-phase vibrational overview spectrum of \textbf{1-Te}. Theoretical spectrum was computed on the DFT level within the double-harmonic approximation and is displayed using an overall scaling factor of the harmonic vibrational wavenumbers of 0.968 and a Gaussian lineshape function with HWHM of \SI{1.5}{\per\centi\meter}. The underlying theoretical stick spectrum is indicated in black. For some fundamentals the approximate character, as obtained by visual inspection of the corresponding normal modes, is indicated.\label{fig:Tevib}}
\end{figure}

\begin{figure}[htbp]
\centering
  \includegraphics[width=1.05\textwidth]{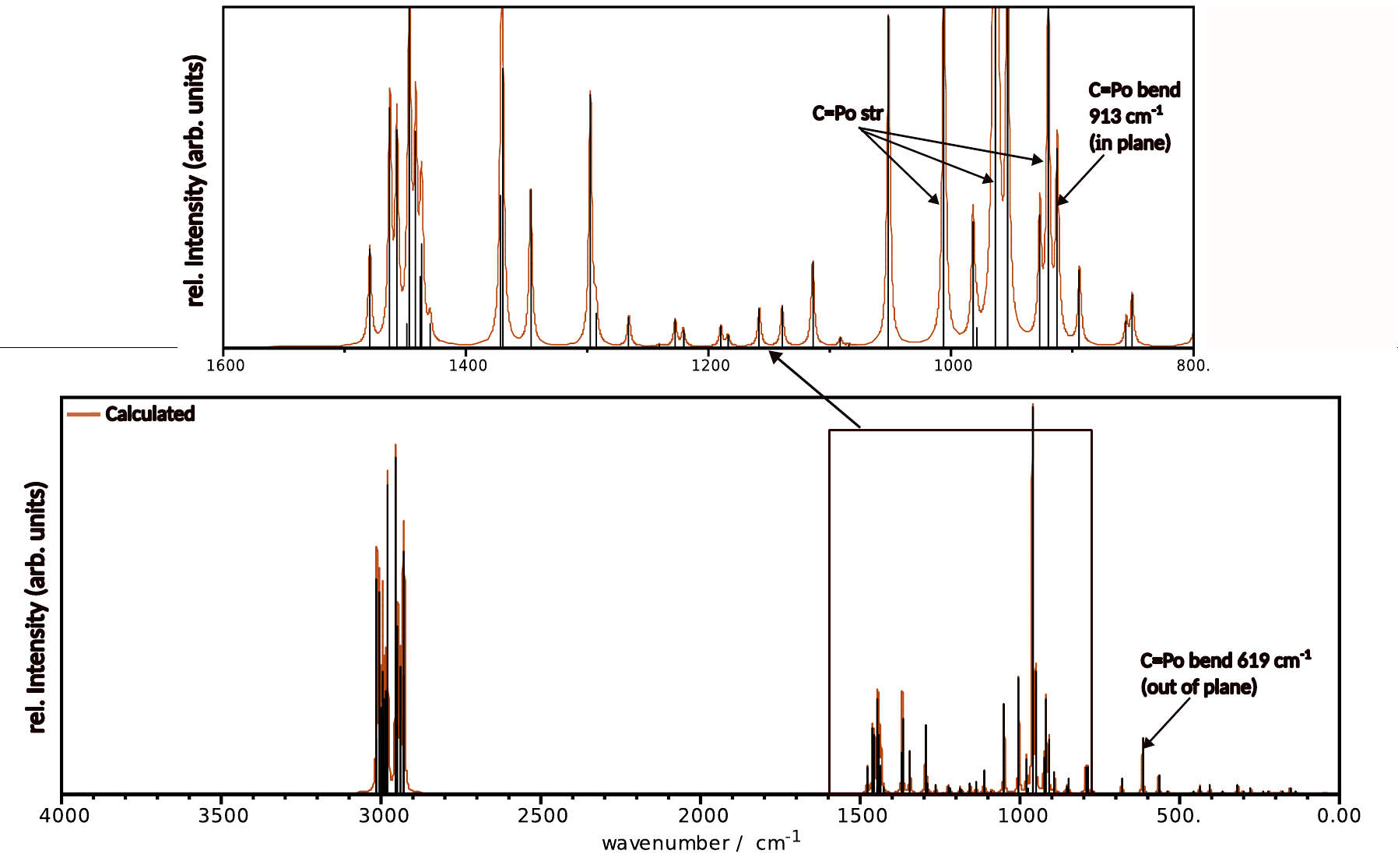}
  \caption{Theoretical gas-phase vibrational overview spectrum of \textbf{1-Po}. Theoretical spectrum was computed on the DFT level within the double-harmonic approximation and is displayed using an overall scaling factor of the harmonic vibrational wavenumbers of 0.968 and a Gaussian lineshape function with HWHM of \SI{1.5}{\per\centi\meter}. The underlying theoretical stick spectrum is indicated in black. For some fundamentals the approximate character, as obtained by visual inspection of the corresponding normal modes, is indicated.
  \label{fig:Povib}}
\end{figure}

Because compounds \textbf{1-Te} and \textbf{1-Po} were not synthesized, only theoretical investigations are presented for these in our study. To the best of our knowledge, no successful synthesis or computational investigation of \textbf{1-Te} and \textbf{1-Po} have been reported yet, so that we present here IR spectra of \textbf{1-Te} (Figure~\ref{fig:Tevib}) and \textbf{1-Po} (Figure~\ref{fig:Povib}) as well as their corresponding band listings (Tables~\ref{tab:Tefundamentals} and \ref{tab:Pofundamentals}) for the first time. 

Following the trend to lower stretching wavenumbers from \textbf{1-O} via \textbf{1-S} to \textbf{1-Se}, C=Te and C=Po stretching modes are predicted at even lower wavenumbers, at \SI{1020}{\per\centi\meter} and \SI{965}{\per\centi\meter}, respectively. Another two bands, to which the C=Te stretching
motion contributes pronouncedly, are \SI{1057}{\per\centi\meter} and \SI{991}{\per\centi\meter}. And yet another band with some contribution from C=Te stretching motions is located at \SI{980}{\per\centi\meter}. Apart from the main C=Po stretching fundamental, which our DFT calculation places at \SI{965}{\per\centi\meter}, there are two more bands at \SI{1008}{\per\centi\meter} and \SI{920}{\per\centi\meter} that involve some contribution from the C=Po stretching motion. Furthermore, C=Te in-plane bending could be assigned at \SI{913}{\per\centi\meter} and out-of-plane at \SI{625}{\per\centi\meter}. Similarly, in-plane bending corresponding to C=Po bond is obtained at \SI{913}{\per\centi\meter} and out-of-plane bending at \SI{619}{\per\centi\meter}. But we shall note here that also the mode at \SI{954}{\per\centi\meter} acquires some in-plane C=Po bending character. 
The umbrella mode doublet, characteristic for the geminal methyl group arrangement, features in \textbf{1-Po} at about \SI{1372}{\per\centi\meter} and \SI{1348}{\per\centi\meter}, with the upper wavenumber band being slightly split into two transitions at \qtylist[list-units=bracket]{1373;1372}{\per\centi\meter} due to coupling with the isolated methyl group.
With almost no major shift in the higher region in comparison to the lighter
derivatives, in \textbf{1-Te} also the stretching modes of \ce{CH3}, \ce{CH2} and CH are predicted at 
\qtyrange[range-phrase=--,range-units=single]{3019}{2931}{\per\centi\meter}, \qtyrange[range-phrase=--,range-units=single]{3009}{2944}{\per\centi\meter} and \SI{2982}{\per\centi\meter}, respectively. 
These modes for \textbf{1-Po} could be found predicted at \qtyrange[range-phrase=--,range-units=single]{3018}{2929}{\per\centi\meter}, \qtyrange[range-phrase=--,range-units=single]{3008}{2943}{\per\centi\meter} and \SI{2981}{\per\centi\meter}, respectively.

\begin{table}[htp]
\footnotesize
\caption{Calculated harmonic (scaled by 0.968) and anharmonic wavenumbers $\Tilde{\nu}$ (cm\textsuperscript{-1}), integrated absorption coefficients $A$ (\SI{}{\kilo\meter\per\mol}) of \textbf{1-Te}.\label{tab:Tefundamentals}}
\begin{tabular}{ccccc}
\hline
Mode & \multicolumn{2}{c}{Harmonic}  & \multicolumn{2}{c}{Anharmonic}  \\ 
\ & $\Tilde{\nu}$   & $A$ & $\Tilde{\nu}$   & $A$  \\
\ &  (cm$^{-1}$)  & (\SI{}{\kilo\meter\per\mol}) & \ (cm$^{-1}$) & (\SI{}{\kilo\meter\per\mol})   \\
\hline
$\nu_{1}$ & 3018.6 &  12.6 & 3039.8 &  23.2 \\
$\nu_{2}$ & 3015.7 &  31.4 & 3019.4 &  13.2 \\
$\nu_{3}$ & 3008.5 &  36.4 & 2951.8 &  15.9 \\
$\nu_{4}$ & 3003.8 &  13.4 & 2941.4 &   5.1 \\
$\nu_{5}$ & 2999.3 &  16.4 & 2974.5 &  20.4 \\
$\nu_{6}$ & 2998.9 &  18.5 & 2925.2 &  19.2 \\
$\nu_{7}$ & 2995.5 &   0.8 & 2934.3 &  22.5 \\
$\nu_{8}$ & 2991.1 &  20.8 & 2908.5 &   9.8 \\
$\nu_{9}$ & 2988.7 &  11.5 & 3001.9 &  46.0 \\
$\nu_{10}$& 2982.3 &  50.4 & 2956.7 &  91.0 \\
$\nu_{11}$& 2958.1 &  52.2 & 2980.9 &  57.6 \\
$\nu_{12}$& 2953.4 &  25.4 & 2900.6 &  10.5 \\
$\nu_{13}$& 2944.3 &  21.5 & 2913.8 &  18.2 \\
$\nu_{14}$& 2935.2 &  36.8 & 2894.4 &  22.1 \\
$\nu_{15}$& 2931.1 &  10.8 & 2866.9 &   0.5 \\
$\nu_{16}$& 2931.0 &  21.4 & 2984.7 &  38.9 \\
$\nu_{17}$& 1482.9 &   4.2 & 1497.6 &   2.4 \\
$\nu_{18}$& 1465.6 &  10.7 & 1458.7 &   3.9 \\
$\nu_{19}$& 1461.6 &   9.4 & 1467.5 &   2.4 \\
$\nu_{20}$& 1451.1 &   6.0 & 1456.4 &   1.5 \\
$\nu_{21}$& 1448.1 &   9.5 & 1456.5 &   9.4 \\
$\nu_{22}$& 1444.5 &   9.9 & 1450.2 &  17.7 \\
$\nu_{23}$& 1440.7 &   1.6 & 1445.8 &   0.5 \\
$\nu_{24}$& 1438.3 &   5.0 & 1447.4 &   2.2 \\
$\nu_{25}$& 1430.1 &   1.4 & 1438.2 &   0.3 \\\hline
\end{tabular}
\end{table}

\begin{table}[htp]
\footnotesize
\begin{tabular}{ccccc}
\hline
Mode & \multicolumn{2}{c}{Harmonic}  & \multicolumn{2}{c}{Anharmonic}  \\ 
\ & $\Tilde{\nu}$   & $A$ & $\Tilde{\nu}$   & $A$  \\
\ &  (cm$^{-1}$)  & (\SI{}{\kilo\meter\per\mol}) & \ (cm$^{-1}$) & (\SI{}{\kilo\meter\per\mol})   \\
\hline
$\nu_{26}$& 1375.1 &   7.0 & 1391.5 &  14.4 \\
$\nu_{27}$& 1372.0 &  10.7 & 1385.4 &  14.3 \\
$\nu_{28}$& 1349.3 &   6.8 & 1367.2 &   4.3 \\
$\nu_{29}$& 1300.8 &   7.8 & 1308.4 &   0.5 \\
$\nu_{30}$& 1295.4 &   1.2 & 1308.9 &   1.2 \\
$\nu_{31}$& 1270.6 &   1.1 & 1281.5 &   0.1 \\
$\nu_{32}$& 1244.8 &   0.8 & 1248.7 &   0.1 \\
$\nu_{33}$& 1228.5 &   1.4 & 1245.1 &   0.3 \\
$\nu_{34}$& 1224.1 &   1.3 & 1234.3 &   1.8 \\
$\nu_{35}$& 1197.9 &   1.7 & 1200.5 &   0.1 \\
$\nu_{36}$& 1188.4 &   0.5 & 1194.2 &   0.2 \\
$\nu_{37}$& 1164.3 &   2.6 & 1169.0 &   2.4 \\
$\nu_{38}$& 1141.3 &   4.1 & 1147.9 &   0.6 \\
$\nu_{39}$& 1120.8 &   5.1 & 1127.6 &   5.5 \\
$\nu_{40}$& 1101.6 &   3.1 & 1107.5 &   1.1 \\
$\nu_{41}$& 1087.8 &   0.2 & 1095.1 &   0.3 \\
$\nu_{42}$& 1057.3 &  23.1 & 1067.7 &  17.7 \\
$\nu_{43}$& 1019.7 &  48.3 & 1027.5 &  33.2 \\
$\nu_{44}$&  991.3 &  24.1 & 1006.4 &  17.4 \\
$\nu_{45}$&  983.3 &   0.9 &  989.5 &   0.6 \\
$\nu_{46}$&  980.1 &  19.3 &  988.9 &   7.9 \\
$\nu_{47}$&  957.2 &   5.5 &  962.3 &   6.5 \\
$\nu_{48}$&  931.9 &   1.8 &  938.5 &   2.2 \\
$\nu_{49}$&  924.3 &   2.2 &  935.3 &   3.3 \\
$\nu_{50}$&  913.8 &   7.7 &  921.3 &   5.8 \\\hline
\end{tabular}
\end{table}

\begin{table}[htp]
\footnotesize
\begin{tabular}{ccccc}
\hline
Mode & \multicolumn{2}{c}{Harmonic}  & \multicolumn{2}{c}{Anharmonic}  \\ 
\ & $\Tilde{\nu}$   & $A$ & $\Tilde{\nu}$   & $A$  \\
\ &  (cm$^{-1}$)  & (\SI{}{\kilo\meter\per\mol}) & \ (cm$^{-1}$) & (\SI{}{\kilo\meter\per\mol})   \\
\hline
$\nu_{51}$&  897.2 &   3.0 &  901.0 &   0.7 \\
$\nu_{52}$&  857.9 &   1.4 &  867.1 &   0.6 \\
$\nu_{53}$&  853.2 &   2.4 &  859.4 &   0.7 \\
$\nu_{54}$&  800.6 &   0.3 &  803.9 &   2.9 \\
$\nu_{55}$&  798.7 &   7.0 &  813.8 &   3.8 \\
$\nu_{56}$&  776.7 &   0.1 &  786.7 &   0.7 \\
$\nu_{57}$&  696.1 &   2.4 &  702.8 &   2.3 \\
$\nu_{58}$&  624.8 &   6.9 &  638.7 &   6.3 \\
$\nu_{59}$&  572.3 &   2.5 &  581.3 &   1.1 \\
$\nu_{60}$&  544.4 &   0.2 &  558.6 &   0.1 \\
$\nu_{61}$&  469.3 &   0.1 &  477.9 &   0.1 \\
$\nu_{62}$&  444.7 &   0.8 &  460.2 &   0.4 \\
$\nu_{63}$&  411.9 &   1.2 &  420.9 &   0.9 \\
$\nu_{64}$&  374.2 &   0.1 &  384.6 &   0.1 \\
$\nu_{65}$&  326.0 &   1.4 &  336.0 &   1.3 \\
$\nu_{66}$&  314.8 &   0.1 &  317.6 &   0.1 \\
$\nu_{67}$&  278.9 &   0.6 &  288.0 &   0.3 \\
$\nu_{68}$&  246.0 &   0.1 &  279.3 &   0.5 \\
$\nu_{69}$&  231.2 &   0.3 &  300.5 &   0.1 \\
$\nu_{70}$&  217.0 &   0.1 &  135.4 &   0.1 \\
$\nu_{71}$&  212.7 &   0.1 &  228.5 &   0.3 \\
$\nu_{72}$&  202.3 &   0.4 &  221.7 &   0.7 \\
$\nu_{73}$&  161.2 &   0.3 &  163.2 &   0.1 \\
$\nu_{74}$&  152.1 &   0.2 &  153.9 &   0.3 \\
$\nu_{75}$&   57.4 &   0.1 &   54.4 &   0.1 \\\hline
\end{tabular}
\end{table}

%------------------------------------------------------------------------------------------
\clearpage
%------------------------------------------------------------------------------------------

\begin{table}[htp]
\footnotesize
\caption{Calculated harmonic (scaled by 0.968) and anharmonic wavenumbers $\Tilde{\nu}$ (cm\textsuperscript{-1}), integrated absorption coefficients $A$ (\SI{}{\kilo\meter\per\mol}) of \textbf{1-Po}.\label{tab:Pofundamentals}}
\begin{tabular}{ccccc}
\hline
Mode & \multicolumn{2}{c}{Harmonic}  & \multicolumn{2}{c}{Anharmonic}  \\ 
\ & $\Tilde{\nu}$   & $A$ & $\Tilde{\nu}$   & $A$  \\
\ &  (cm$^{-1}$)  & (\SI{}{\kilo\meter\per\mol}) & \ (cm$^{-1}$) & (\SI{}{\kilo\meter\per\mol})   \\
\hline
$\nu_{ 1}$& 3018.5 &   8.4 & 2950.5 &  37.7 \\
$\nu_{ 2}$& 3016.6 &  34.1 & 2991.6 &  56.1 \\
$\nu_{ 3}$& 3007.8 &  31.4 & 2964.8 &  34.9 \\
$\nu_{ 4}$& 3002.5 &  18.8 & 3014.7 & 103.5 \\
$\nu_{ 5}$& 2998.7 &  17.2 & 2917.6 &   7.0 \\
$\nu_{ 6}$& 2998.1 &  15.6 & 3006.8 &  48.1 \\
$\nu_{ 7}$& 2994.7 &   1.2 & 2933.8 &  24.7 \\
$\nu_{ 8}$& 2990.5 &  21.3 & 2981.3 & 114.8 \\
$\nu_{ 9}$& 2987.8 &  10.8 & 2910.8 &   8.2 \\
$\nu_{10}$& 2980.6 &  53.3 & 2951.5 &   8.1 \\
$\nu_{11}$& 2957.1 &  51.6 & 2910.1 &  39.2 \\
$\nu_{12}$& 2949.7 &  28.6 & 2876.4 &  29.0 \\
$\nu_{13}$& 2943.2 &  21.0 & 2899.0 &   8.1 \\
$\nu_{14}$& 2934.2 &  40.5 & 2894.5 &  10.7 \\
$\nu_{15}$& 2930.0 &  10.2 & 2865.9 &   0.9 \\
$\nu_{16}$& 2929.8 &  21.3 & 2974.9 &  38.1 \\
$\nu_{17}$& 1482.0 &   4.2 & 1503.0 &   2.8 \\
$\nu_{18}$& 1465.0 &  10.3 & 1462.2 &  10.7 \\
$\nu_{19}$& 1460.3 &   9.3 & 1470.3 &   7.2 \\
$\nu_{20}$& 1450.9 &   3.8 & 1464.9 &  52.5 \\
$\nu_{21}$& 1448.2 &  13.2 & 1459.5 &   2.8 \\
$\nu_{22}$& 1443.6 &   9.0 & 1454.1 &  10.5 \\
$\nu_{23}$& 1439.8 &   2.0 & 1447.7 &   8.1 \\
$\nu_{24}$& 1438.6 &   5.9 & 1445.8 &   7.8 \\
$\nu_{25}$& 1431.2 &   1.2 & 1444.2 &   0.6 \\\hline
\end{tabular}
\end{table}

\begin{table}[htp]
\footnotesize
\begin{tabular}{ccccc}
\hline
Mode & \multicolumn{2}{c}{Harmonic}  & \multicolumn{2}{c}{Anharmonic}  \\ 
\ & $\Tilde{\nu}$   & $A$ & $\Tilde{\nu}$   & $A$  \\
\ &  (cm$^{-1}$)  & (\SI{}{\kilo\meter\per\mol}) & \ (cm$^{-1}$) & (\SI{}{\kilo\meter\per\mol})   \\
\hline
$\nu_{26}$& 1373.2 &   3.8 & 1371.2 &   5.9 \\
$\nu_{27}$& 1372.1 &  15.2 & 1399.8 &   7.0 \\
$\nu_{28}$& 1348.3 &   7.0 & 1359.2 &   4.4 \\
$\nu_{29}$& 1298.5 &  11.1 & 1306.2 &   1.6 \\
$\nu_{30}$& 1293.8 &   1.4 & 1298.5 &  10.0 \\
$\nu_{31}$& 1267.3 &   1.4 & 1270.7 &   2.1 \\
$\nu_{32}$& 1241.4 &   0.1 & 1242.6 &   0.1 \\
$\nu_{33}$& 1228.6 &   1.2 & 1189.7 &   0.1 \\
$\nu_{34}$& 1222.3 &   0.8 & 1235.7 &   0.2 \\
$\nu_{35}$& 1191.1 &   0.9 & 1206.4 &   0.7 \\
$\nu_{36}$& 1185.2 &   0.5 & 1193.6 &   0.3 \\
$\nu_{37}$& 1159.4 &   1.7 & 1167.2 &   0.1 \\
$\nu_{38}$& 1140.6 &   1.7 & 1144.8 &   2.2 \\
$\nu_{39}$& 1115.1 &   3.8 & 1114.5 &   0.4 \\
$\nu_{40}$& 1091.8 &   0.3 & 1080.8 &   0.6 \\
$\nu_{41}$& 1086.4 &   0.1 & 1097.1 &   1.0 \\
$\nu_{42}$& 1053.4 &  14.5 & 1055.4 &   6.3 \\
$\nu_{43}$& 1007.6 &  20.3 & 1019.2 &  20.9 \\
$\nu_{44}$&  984.5 &   5.1 & 1003.7 &   6.2 \\
$\nu_{45}$&  980.6 &   1.0 &  982.1 &   7.0 \\
$\nu_{46}$&  965.3 &  59.4 &  977.8 &  46.9 \\
$\nu_{47}$&  954.1 &  24.0 &  957.9 &  27.4 \\
$\nu_{48}$&  927.7 &   5.0 &  933.4 &   1.5 \\
$\nu_{49}$&  920.3 &  14.3 &  926.5 &  11.5 \\
$\nu_{50}$&  912.6 &   9.4 &  922.0 &   5.0 \\\hline
\end{tabular}
\end{table}

\begin{table}[htp]
\footnotesize
\begin{tabular}{ccccc}
\hline
Mode & \multicolumn{2}{c}{Harmonic}  & \multicolumn{2}{c}{Anharmonic}  \\ 
\ & $\Tilde{\nu}$   & $A$ & $\Tilde{\nu}$   & $A$  \\
\ &  (cm$^{-1}$)  & (\SI{}{\kilo\meter\per\mol}) & \ (cm$^{-1}$) & (\SI{}{\kilo\meter\per\mol})   \\
\hline
$\nu_{51}$&  894.5 &   3.2 &  897.3 &   2.9 \\
$\nu_{52}$&  856.3 &   1.3 &  865.5 &   0.2 \\
$\nu_{53}$&  851.0 &   2.1 &  858.1 &   0.0 \\
$\nu_{54}$&  798.2 &   6.0 &  814.8 &   1.6 \\
$\nu_{55}$&  792.2 &   2.2 &  800.8 &   4.4 \\
$\nu_{56}$&  772.8 &   0.2 &  774.2 &   1.5 \\
$\nu_{57}$&  684.9 &   2.5 &  690.4 &   1.8 \\
$\nu_{58}$&  619.0 &   8.9 &  628.5 &   8.6 \\
$\nu_{59}$&  568.8 &   3.0 &  577.5 &   1.6 \\
$\nu_{60}$&  540.9 &   0.3 &  553.5 &   0.3 \\
$\nu_{61}$&  464.4 &   0.1 &  466.6 &   0.2 \\
$\nu_{62}$&  442.1 &   1.1 &  443.3 &   0.8 \\
$\nu_{63}$&  409.7 &   1.3 &  413.4 &   0.4 \\
$\nu_{64}$&  369.2 &   0.2 &  378.6 &   0.1 \\
$\nu_{65}$&  323.4 &   1.4 &  322.4 &   1.2 \\
$\nu_{66}$&  306.8 &   0.2 &  304.3 &   0.8 \\
$\nu_{67}$&  281.1 &   0.8 &  283.5 &   0.8 \\
$\nu_{68}$&  246.4 &   0.2 &  275.4 &   0.2 \\
$\nu_{69}$&  228.8 &   0.3 &  115.2 &   0.2 \\
$\nu_{70}$&  216.0 &   0.1 &  215.4 &   0.8 \\
$\nu_{71}$&  203.4 &   0.1 &  295.4 &   0.1 \\
$\nu_{72}$&  183.0 &   0.3 &  177.6 &   0.4 \\
$\nu_{73}$&  160.0 &   0.8 &  143.0 &   0.5 \\
$\nu_{74}$&  141.7 &   0.2 &  133.6 &   0.2 \\
$\nu_{75}$&   56.6 &   0.0 &   30.6 &   0.0 \\\hline
\end{tabular}
\end{table}

%------------------------------------------------------------------------------------------
\clearpage
%------------------------------------------------------------------------------------------

\begin{figure}[hbt]
\centering
  \includegraphics[angle=-90,width=0.7\textwidth]{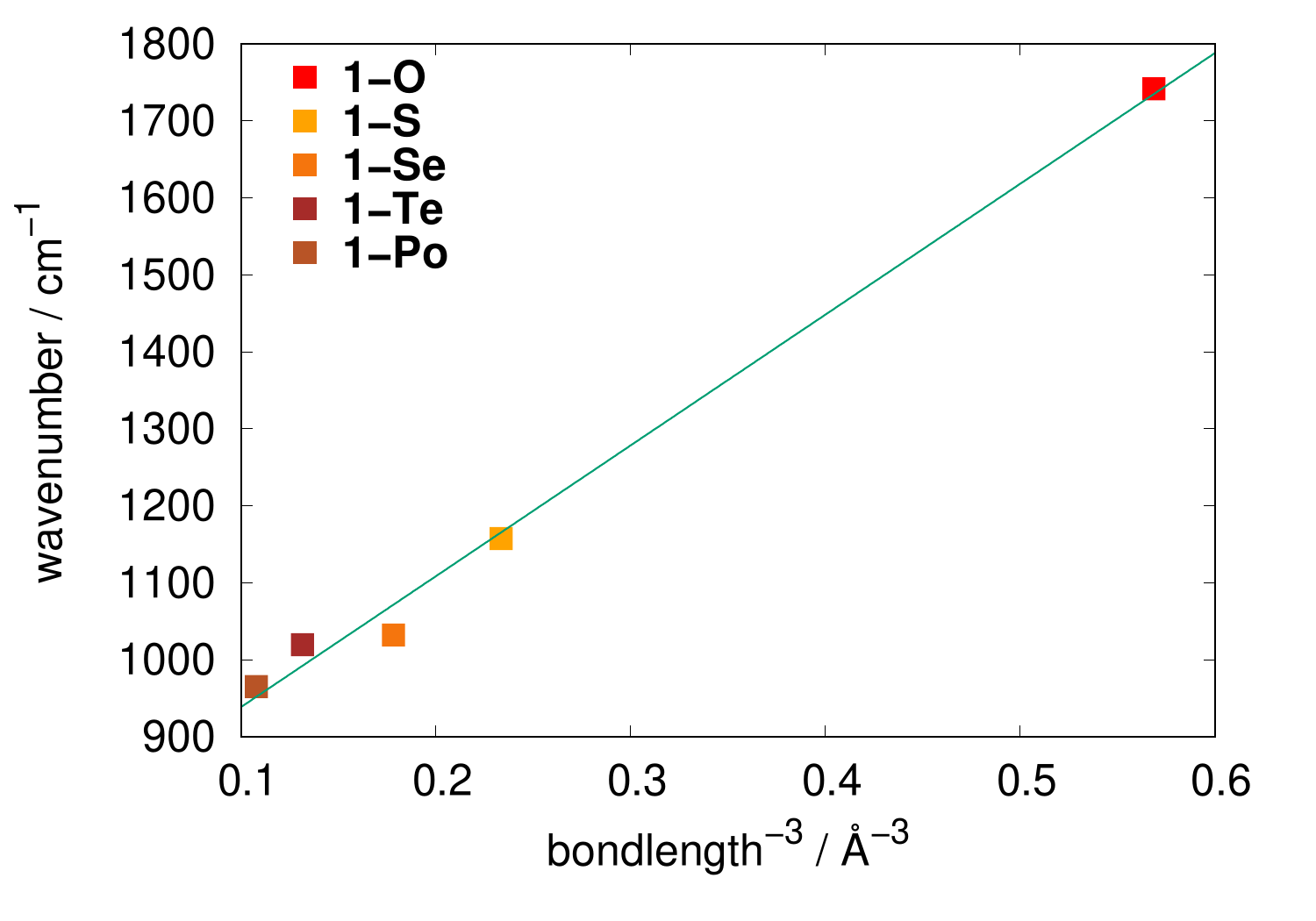}
  \caption{Change in scaled harmonic vibrational wavenumber of the C=X stretching mode in association to the cubed inverse of the internuclear distance as computed for \textbf{1-X}. Data points from left to right correspond to \textbf{1-Po}, \textbf{1-Te}, \textbf{1-Se}, \textbf{1-S} and \textbf{1-O}, for which in each case the most intense transition with C-X stretching contribution was selected for this plot. The slope of the regression line corresponds to approximately \SI{1700}{\angstrom^3\centi\meter^{-1}}, the y-intercept to about \SI{770}{\centi\meter^{-1}}. \label{fig:Xstretchtrend}}
\end{figure}

It is very interesting to compare the featured vibrational modes in these benchmark systems. Following early work of Morse \cite{morse:1929} on diatomic molecules, we plotted in Figure~\ref{fig:Xstretchtrend} the calculated (scaled) harmonic vibrational wavenumbers, which are specifically related to the heavy nucleus (X), against the cubed inverse of the internuclear distance between X and its bonding partner. The internuclear distance of C=X increases significantly while moving from oxygen to polonium down the group which is expected with the increase in the atom size. This causes a bathochromic shift in the relevant vibrations, which is best shown in Figure 8 and Tables \ref{tab:Xstretch} and \ref{tab:Xforce} of the Appendix. An overall lowering from \SI{1741}{\per\centi\meter} to \SI{964}{\per\centi\meter} in C=X stretching is reported. As evident in Figure~\ref{fig:Xstretchtrend}, the expected largest shift is among C=O and C=S stretching modes and the internuclear distance change among these two is also the most. Largely different force constants for C=X bonds are observed in concordance to the force of the bond (Appendix Table \ref{tab:Xforce}). The smallest shifts are
seen while moving from C=Se to C=Te and from C=Te to C=Po, with the change in the internuclear distance in these cases being also the least.
On the other hand a respective shift in the bending vibrational modes is very small after thiofenchone
for in-plane modes (915-913 cm\textsuperscript{-1} from \textbf{1-S} to \textbf{1-Po}) which further increases for the out-of-plane modes
(653-618 cm\textsuperscript{-1} from \textbf{1-S} to \textbf{1-Po}).

Apart from band positions, the respective intensity of the above mentioned vibrations is also changing in a systematic pattern. The C=O stretching is significantly pronounced in comparison to all other C=X stretches. The intensity of the bending modes is moving systematically in an overall trend. It should be noted that this only corresponds to the harmonic picture. In the anharmonic calculations, band strengths for the C=O stretching mode become reduced. Both stretching and in-plane bending (\#) intensities are getting lower while moving to the heavier chalcogenoketone. On
contrary, the out-of-plane bending (§) is gaining intensity when moving in the same order.

\begin{figure}[htbp]
\centering
  \includegraphics[width=0.99\textwidth]{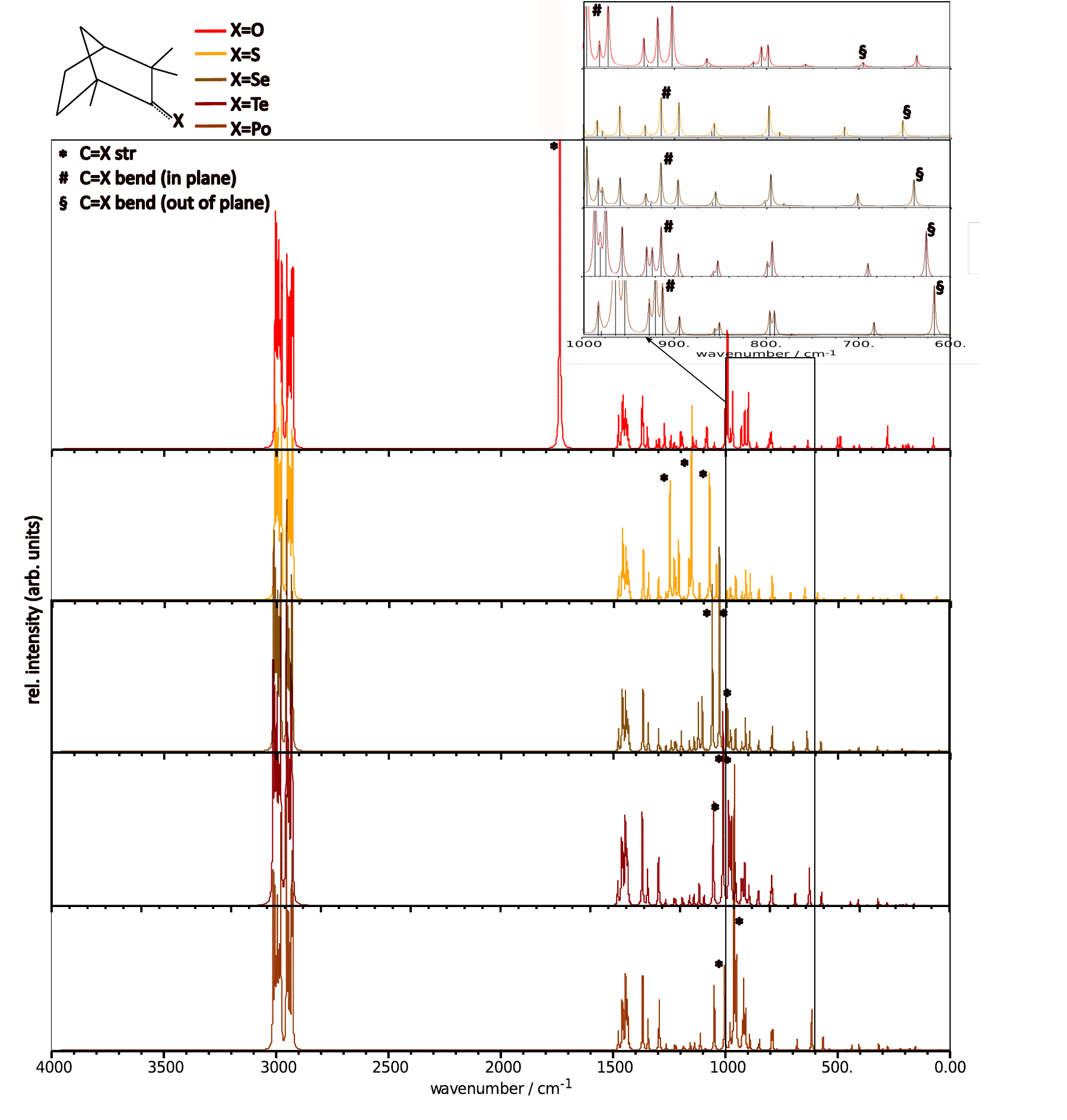}
  \caption{Comparison of calculated harmonic vibrational spectra of chalcogenofenchones. From top to bottom: \textbf{1-O} (Fenchone, top trace), \textbf{1-S} (Thiofenchone), \textbf{1-Se} (Selenofenchone, middle trace), \textbf{1-Te} (Tellurofenchone), \textbf{1-Po} (Polonofenchone, bottom trace). A systematic lowering in the wavenumbers, associated to the selected vibrational motions (C=X stretch and bend), with the increase in bond lengths and masses is presented in the overview or inset for each case. Computed spectra are displayed using an overall scaling factor of the harmonic vibrational wavenumbers of 0.968 and a Gaussian lineshape function with HWHM of \SI{1.5}{\per\centi\meter}.\label{fig:Xharmtrend}}
\end{figure}

\clearpage

\subsubsection{Vibrational spectra based on anharmonic force fields\label{sec:anharm}}

Both band positions, for which an overall scaling factor of the harmonic vibrational wavenumbers of 0.968 was used in the previous section, and intensities should be treated computationally beyond the harmonic approximation to achieve more accurate spectral profiles devoid of overall scaling factors. 
To exceed the harmonic approximation, we applied 
second-order vibrational perturbation theory (VPT2) based on DFT-level anharmonic force fields which included the full cubic force field together with the semi-diagonal part of the quartic force field. 

A statement about resonance situations, which are one of the major pitfalls in VPT2 approaches, at this point seems appropriate. VPT2 was used at the default settings, wherein for some modes---namely three modes of the dense set of asymmetric methyl \ce{CH}-bending and methylene scissoring modes in \textbf{1-O}, two modes of the same type in \textbf{1-S} and also two modes of this type together with two of the asymmetric methyl C-H stretching modes in \textbf{1-Se}---resonance situations that led to a lack of sufficient overlap between deperturbed and variational states were removed by keeping corresponding modes frozen. Wherever applicable, these modes are highlighted in blue in the spectra plotted from VPT2 and displayed at their (unscaled) harmonic vibrational wavenumber and the harmonically computed intensity to facilitate their simplified identification. 

Overall, experimentally observed IR features appear typically well reproduced in our theoretical spectra obtained from the anharmonic force field treatment. Thus, comparisons between experimental and VPT2-based spectra allow for elaborated assignment of the bands observed in the former cases and are presented in this section along with the spectra obtained within the double-harmonic approximation. 
The prediction of intensities seems more demanding, in general, than band positions. We found that with an improvement in intensities, the selected level of theory shows good agreement with experimental results, which renders it a suitable choice for the relevant systems, in particular when aiming at predictions of IR features of the as of yet unobserved chalcogenofenchones \textbf{1-Te} and \textbf{1-Po}. 

\begin{figure}[htbp]
\centering
  \includegraphics[width=0.8\textwidth]{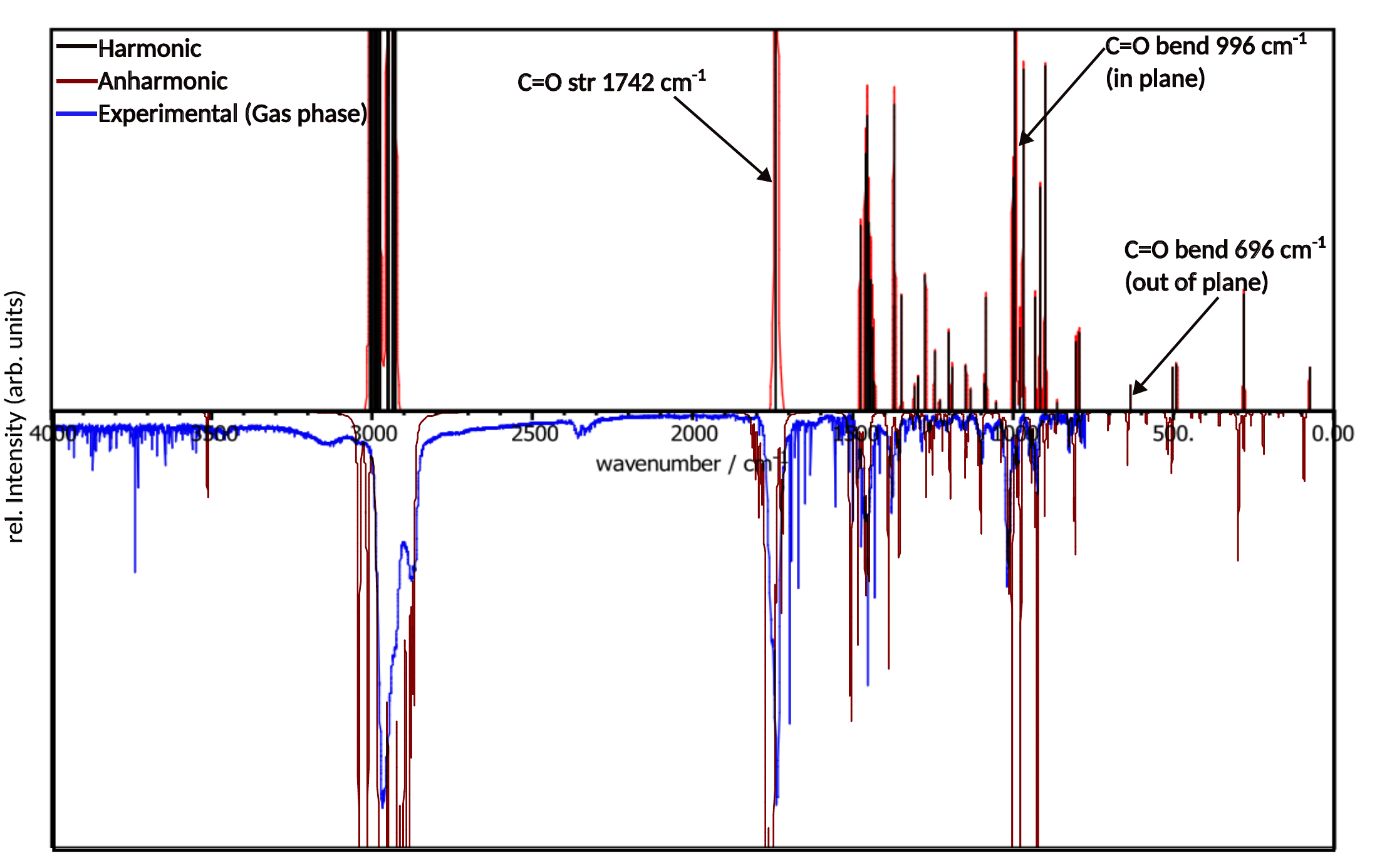}
  \caption{Comparison of theoretical harmonic (red envelope, black stick representation; directed upwards) and anharmonic (dark red envelope, various red shades for stick representation; directed downwards) IR spectra as well as experimental gas-phase (blue; directed downwards) IR spectrum of \textbf{1-O}. Theoretical harmonic spectrum was computed on the DFT level within the double-harmonic approximation and is displayed using an overall scaling factor of the harmonic vibrational wavenumbers of 0.968 and a Lorentzian lineshape function with HWHM of \SI{1.5}{\per\centi\meter}. The anharmonic contributions were calculated using second-order vibrational perturbation theory (VPT2) at the same level wherein some selected modes were kept frozen in order to treat the resonance problem of poorly overlapped states. For those modes kept frozen, the intensity calculated for the corresponding harmonic modes is then used also for the anharmonic case. A Lorentzian lineshape function with HWHM of \SI{1.5}{\per\centi\meter} is also used when displaying the anharmonic case. For a better visualization of the overall pattern, intensity (y-max) in the anharmonic spectrum is chosen in such a way that C=O stretching in anharmonic and experimental spectra align. Otherwise, the scale of harmonic and anharmonic spectra (y-max) is always kept same and is directly comparable. See text description of more technical details.\label{fig:Ovibanharmoverview}}
\end{figure}

\begin{figure}[htbp]
\centering
  \includegraphics[width=0.8\textwidth]{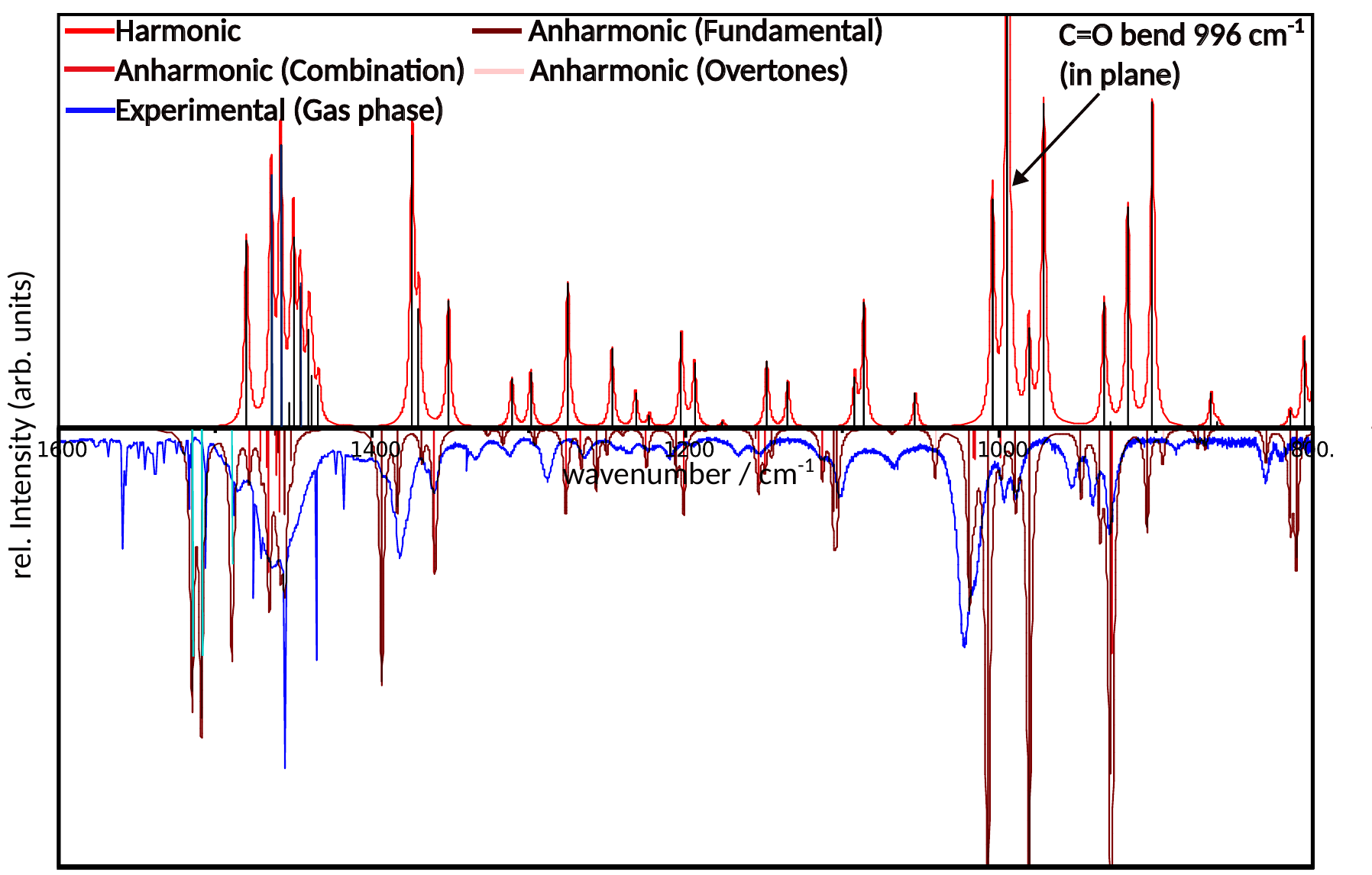}
  \caption{Theoretical harmonic (red envelope, black stick representation; directed upwards) and anharmonic (dark red envelope, various red shades for stick representations; directed downwards) IR spectra as well as (low-resolution) experimental gas-phase (blue; directed downwards) IR spectrum of \textbf{1-O} in the range of \qtyrange[range-phrase=--,range-units=single]{1600}{800}{\per\centi\meter}. In this version, the intensity scale (y-max) is chosen to be different from the parental overview spectrum (Figure~\ref{fig:Ovibanharmoverview}) to enhance the visualization clarity. Theoretical harmonic spectrum was computed on the DFT level within the double-harmonic approximation and is displayed using an overall scaling factor of the harmonic vibrational wavenumbers of 0.968 and a Lorentzian lineshape function with HWHM of \SI{1.5}{\per\centi\meter}. The same lineshape function was used in the anharmonic case. These anharmonic contributions were calculated using second-order vibrational perturbation theory (VPT2) at the same level wherein the selected modes (colored in cyan in the stick representation; directed downwards) were kept frozen in order to treat the resonance problem of poorly overlapped states. For the frozen modes, the intensity calculated for the corresponding harmonic modes is then used also for the anharmonic case. Different colors in the stick representation distinguish fundamentals from overtones and combination bands. See text description for more technical details.\label{fig:Ovibanharmdetail}}
\end{figure}

\begin{figure}[htbp]
\centering
  \includegraphics[width=0.8\textwidth]{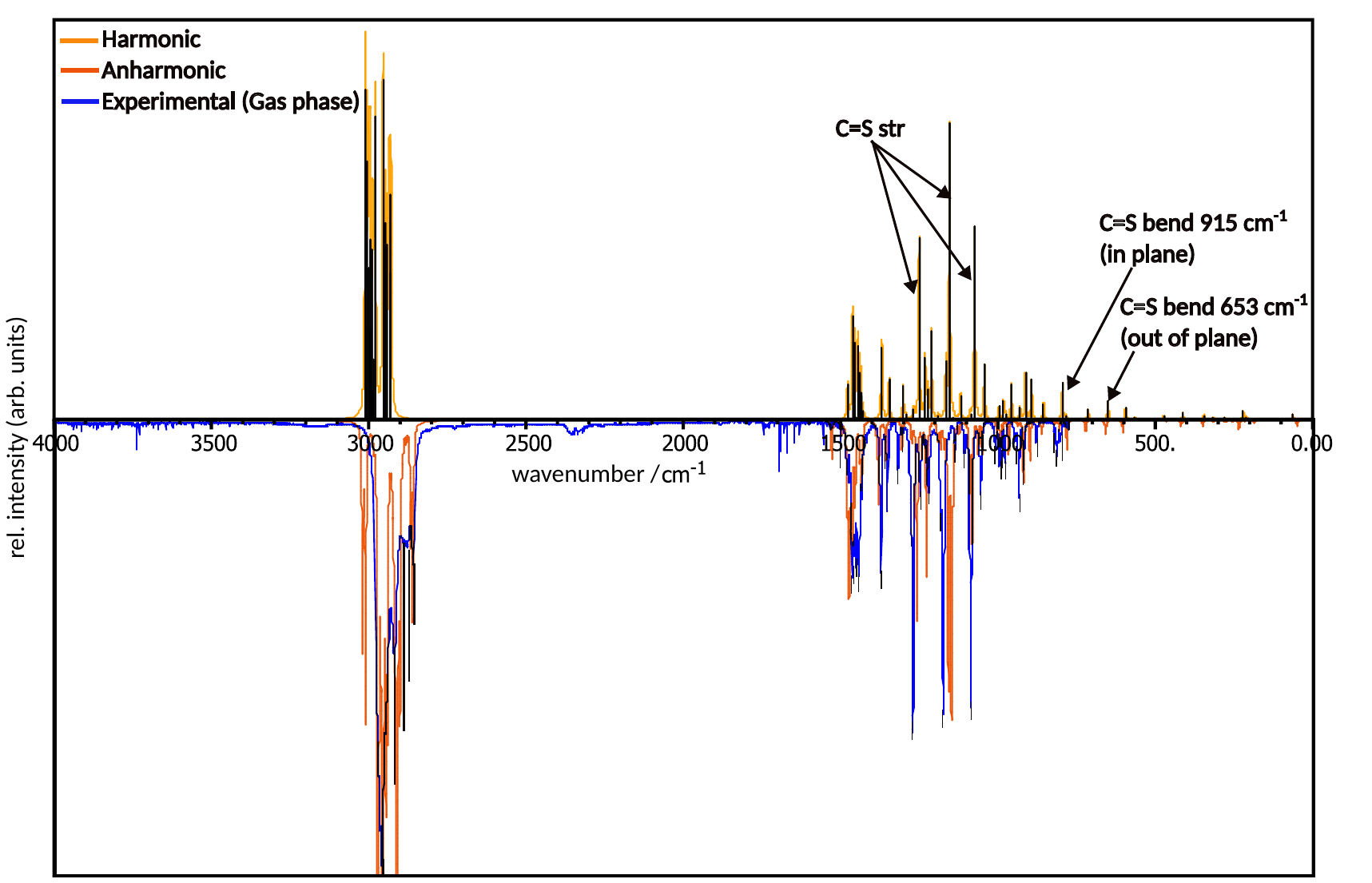}
  \caption{Comparison of theoretical harmonic (orange envelope, black stick representation; directed upwards) and anharmonic (dark orange envelope; directed downwards) IR spectra as well as experimental gas-phase (blue; directed downwards) IR spectrum of \textbf{1-S}. Theoretical harmonic spectrum was computed on the DFT level within the double-harmonic approximation and is displayed using an overall scaling factor of the harmonic vibrational wavenumbers of 0.968 and a Lorentzian lineshape function with HWHM of \SI{1.5}{\per\centi\meter}. The anharmonic contributions were calculated using second-order vibrational perturbation theory (VPT2) at the same level wherein some selected modes were kept frozen in order to treat the resonance problem of poorly overlapped states. For those frozen modes, the intensity calculated for the corresponding harmonic modes is then used also for the anharmonic case. A Lorentzian lineshape function with HWHM of \SI{1.5}{\per\centi\meter} is also used when displaying the anharmonic case. For a better visualization of the overall pattern, intensity (y-max) in the anharmonic spectrum is chosen in such a way that C=S stretching bands in anharmonic and experimental spectra align. Otherwise, the scale of harmonic and anharmonic spectra (y-max) is always kept same and is directly comparable. See text description of more technical details.\label{fig:Svibanharmoverview}}
\end{figure}

\begin{figure}[htbp]
\centering
  \includegraphics[width=0.8\textwidth]{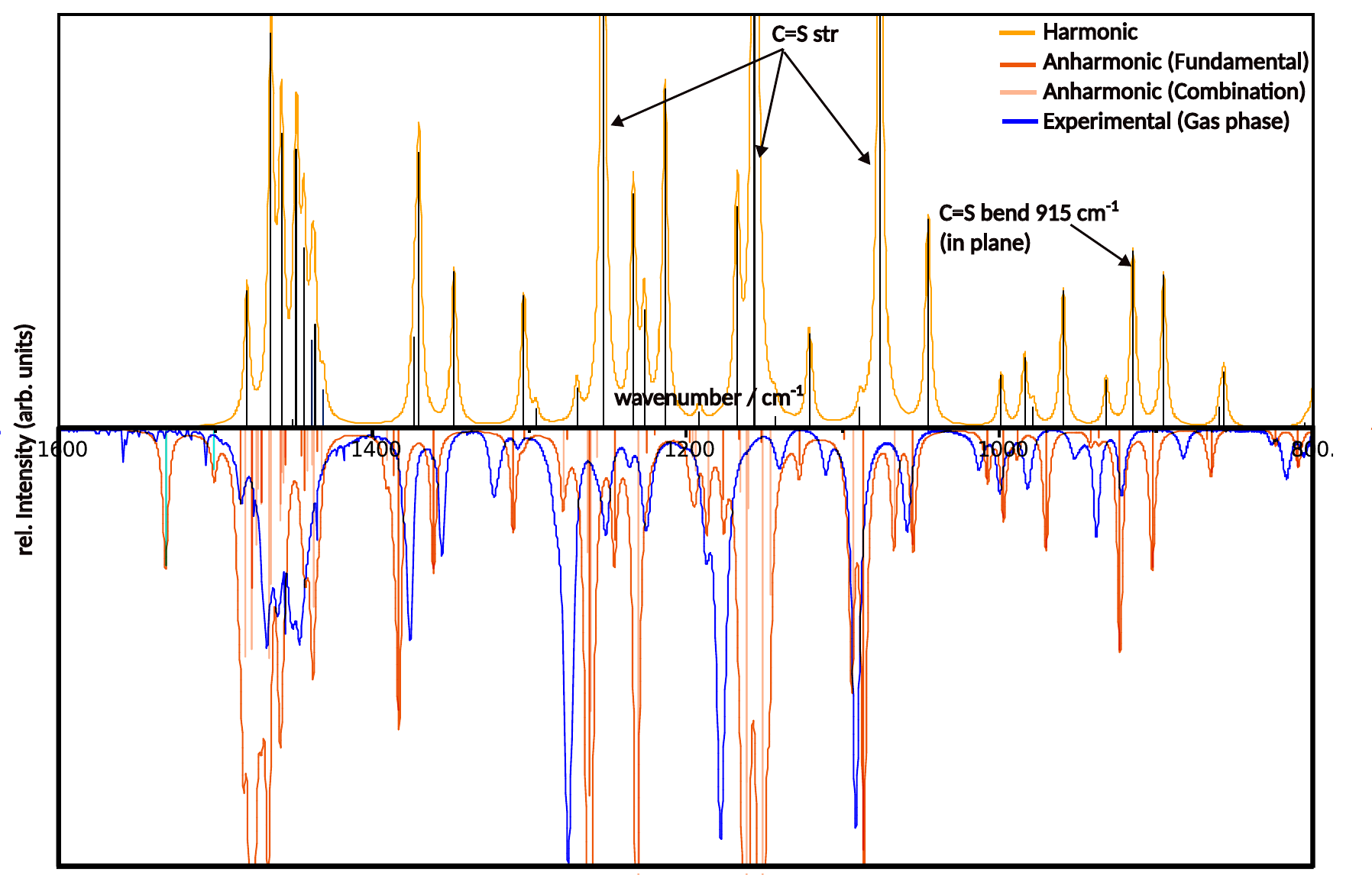}
  \caption{Theoretical harmonic (orange envelope, black stick representation; directed upwards) and anharmonic (dark orange, varying shades for stick representations; directed downwards) IR spectra as well as experimental gas-phase (blue; directed downwards) IR spectrum of \textbf{1-S} in the range of \qtyrange[range-phrase=--,range-units=single]{1600}{800}{\per\centi\meter}. In this version, the intensity scale (y-max) is chosen to be different from the parental overview spectrum (Figure~\ref{fig:Svibanharmoverview}) to enhance the visualization clarity. Theoretical harmonic spectrum was computed on the DFT level within the double-harmonic approximation and is displayed using an overall scaling factor of the harmonic vibrational wavenumbers of 0.968 and a Lorentzian lineshape function with HWHM of \SI{1.5} {\per\centi\meter}. The same lineshape function was used in the anharmonic case. These anharmonic contributions were calculated using second-order vibrational perturbation theory (VPT2) at the same level wherein the selected modes (colored in cyan in the stick representation; directed downwards) were kept frozen in order to treat the resonance problem of poorly overlapped states. For the frozen modes, the intensity calculated for the corresponding harmonic modes is then used also for the anharmonic case. Different colors in the stick representation distinguish fundamentals from overtones and combination bands. See text description of more technical details.\label{fig:Svibanharmdetails}}
\end{figure}

\begin{figure}[htbp]
\centering
  \includegraphics[width=1.0\textwidth]{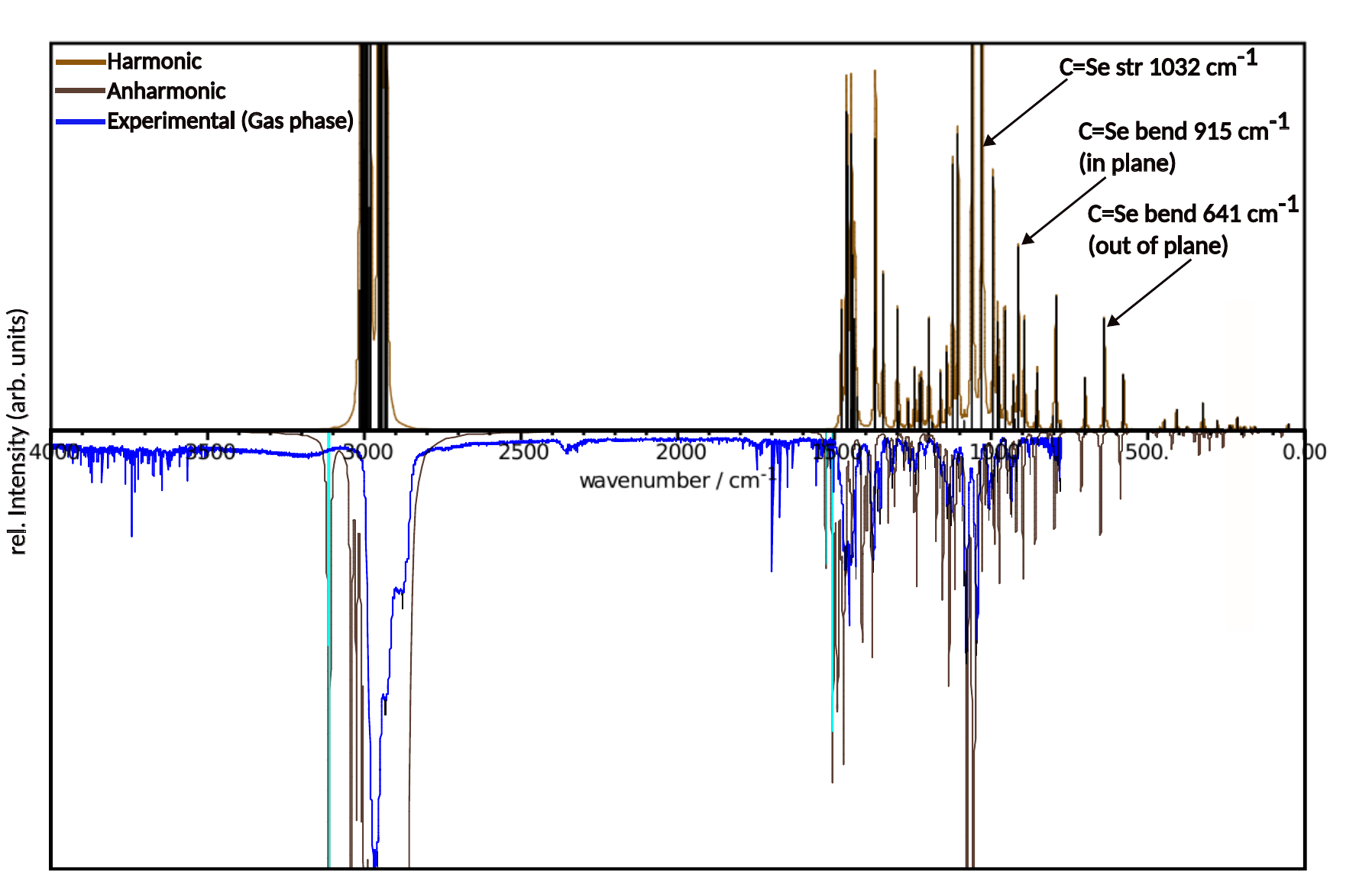}
  \caption{Comparison of theoretical harmonic (envelope and stick representation; directed upwards) and anharmonic (dark brown envelope; directed downwards) IR spectra as well as experimental gas-phase (blue; directed downwards) IR spectrum of 1-Se. Theoretical harmonic spectrum was computed on the DFT level within the double-harmonic approximation and is displayed using an overall scaling factor of the harmonic vibrational wavenumbers of 0.968 and a Lorentzian lineshape function with HWHM of \SI{1.5} {\per\centi\meter}. The anharmonic contributions were calculated using second-order vibrational perturbation theory (VPT2) at the same level wherein some selected modes were kept frozen in order to treat the resonance problem of poorly overlapped states. For the frozen modes, the intensity calculated for the corresponding harmonic modes is then used also for the anharmonic case (colored in cyan in the stick representation; directed downwards). A Lorentzian lineshape function with HWHM of \SI{1.5} {\per\centi\meter} is also used when displaying the anharmonic case. For a better visualization of the overall pattern, intensity (y-max) in the anharmonic spectrum is chosen in such a way that C=Se stretching in anharmonic and experimental spectra align. Otherwise, the scale of harmonic and anharmonic (y-max) is always kept same and is directly comparable. See text description of more technical details.\label{fig:Sevibanharmoverview}}
\end{figure}

\begin{figure}[htbp]
\centering
  \includegraphics[width=0.8\textwidth]{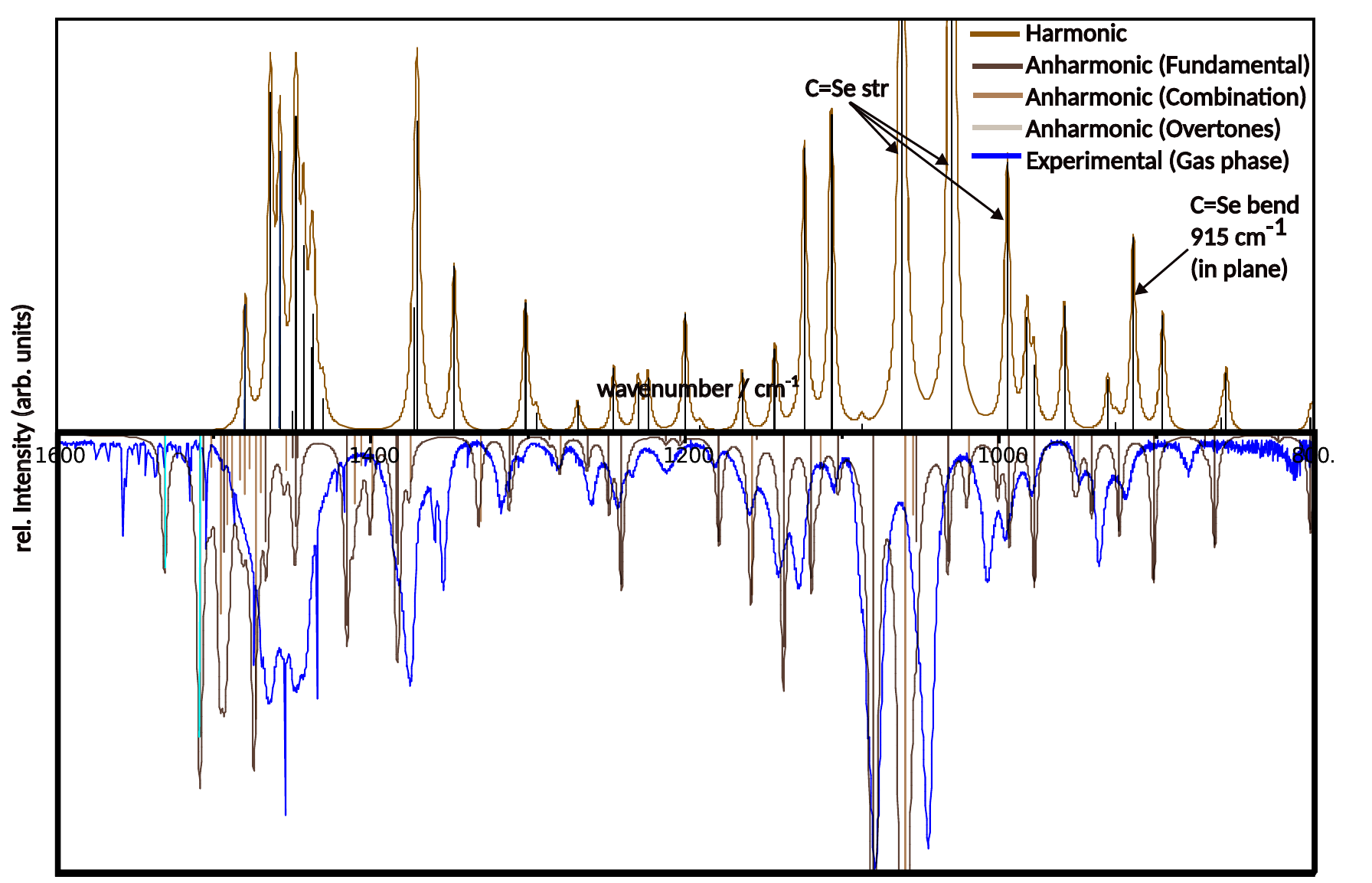}
  \caption{Theoretical harmonic (brown envelope, black stick representation; directed upwards) and anharmonic (dark brown envelope, various color shades for stick representation; directed downwards) IR spectra as well as experimental gas-phase (blue; directed downwards) IR spectrum of \textbf{1-Se} in the range of \qtyrange[range-phrase=--,range-units=single]{1600}{800}{\per\centi\meter}. In this version, the intensity scale (y-max) is chosen to be different from the parental overview spectrum (Figure~\ref{fig:Sevibanharmoverview}) to enhance the visualization clarity. Theoretical harmonic spectrum was computed on the DFT level within the double-harmonic approximation and is displayed using an overall scaling factor of the harmonic vibrational wavenumbers of 0.968 and a Lorentzian lineshape function with HWHM of \SI{1.5} {\per\centi\meter}. The same lineshape function was used in the anharmonic case. These anharmonic contributions were calculated using second-order vibrational perturbation theory (VPT2) at the same level wherein the selected modes (colored in cyan in the stick representation; directed downwards) were kept frozen in order to treat the resonance problem of poorly overlapped states. For the frozen modes, the intensity calculated for the corresponding harmonic modes is then used also for the anharmonic case. Different colors in the stick representation distinguish fundamentals from overtones and combination bands. See text description of more technical details.\label{fig:Sevibanharmdetails}}
\end{figure}

In the harmonic approximation, the calculated resonance wavenumbers are typically slightly on the higher side than experimentally observed fundamentals, and as such benefit from overall scaling. In contrast, the anharmonic approximations are much closer to the experimental wavenumbers, so no scaling of the calculated wavenumbers appears to be required.
We find for instance in all chalcogenofenchones except \textbf{1-Se} that in the VPT2 approach only two or three among the  C-H stretching fundamentals exceed the typical mark of \SI{3000}{\per\centi\meter}, below which aliphatic C-H stretching fundamentals remain, and then even mildly so by at most \SI{50}{\per\centi\meter}. Also the lower wavenumber end of the C-H stretching fundamentals is well reproduced with the lowest wavenumber being about \SI{5}{\per\centi\meter} higher than observed for \textbf{1-O} and \textbf{1-S}. The exceptional case of \textbf{1-Se} is expected to result primarily just from having kept two of the C-H stretching modes frozen in the VPT2 treatment.
When comparing with experimental spectra, the intensities are also found in general to be a little overestimated in the harmonic approximations (a direct comparison of harmonic and
anharmonic intensities is presented in all the cases, Figures~\ref{fig:Ovibanharmoverview}--\ref{fig:Sevibanharmdetails}).

An overview and an expanded spectrum of \textbf{1-O} is presented in Figures~\ref{fig:Ovibanharmoverview} and \ref{fig:Ovibanharmdetail}, respectively, whereas a selection of most prominent first overtones and combination bands is provided in Table~\ref{tab:O-overtones}. 

The C=O stretching band observed at \SI{1742}{\per\centi\meter} comprises of both fundamental and combination contributions with almost minimal presence of prominent overtones near this wavenumber region. Thus, the experimentally observed shoulder (peak 31 in Figure~\ref{fig:Ovib} and Table~\ref{tab:Ofundamentals}) can tentatively be assigned as arising from the combination band $\nu_{55}+\nu_{48}$. Peak 38, in contrast, is ascribed to the first overtone of the C=O stretching mode. Further
overtones are pronounced in the CH stretching region, but give  negligible contributions in the fingerprint part of the spectrum. On the other hand, combination bands are present in both CH stretching and lower wavenumber regions with accountable intensities. In particular worth mentioning is the combination band $\nu_{71}+\nu_{39}$ that accounts for the otherwise missing fourth peak of the peak group 25--28 in the region of the characteristic doublet caused by the \ce{CH3}-umbrella motions of the geminal dimethyl substituents. As per the redistributed intensity pattern within the VPT2 approach (see Table~\ref{tab:Ofundamentals}) and the more favorable alignment of the transition wavenumbers, one could also tend to assign peak 4 to $\nu_{50}$ and include the comparatively intense combination band $\nu_{62}+\nu_{64}$ as a candidate for peak 5. For the frozen mode fundamentals $\nu_{19}, \nu_{20}, \nu_{24}$, the intensities calculated for the corresponding harmonic modes are included for the anharmonic case in Figures~\ref{fig:Ovibanharmoverview} and \ref{fig:Ovibanharmdetail}. 

\newrobustcmd{\B}{\color{blue}}
\begin{table}[htp]
\footnotesize
\caption{Transition wavenumbers ($\tilde{\nu}_\mathrm{theo}$) and integrated absorption coefficients ($A_\mathrm{theo}$) of prominent first overtones (\textcolor{blue}{blue}) and combination bands (black) of \textbf{1-O} computed at DFT/B3LYP within second-order vibrational perturbation theory.\label{tab:O-overtones}}
\begin{tabular}{l S[detect-weight,round-mode=places,round-precision=1,group-digits = none] S[round-mode=places,round-precision=1,group-digits = none] l S[round-mode=places,round-precision=1,group-digits = none] S[detect-weight,round-mode=places,round-precision=1,group-digits = none]}
\hline
Mode & $\tilde{\nu}_\mathrm{theo}$ & $A_\mathrm{theo}$ & 
Mode & $\tilde{\nu}_\mathrm{theo}$ & $A_\mathrm{theo}$ \\
     &  \multicolumn{1}{c}{(cm$^{-1}$)} & \multicolumn{1}{c}{(\SI{}{\kilo\meter\per\mol})} &
     &  \multicolumn{1}{c}{(cm$^{-1}$)} & \multicolumn{1}{c}{(\SI{}{\kilo\meter\per\mol})} \\
\hline
\textcolor{blue}{$2\nu_{17}$} & \B 3522.313 & \B   2.95024445 &   
\textcolor{blue}{$2\nu_{18}$} & \B 2974.431 & \B 128.31759380 \\
\textcolor{blue}{$2\nu_{21}$} & \B 2956.512 & \B  11.93223428 &   
\textcolor{blue}{$2\nu_{22}$} & \B 2917.652 & \B   8.06442071 \\
\textcolor{blue}{$2\nu_{23}$} & \B 2910.864 & \B   6.13919242 &   
\textcolor{blue}{$2\nu_{25}$} & \B 2884.730 & \B   8.37512937 \\
\textcolor{blue}{$2\nu_{26}$} & \B 2880.618 & \B   6.52903034 &   
\textcolor{blue}{$2\nu_{51}$} & \B 1811.255 & \B   2.03467638 \\\hline
$\nu_{37}$+$\nu_{17}$ & 2976.123 &   2.32371134 & 
$\nu_{22}$+$\nu_{18}$ & 2953.489 &  12.41750196 \\ 
$\nu_{21}$+$\nu_{18}$ & 2949.206 &  94.72285987 & 
$\nu_{26}$+$\nu_{21}$ & 2947.219 &  17.89618925 \\ 
$\nu_{23}$+$\nu_{18}$ & 2944.035 &   4.00630187 & 
$\nu_{25}$+$\nu_{18}$ & 2934.633 &  17.38356562 \\ 
$\nu_{22}$+$\nu_{21}$ & 2915.059 &  13.15452860 & 
$\nu_{23}$+$\nu_{22}$ & 2895.028 &  35.80690686 \\ 
$\nu_{55}$+$\nu_{47}$ & 1800.326 &   2.23345373 & 
$\nu_{57}$+$\nu_{44}$ & 1790.912 &   2.23291831 \\ 
$\nu_{52}$+$\nu_{51}$ & 1779.745 &   3.81743865 & 
$\nu_{61}$+$\nu_{33}$ & 1777.829 &   5.81086087 \\ 
$\nu_{53}$+$\nu_{51}$ & 1774.521 &   4.47617387 & 
$\nu_{63}$+$\nu_{30}$ & 1773.688 &  10.40668567 \\ 
$\nu_{61}$+$\nu_{34}$ & 1771.823 &  10.05883475 & 
$\nu_{54}$+$\nu_{48}$ & 1766.888 &   2.02182340 \\ 
$\nu_{57}$+$\nu_{46}$ & 1765.705 &   3.78547440 & 
$\nu_{62}$+$\nu_{33}$ & 1764.810 &   3.05376994 \\ 
$\nu_{67}$+$\nu_{21}$ & 1763.013 &   2.96692886 & 
$\nu_{55}$+$\nu_{48}$ & 1758.382 & 103.03531412 \\ 
$\nu_{54}$+$\nu_{50}$ & 1752.953 &  15.00316156 & 
$\nu_{55}$+$\nu_{50}$ & 1745.637 &   2.30966100 \\ 
$\nu_{54}$+$\nu_{51}$ & 1729.284 &   6.16324558 & 
$\nu_{55}$+$\nu_{51}$ & 1722.460 &   2.91614353 \\ 
$\nu_{74}$+$\nu_{32}$ & 1466.122 &   4.89388864 & 
$\nu_{70}$+$\nu_{34}$ & 1458.896 &   2.86455539 \\ 
$\nu_{71}$+$\nu_{39}$ & 1360.540 &   4.92607829 & 
$\nu_{73}$+$\nu_{41}$ & 1277.061 &   2.90639948 \\ 
$\nu_{70}$+$\nu_{46}$ & 1201.971 &   2.81094881 & 
$\nu_{64}$+$\nu_{62}$ & 928.452 &    7.70957816 \\
\hline 
\end{tabular}
\end{table}

\clearpage

With this we move on to thiofenchone's case: Table~\ref{tab:Sfundamentals} contains besides the harmonic also the anharmonic fundamental wavenumbers  and dipole strengths predicted for \textbf{1-S}. Quantum chemically computed values are compared with the experimental wavenumbers, where available, and a vibrational mode assignment is provided. As becomes evident from Table~\ref{tab:Sfundamentals}, in addition to the expected most intense bands of the C-H stretching region, further intense bands are present in the range of \qtyrange[range-phrase=--,range-units=single]{1500}{1000}{\per\centi\meter} for \textbf{1-S}. The anharmonic approximation no longer calls for scaling the transition wavenumbers, which was needed within the harmonic approximation.

Overtones and combination bands are not included in Table~\ref{tab:Sfundamentals} and it is not surprising that such modes are not readily apparent in the measured spectrum. Nevertheless, most of the prominent fundamental modes from the asymmetric methyl bending and methylene scissoring region near \SI{1500}{\per\centi\meter} with large intensities are predicted to have also comparatively intense overtones that acquire intensity and show up in the C-H stretching region below \SI{3000}{\per\centi\meter} as compiled in Table~\ref{tab:S-overtones}. Other than these modes, overtones with much smaller intensity are not listed in the table. A full assignment of the spectra in these two regions is cumbersome because of several possible interactions, however, a direct comparison of the computed IR spectra and the experimental of \textbf{1-S} is presented in Figures~\ref{fig:Svibanharmoverview} and \ref{fig:Svibanharmdetails}.

The vibrational bands present in the C-H str region are expectantly most intense. However, the fingerprint section 
below \SI{1300}{\per\centi\meter} is dominated by signals from C=S stretching motions which needs to be discussed in details.

As evident from Figures~\ref{fig:Svibanharmoverview} and \ref{fig:Svibanharmdetails}, there are no overtones with significant electric dipole absorption cross section in the lower wavenumber region and most
prominent bands there are assigned clearly to fundamental and combination bands. As a
common trend, the overtones with accountable intensity are mainly found in the C-H stretching region in all
chalcogenofenchones including \textbf{1-S}. Further, it follows from Figure~\ref{fig:Svibanharmdetails} and Table~\ref{tab:S-overtones} that there emerges specifically a notable intensity redistribution from bands with C=S stretching character to combination bands in the VPT2-based description of the anharmonic spectrum. The intense harmonic mode $\nu_{38}$ specifically corresponds C=S stretching in
the vibrational assignment. The corresponding most intense band in this region, observed in experiment at about \SI{1179}{\per\centi\meter}, would according to VPT2 then majorly benefiting from combination band ($\nu_{73}+\nu_{47}$) whereas two other strong bands at \qtylist[list-units=bracket]{1062;995}{\per\centi\meter} would in this framework be ascribed to nearly equal contributions from fundamental and combination bands. In contrast to \textbf{1-O},
the intensity of combination bands appears thus greatly amplified with respect to the fundamental band's intensity
in the case of \textbf{1-S} within the perturbative anharmonic treatment.

\begin{table}[htp]
\footnotesize
\caption{Transition wavenumbers ($\tilde{\nu}_\mathrm{theo}$) and integrated absorption coefficients ($A_\mathrm{theo}$) of most prominent first overtones (\textcolor{blue}{blue}) and combination bands (black) of \textbf{1-S} computed at DFT/B3LYP within second-order vibrational perturbation theory.\label{tab:S-overtones}}
\begin{tabular}{l S[detect-weight,round-mode=places,round-precision=1,group-digits = none] S[round-mode=places,round-precision=1,group-digits = none] l S[round-mode=places,round-precision=1,group-digits = none] S[detect-weight,round-mode=places,round-precision=1,group-digits = none]}
\hline
Mode & $\tilde{\nu}_\mathrm{theo}$ & $A_\mathrm{theo}$ & 
Mode & $\tilde{\nu}_\mathrm{theo}$ & $A_\mathrm{theo}$ \\
     &  \multicolumn{1}{c}{(cm$^{-1}$)} & \multicolumn{1}{c}{(\SI{}{\kilo\meter\per\mol})} &
     &  \multicolumn{1}{c}{(cm$^{-1}$)} & \multicolumn{1}{c}{(\SI{}{\kilo\meter\per\mol})} \\
\hline
\textcolor{blue}{$2\nu_{18}$} & \B 2969.5 & \B 21.8 & 
\textcolor{blue}{$2\nu_{19}$} & \B 2927.2 & \B  5.2 \\
\textcolor{blue}{$2\nu_{21}$} & \B 2909.4 & \B 10.6 & 
\textcolor{blue}{$2\nu_{23}$} & \B 2884.2 & \B  1.4 \\
\textcolor{blue}{$2\nu_{24}$} & \B 2863.2 & \B 22.6 &
\textcolor{blue}{$2\nu_{25}$} & \B 2859.0 & \B  2.9 \\\hline
$\nu_{19}$+$\nu_{18}$ & 2948.8 & 37.3 &  
$\nu_{21}$+$\nu_{19}$ & 2922.8 &  4.2 \\
$\nu_{22}$+$\nu_{18}$ & 2921.8 &  8.5 & 
$\nu_{22}$+$\nu_{19}$ & 2919.1 &  2.9 \\
$\nu_{23}$+$\nu_{18}$ & 2918.6 &  3.0 & 
$\nu_{23}$+$\nu_{19}$ & 2904.5 &  5.1 \\
$\nu_{25}$+$\nu_{21}$ & 2902.8 & 12.1 & 
$\nu_{61}$+$\nu_{45}$ & 1485.1 &  2.6 \\
$\nu_{60}$+$\nu_{51}$ & 1482.2 &  7.9 & 
$\nu_{63}$+$\nu_{43}$ & 1477.8 &  3.8 \\
$\nu_{62}$+$\nu_{44}$ & 1477.7 &  7.6 & 
$\nu_{66}$+$\nu_{39}$ & 1474.6 &  4.0 \\
$\nu_{58}$+$\nu_{55}$ & 1471.1 &  2.7 & 
$\nu_{62}$+$\nu_{45}$ & 1466.7 &  7.9 \\
$\nu_{64}$+$\nu_{43}$ & 1465.8 &  5.4 &
$\nu_{61}$+$\nu_{47}$ & 1459.3 &  3.2 \\
$\nu_{74}$+$\nu_{31}$ & 1437.7 &  6.2 &
$\nu_{66}$+$\nu_{48}$ & 1263.1 &  4.3 \\
$\nu_{62}$+$\nu_{56}$ & 1261.8 &  9.8 &
$\nu_{63}$+$\nu_{55}$ & 1231.4 & 18.7 \\
$\nu_{60}$+$\nu_{59}$ & 1186.5 &  3.0 & 
$\nu_{70}$+$\nu_{48}$ & 1161.9 & 12.2 \\
$\nu_{65}$+$\nu_{64}$ & 1161.7 & 16.2 &
$\nu_{74}$+$\nu_{45}$ & 1161.4 &  2.8 \\
$\nu_{73}$+$\nu_{47}$ & 1152.1 & 33.9 &
$\nu_{69}$+$\nu_{50}$ & 1146.6 &  5.7 \\
$\nu_{61}$+$\nu_{59}$ & 1094.2 &  8.0 & 
$\nu_{61}$+$\nu_{60}$ & 1067.3 &  3.8 \\
$\nu_{73}$+$\nu_{59}$ &  785.3 &  2.8 &
                      &        &      \\ 
\hline
\end{tabular}
\end{table}

\clearpage

The case of selenofenchone \textbf{1-Se} resembles in several aspects more thiofenchone than the parent compound of this series, fenchone. This is mostly caused by the fact that C=X stretching modes in \textbf{1-Se} and \textbf{1-S} dive into the fingerprint part of the spectrum, in the region of skeleton modes and the \ce{CH2} twisting and wagging fundamentals, and thus lead to pronounced couplings to these, whereas the C=X stretching mode in \textbf{1-O} resides in an otherwise isolated region of the spectrum outside the fingerprint range.

Above, we have already alluded to some of the caveats of our VPT2 treatment, which led us to keep four vibrational modes frozen as a remedy to cope with otherwise rather poor overlap between deperturbed states on the variational ones. This certainly limits the quality of the description in the C-H stretching region (see overview spectrum in Figure~\ref{fig:Sevibanharmoverview}) and also affects the spectral region populated by \ce{CH2} scissoring and asymmetric \ce{CH3} bending modes (see Figure~\ref{fig:Sevibanharmdetails}). In the case of \textbf{1-Se}, this seems to impact also slightly unfavorably on the spectral representation of the nearby doublet for the geminal dimethyl substituents that is predicted at too high a wavenumber (about \SI{30}{\per\centi\meter} off).

On the other hand, the region of peaks 18--14 (see Figure~\ref{fig:Sevib} for the numbering) in the range \qtyrange[range-phrase=--,range-units=single]{1200}{1100}{\per\centi\meter} appears greatly improved in the VPT2 treatment and features also two prominent combination bands that are listed besides other combination bands and first overtones with appreciable intensity in Table~\ref{tab:Se-overtones}.

As for \textbf{1-S}, we also notice considerable redistribution from intensity of prominent fundamentals to combination bands. Most interesting is here the behavior of modes with C=Se stretching character: In the double-harmonic approximation, mode $\nu_{43}$ at \SI{1032}{\per\centi\meter} (after scaling) gives rise to the strongest signal in the fingerprint region. But in the VPT2 treatment, which shifts this fundamental to \SI{1053}{\per\centi\meter}, its intensity is predicted to be mostly redistributed to the nearly degenerate combination bands $\nu_{69}+\nu_{55}$ and $\nu_{65}+\nu_{57}$ emerging at \SI{1061}{\per\centi\meter}. This points to the limitation of using only the semi-diagonal part of the quartic force field in the VPT2 that lacks the coupling required to further split the two combination bands and reshuffle intensity as to reproduce the experimental pattern of peaks 12--9, with the two shoulders observed as peaks 12 and 10 at \SI{1085}{\per\centi\meter} and \SI{1053}{\per\centi\meter}, respectively. 

In the region of peaks 5--2, in the VPT2 treatment $\nu_{49}$ at \SI{953}{\per\centi\meter} and the combination band $\nu_{72}+\nu_{57}$ can tentatively be attributed to account for peak 5 observed at \SI{950}{\per\centi\meter}, whereas the in-plane C=Se bending mode $\nu_{50}$ at \SI{924}{\per\centi\meter} may be shifted slightly by \SI{14}{\per\centi\meter} to merge with $\nu_{48}$ predicted at \SI{942}{\per\centi\meter} to become peak 4 at \SI{938}{\per\centi\meter}, $\nu_{51}$ at \SI{902}{\per\centi\meter} may be shifted by \SI{19}{\per\centi\meter} to become peak 3 at \SI{922}{\per\centi\meter} and finally $\nu_{52}$ and $\nu_{53}$ at \SI{870}{\per\centi\meter} and \SI{863}{\per\centi\meter}, respectively could be shifted by about \SI{12}{\per\centi\meter} to become peak 2 at \SI{882}{\per\centi\meter}. This indicates that also for the region involving the in-plane C=Se bending mode intensity becomes redistributed, including combination bands.

\clearpage

\begin{table}[htp]
\footnotesize
\caption{Transition wavenumbers ($\tilde{\nu}_\mathrm{theo}$) and integrated absorption coefficients ($A_\mathrm{theo}$) of some of the most prominent first overtones (\textcolor{blue}{blue}) and combination bands (black) of \textbf{1-Se} computed at DFT/B3LYP within second-order vibrational perturbation theory.\label{tab:Se-overtones}}
\begin{tabular}{l S[detect-weight,round-mode=places,round-precision=1,group-digits = none] S[round-mode=places,round-precision=1,group-digits = none] l S[round-mode=places,round-precision=1,group-digits = none] S[detect-weight,round-mode=places,round-precision=1,group-digits = none]}
\hline
Mode & $\tilde{\nu}_\mathrm{theo}$ & $A_\mathrm{theo}$ & 
Mode & $\tilde{\nu}_\mathrm{theo}$ & $A_\mathrm{theo}$ \\
     &  \multicolumn{1}{c}{(cm$^{-1}$)} & \multicolumn{1}{c}{(\SI{}{\kilo\meter\per\mol})} &
     &  \multicolumn{1}{c}{(cm$^{-1}$)} & \multicolumn{1}{c}{(\SI{}{\kilo\meter\per\mol})} \\
\hline
\textcolor{blue}{$2\nu_{18}$} & \B 3003.175 & \B  10.85350429 &
\textcolor{blue}{$2\nu_{21}$} & \B 2946.384 & \B 181.94746856 \\
\textcolor{blue}{$2\nu_{20}$} & \B 2938.371 & \B   5.08731324 &
\textcolor{blue}{$2\nu_{22}$} & \B 2907.560 & \B 212.40007754 \\
\textcolor{blue}{$2\nu_{25}$} & \B 2890.424 & \B   5.05603038 &
\textcolor{blue}{$2\nu_{24}$} & \B 2872.361 & \B 496.09692290 \\\hline
$\nu_{75}$+$\nu_{13}$ & 3011.940 &  9.13650643 &
$\nu_{23}$+$\nu_{18}$ & 2984.900 & 60.11537763 \\
$\nu_{20}$+$\nu_{18}$ & 2965.359 & 14.72024984 &
$\nu_{22}$+$\nu_{18}$ & 2954.797 &  5.29433487 \\
$\nu_{21}$+$\nu_{20}$ & 2935.010 & 10.29181260 &
$\nu_{22}$+$\nu_{20}$ & 2918.768 & 63.65534028 \\
$\nu_{25}$+$\nu_{21}$ & 2898.558 & 55.88419513 &
$\nu_{62}$+$\nu_{44}$ & 1496.496 &  6.21226941 \\
$\nu_{72}$+$\nu_{32}$ & 1492.508 &  3.17220712 &
$\nu_{61}$+$\nu_{46}$ & 1490.279 &  2.22707891 \\
$\nu_{67}$+$\nu_{37}$ & 1481.688 &  2.15886429 &
$\nu_{60}$+$\nu_{51}$ & 1474.938 &  9.81369508 \\
$\nu_{58}$+$\nu_{54}$ & 1471.708 &  2.09726643 &
$\nu_{64}$+$\nu_{44}$ & 1411.608 &  2.47841490 \\
$\nu_{63}$+$\nu_{47}$ & 1400.848 &  2.97072915 &
$\nu_{69}$+$\nu_{42}$ & 1331.952 &  3.08056259 \\
$\nu_{75}$+$\nu_{42}$ & 1158.584 &  5.57370804 &
$\nu_{69}$+$\nu_{55}$ & 1061.473 & 22.84065749 \\
$\nu_{65}$+$\nu_{57}$ & 1061.078 & 30.32463242 &
$\nu_{61}$+$\nu_{60}$ & 1056.265 &  2.84200600 \\
$\nu_{67}$+$\nu_{57}$ & 1021.258 &  2.14713169 &
$\nu_{72}$+$\nu_{57}$ &  951.811 &  1.47734485 \\
\hline
\end{tabular}
\end{table}

\clearpage

\begin{figure}[htbp]
\centering
  \includegraphics[width=1.0\textwidth]{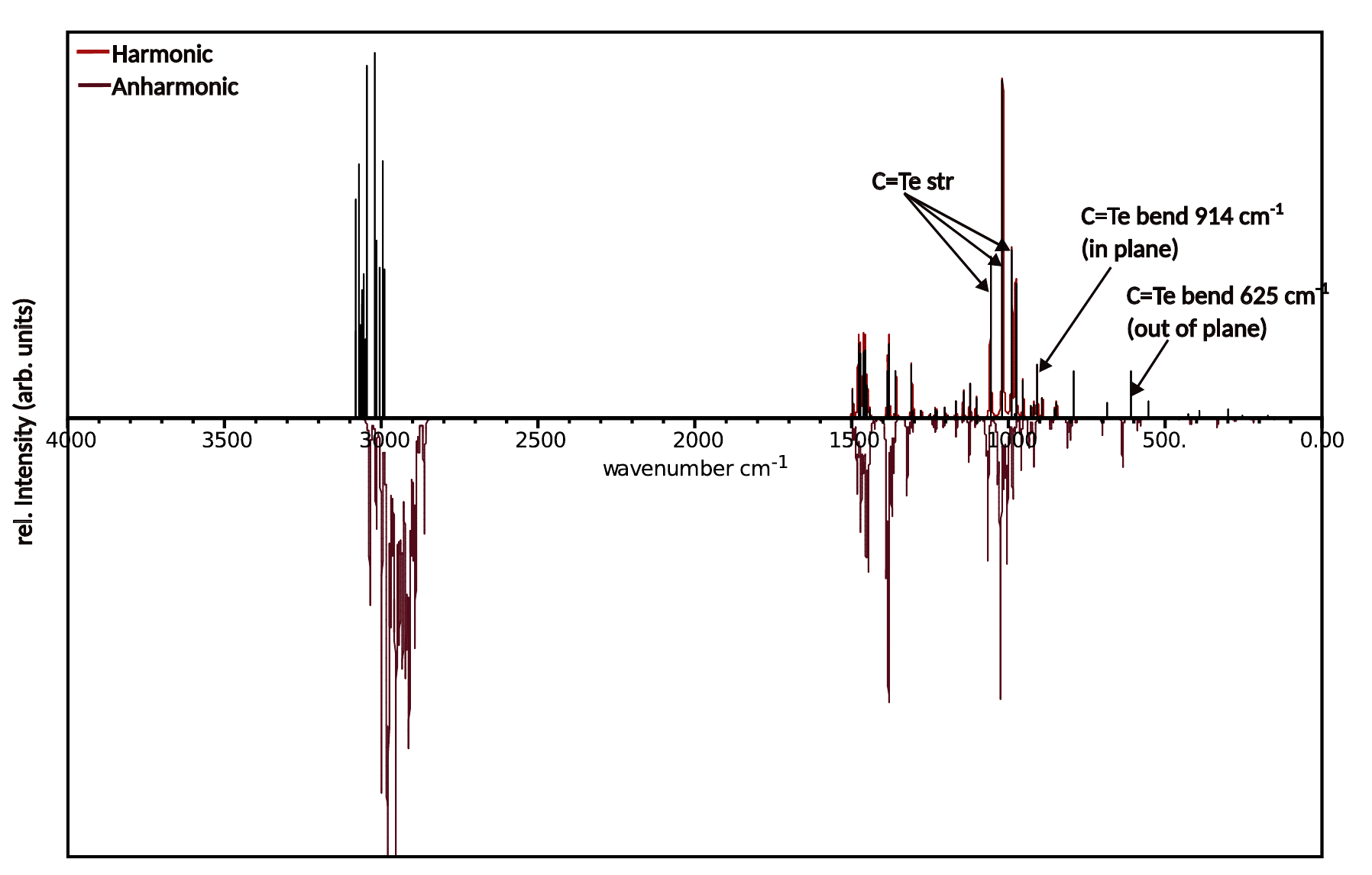}
  \caption{Comparison of theoretical harmonic (envelope and stick representation; directed upwards) and anharmonic (envelope; directed downwards) IR spectra of \textbf{1-Te}. Theoretical harmonic spectrum was computed on the DFT level within the double-harmonic approximation and is displayed using an overall scaling factor of the harmonic vibrational wavenumbers of 0.968 and a Lorentzian lineshape function with HWHM of \SI{1.5} {\per\centi\meter}. The anharmonic contributions were calculated using second-order vibrational perturbation theory (VPT2) and are plotted with the same lineshape function.\label{fig:Tevibanharmoverview}}
  \end{figure}

\begin{figure}[htbp]
\centering
  \includegraphics[width=0.8\textwidth]{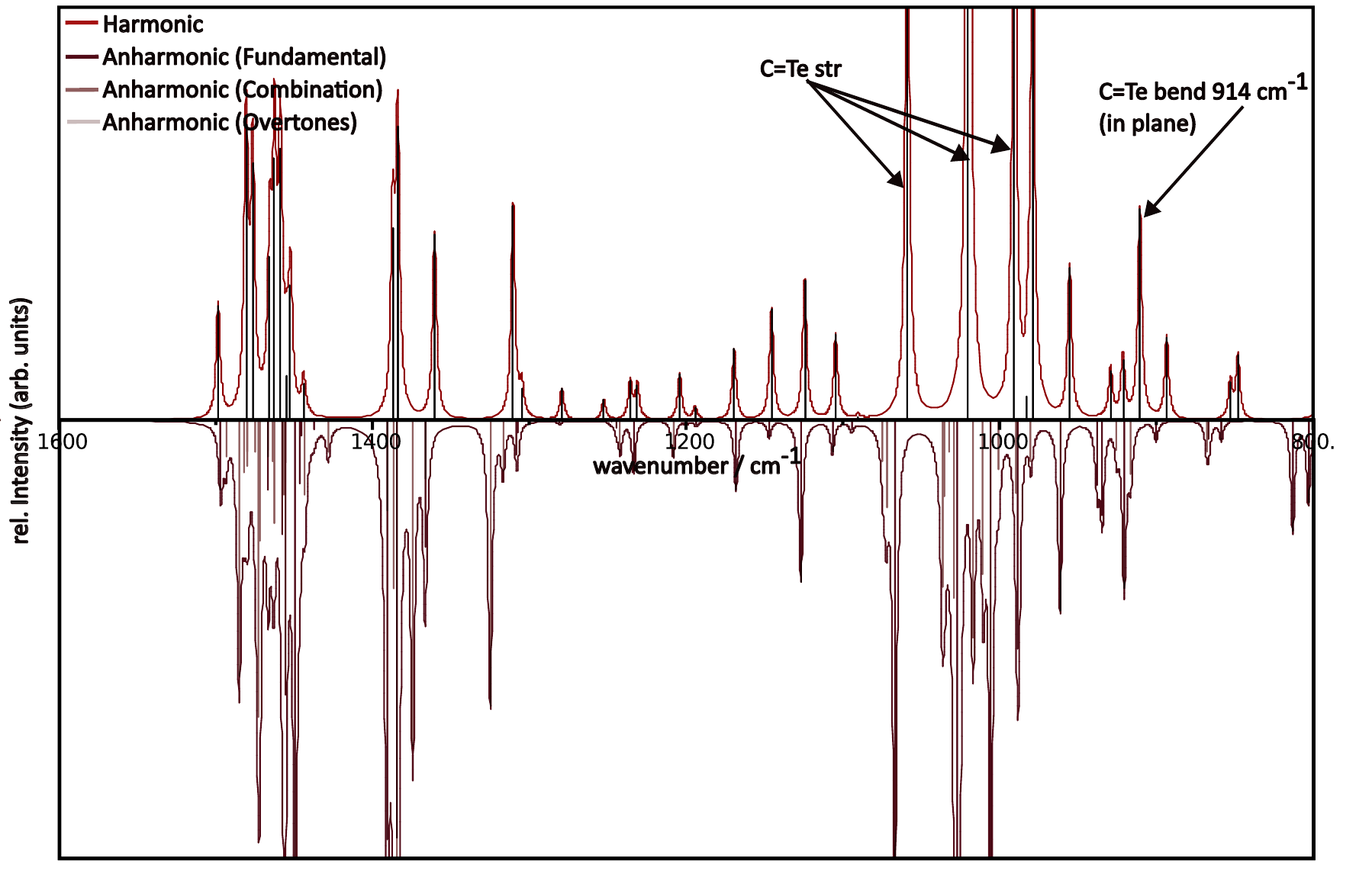}
  \caption{Theoretical harmonic (envelope and stick representation; directed upwards) and anharmonic (envelope and stick representation; directed downwards) IR spectra of \textbf{1-Te} in the range of \qtyrange[range-phrase=--,range-units=single]{1600}{800}{\per\centi\meter}. In this version, the intensity scale (y-max) is chosen to be different from the parental overview spectrum (Figure~\ref{fig:Tevibanharmoverview}) to enhance the visualization clarity. Theoretical harmonic spectrum was computed on the DFT level within the double-harmonic approximation and is displayed using an overall scaling factor of the harmonic vibrational wavenumbers of 0.968 and a Lorentzian lineshape function with HWHM of \SI{1.5} {\per\centi\meter}. The same lineshape function was used in the anharmonic case. These anharmonic contributions were calculated using second-order vibrational perturbation theory (VPT2) at the same level. See text description of more technical details.\label{fig:Tevibanharmdetails}}
\end{figure}

\begin{figure}[htbp]
\centering
  \includegraphics[width=1.0\textwidth]{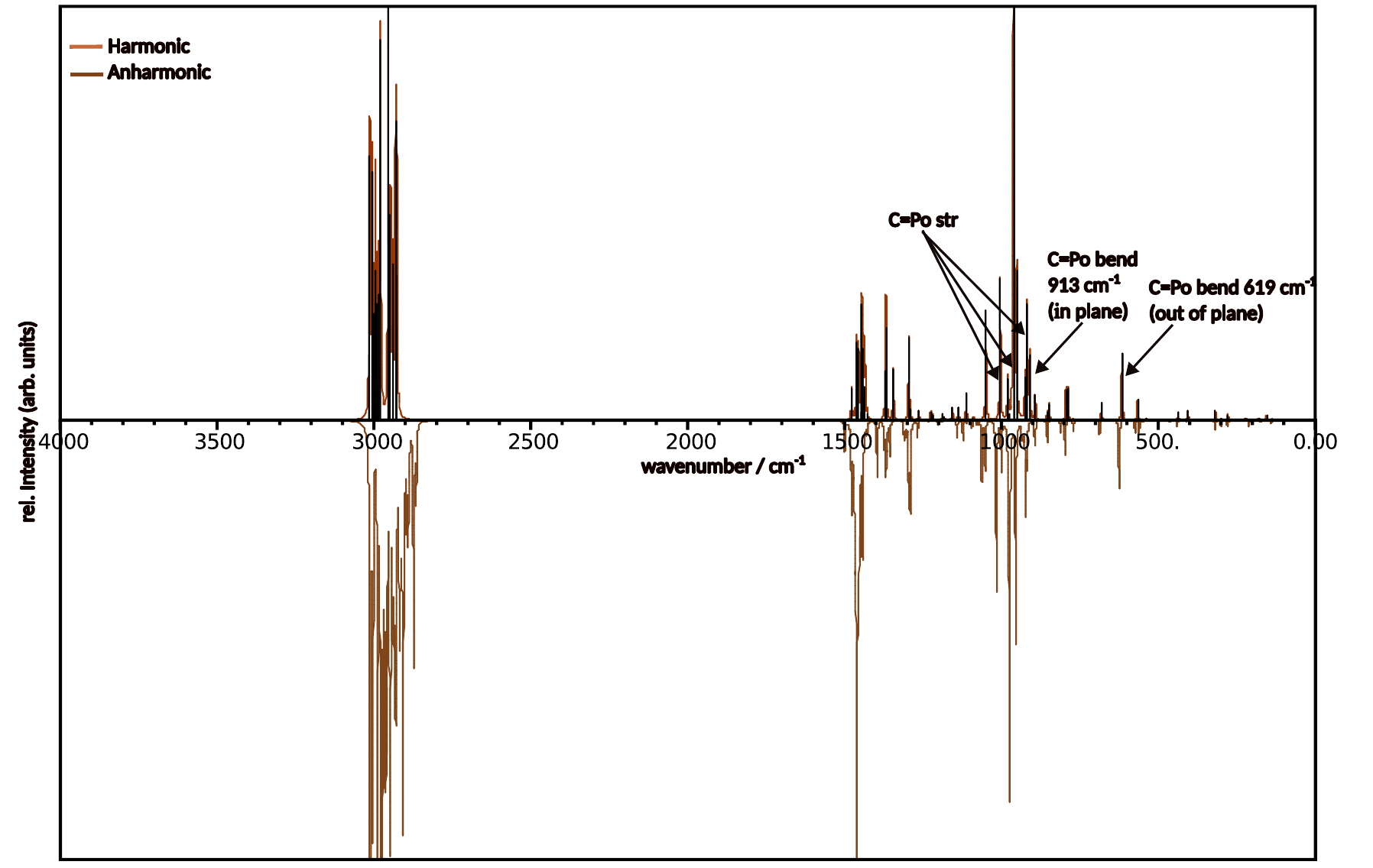}
  \caption{Comparison of theoretical harmonic (envelope and stick representation; directed upwards) and anharmonic (envelope; directed downwards) IR spectra of \textbf{1-Po}. Theoretical harmonic spectrum was computed on the DFT level within the double-harmonic approximation and is displayed using an overall scaling factor of the harmonic vibrational wavenumbers of 0.968 and a Lorentzian lineshape function with HWHM of \SI{1.5} {\per\centi\meter}. The anharmonic contributions were calculated using second-order vibrational perturbation theory (VPT2) and are plotted with the same lineshape function. \label{fig:Povibanharmoverview}}
  \end{figure}

\begin{figure}[htbp]
\centering
  \includegraphics[width=0.8\textwidth]{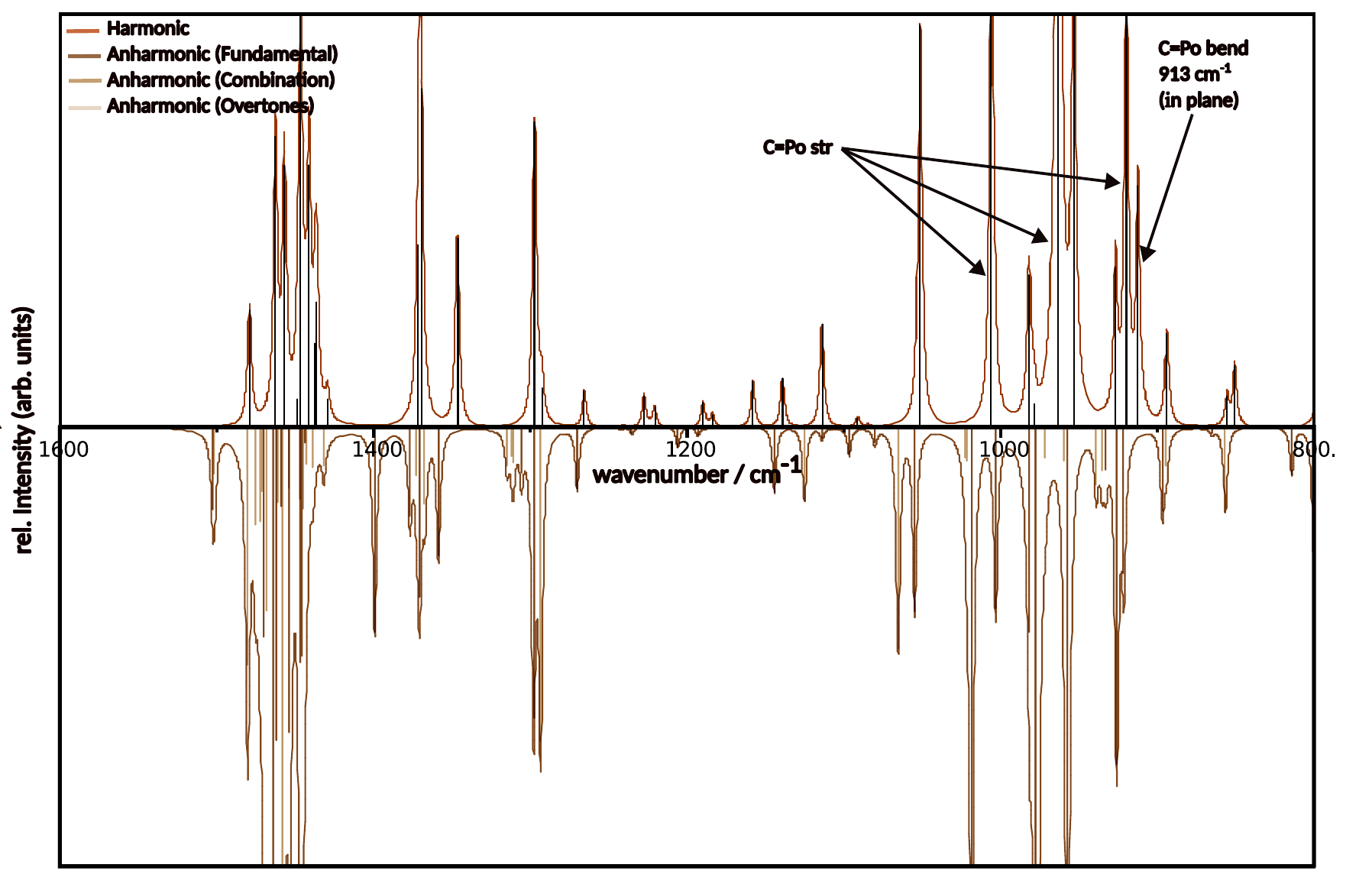}
  \caption{Theoretical harmonic (envelope and stick representation; directed upwards) and anharmonic (envelope and stick representation; directed downwards) IR spectra of \textbf{1-Po} in the range of \qtyrange[range-phrase=--,range-units=single]{1600}{800}{\per\centi\meter}. In this version, the intensity scale (y-max) is chosen to be different from the parental overview spectrum (Figure~\ref{fig:Povibanharmoverview}) to enhance the visualization clarity. Theoretical harmonic spectrum was computed on the DFT level within the double-harmonic approximation and is displayed using an overall scaling factor of the harmonic vibrational wavenumbers of 0.968 and a Lorentzian lineshape function with HWHM of \SI{1.5} {\per\centi\meter}. The anharmonic contributions were calculated using second-order vibrational perturbation theory (VPT2) at the same level and are plotted with the same lineshape function. See text description of more technical details.\label{fig:Povibanharmdetails}}
\end{figure}

For the other two molecules (\textbf{1-Te} and \textbf{1-Po}), all modes could remain active in the VPT2 treatment to obtain theoretical IR spectra beyond the harmonic approximation, in which also the dense spectral region \qtyrange[range-phrase=--,range-units=single]{1500}{1400}{\per\centi\meter} can be covered completely. As per the absence of corresponding spectra obtained experimentally, we compare them in Figures~\ref{fig:Tevibanharmoverview} and \ref{fig:Tevibanharmdetails} as well as \ref{fig:Povibanharmoverview} and \ref{fig:Povibanharmdetails} only to the ones predicted within the double-harmonic approximation.

Quite remarkable is that for both molecules, the VPT2 approach predicts that the region of the doublet for the umbrella motions of the geminal dimethyl substituents becomes significantly more feature-rich due to overtones and combination bands, which acquire intensity in this region. Lists of some of the most prominent among these can be found in Tables~\ref{tab:Te-overtones} and \ref{tab:Po-overtones}.

Overall, however, changes between predicted IR spectra from the (scaled) double-harmonic approximation and the VPT2 anharmonic description appear to be relatively mild, although some of the details are altered, in particular for a subset of those modes, in which motions of the C=Te and C=Po group participate. We provide our computed spectra here for future reference to facilitate identification and assignment, once the first synthesis of \textbf{1-Te} could be achieved.

\begin{table}[htp]
\footnotesize
\caption{Transition wavenumbers ($\tilde{\nu}_\mathrm{theo}$) and intensities ($A_\mathrm{theo}$) of most prominent first overtones (\textcolor{blue}{blue}) and combination bands (black) of \textbf{1-Te} computed at DFT/B3LYP within second-order vibrational perturbation theory.\label{tab:Te-overtones}}
\begin{tabular}{l S[detect-weight,round-mode=places,round-precision=1,group-digits = none] S[round-mode=places,round-precision=1,group-digits = none] l S[round-mode=places,round-precision=1,group-digits = none] S[detect-weight,round-mode=places,round-precision=1,group-digits = none]}
\hline
Mode & $\tilde{\nu}_\mathrm{theo}$ & $A_\mathrm{theo}$ & 
Mode & $\tilde{\nu}_\mathrm{theo}$ & $A_\mathrm{theo}$ \\
     &  \multicolumn{1}{c}{(cm$^{-1}$)} & \multicolumn{1}{c}{(\SI{}{\kilo\meter\per\mol})} &
     &  \multicolumn{1}{c}{(cm$^{-1}$)} & \multicolumn{1}{c}{(\SI{}{\kilo\meter\per\mol})} \\
\hline
\textcolor{blue}{$2\nu_{17}$} & \B 2962.168 & \B 17.75055138 &
\textcolor{blue}{$2\nu_{18}$} & \B 2946.624 & \B 13.04604108 \\
\textcolor{blue}{$2\nu_{21}$} & \B 2928.060 & \B  5.15204486 &
\textcolor{blue}{$2\nu_{19}$} & \B 2920.267 & \B 16.59361151 \\
\textcolor{blue}{$2\nu_{20}$} & \B 2916.422 & \B 18.39803366 &
\textcolor{blue}{$2\nu_{24}$} & \B 2912.045 & \B  4.61623725 \\
\textcolor{blue}{$2\nu_{22}$} & \B 2893.125 & \B  3.89197720 &
\textcolor{blue}{$2\nu_{23}$} & \B 2883.259 & \B  3.36299274 \\
\cline{4-6}
\textcolor{blue}{$2\nu_{25}$} & \B 2865.875 & \B 14.00216497 &
$\nu_{19}$+$\nu_{17}$ & 2968.349 & 11.96138691 \\
\cline{1-3}
$\nu_{18}$+$\nu_{17}$ & 2950.494 &  2.75684453 &
$\nu_{20}$+$\nu_{17}$ & 2944.605 & 18.34900824 \\
$\nu_{21}$+$\nu_{17}$ & 2939.205 &  6.67990447 &
$\nu_{22}$+$\nu_{17}$ & 2937.827 &  2.49176622 \\
$\nu_{19}$+$\nu_{18}$ & 2935.914 &  9.08500590 &
$\nu_{20}$+$\nu_{18}$ & 2924.195 &  8.64652019 \\
$\nu_{21}$+$\nu_{18}$ & 2922.882 &  2.30501067 &
$\nu_{20}$+$\nu_{19}$ & 2921.501 &  2.99088421 \\
$\nu_{22}$+$\nu_{18}$ & 2915.903 &  7.34513592 &
$\nu_{23}$+$\nu_{17}$ & 2915.140 &  6.82812541 \\
$\nu_{22}$+$\nu_{19}$ & 2913.366 &  2.93811222 &
$\nu_{23}$+$\nu_{18}$ & 2906.547 &  5.43346879 \\
$\nu_{23}$+$\nu_{19}$ & 2899.373 &  5.53692288 &
$\nu_{62}$+$\nu_{43}$ & 1485.918 &  8.77448354 \\
$\nu_{70}$+$\nu_{32}$ & 1473.528 & 10.20681674 &
$\nu_{61}$+$\nu_{45}$ & 1472.480 &  4.15368054 \\
$\nu_{72}$+$\nu_{32}$ & 1463.672 &  3.54431533 &
$\nu_{70}$+$\nu_{34}$ & 1457.498 &  2.14412390 \\
$\nu_{67}$+$\nu_{37}$ & 1457.227 &  2.55846682 &
$\nu_{66}$+$\nu_{39}$ & 1443.828 &  2.55929765 \\
$\nu_{59}$+$\nu_{55}$ & 1391.166 &  3.11373780 &
$\nu_{59}$+$\nu_{54}$ & 1387.804 &  5.80095011 \\
$\nu_{66}$+$\nu_{42}$ & 1385.397 & 18.42037242 &
$\nu_{69}$+$\nu_{38}$ & 1375.170 & 11.16921759 \\
$\nu_{59}$+$\nu_{56}$ & 1367.395 &  2.04305756 &
$\nu_{71}$+$\nu_{41}$ & 1325.705 &  9.78973549 \\
$\nu_{72}$+$\nu_{53}$ & 1073.074 &  3.03249323 &
$\nu_{65}$+$\nu_{57}$ & 1037.296 &  5.73416025 \\
$\nu_{61}$+$\nu_{60}$ & 1035.710 &  2.63300941 &
$\nu_{70}$+$\nu_{54}$ & 1030.091 &  6.11314219 \\
$\nu_{62}$+$\nu_{60}$ & 1017.776 &  7.48677805 &
$\nu_{70}$+$\nu_{56}$ & 1011.254 &  5.27472222 \\
$\nu_{75}$+$\nu_{49}$ &  990.189 &  2.49838837 &
                      &          &             \\
\hline
\end{tabular}
\end{table}

\clearpage

\begin{table}[htp]
\footnotesize
\caption{Transition wavenumbers ($\tilde{\nu}_\mathrm{theo}$) and integrated absorption coefficients ($A_\mathrm{theo}$) of most prominent first overtones (\textcolor{blue}{blue}) and combination bands (black) of \textbf{1-Po} computed at DFT/B3LYP within second-order vibrational perturbation theory.\label{tab:Po-overtones}}
\begin{tabular}{l S[detect-weight,round-mode=places,round-precision=1,group-digits = none] S[round-mode=places,round-precision=1,group-digits = none] l S[round-mode=places,round-precision=1,group-digits = none] S[detect-weight,round-mode=places,round-precision=1,group-digits = none]}
\hline
Mode & $\tilde{\nu}_\mathrm{theo}$ & $A_\mathrm{theo}$ & 
Mode & $\tilde{\nu}_\mathrm{theo}$ & $A_\mathrm{theo}$ \\
     &  \multicolumn{1}{c}{(cm$^{-1}$)} & \multicolumn{1}{c}{(\SI{}{\kilo\meter\per\mol})} &
     &  \multicolumn{1}{c}{(cm$^{-1}$)} & \multicolumn{1}{c}{(\SI{}{\kilo\meter\per\mol})} \\
\hline
\textcolor{blue}{$2\nu_{17}$} & \B 2983.949 & \B  5.35911540 &
\textcolor{blue}{$2\nu_{18}$} & \B 2953.116 & \B 22.41025019 \\
\textcolor{blue}{$2\nu_{21}$} & \B 2931.478 & \B  5.43896683 &
\textcolor{blue}{$2\nu_{19}$} & \B 2923.005 & \B 11.18630945 \\
\textcolor{blue}{$2\nu_{20}$} & \B 2919.472 & \B  7.96834107 &
\textcolor{blue}{$2\nu_{22}$} & \B 2893.533 & \B  3.00143336 \\
\textcolor{blue}{$2\nu_{23}$} & \B 2889.702 & \B  2.59855191 &
\textcolor{blue}{$2\nu_{25}$} & \B 2868.973 & \B  9.02228050 \\
\cline{4-6}
\textcolor{blue}{$2\nu_{57}$} & \B 1377.328 & \B  3.07480582 &
$\nu_{18}$+$\nu_{17}$ & 2969.521 & 36.07047698 \\
\cline{1-3}
$\nu_{19}$+$\nu_{17}$ & 2961.328 & 31.23843492 &
$\nu_{20}$+$\nu_{17}$ & 2954.864 &  6.46486824 \\
$\nu_{22}$+$\nu_{17}$ & 2944.075 & 16.55946119 &
$\nu_{21}$+$\nu_{17}$ & 2943.228 &  3.64996718 \\
$\nu_{19}$+$\nu_{18}$ & 2940.692 & 12.01601611 &
$\nu_{22}$+$\nu_{20}$ & 2937.847 & 27.18285609 \\
$\nu_{20}$+$\nu_{18}$ & 2931.378 & 12.26976659 &
$\nu_{20}$+$\nu_{19}$ & 2927.689 &  4.03971617 \\
$\nu_{22}$+$\nu_{18}$ & 2922.064 &  5.85278950 &
$\nu_{23}$+$\nu_{17}$ & 2921.478 &  7.66986210 \\
$\nu_{22}$+$\nu_{19}$ & 2917.357 &  3.28812291 &
$\nu_{21}$+$\nu_{20}$ & 2914.586 &  7.90533069 \\
$\nu_{23}$+$\nu_{18}$ & 2912.815 &  4.97969417 &
$\nu_{23}$+$\nu_{19}$ & 2905.137 &  4.37603382 \\
$\nu_{23}$+$\nu_{22}$ & 2898.372 &  2.00976079 &
$\nu_{25}$+$\nu_{21}$ & 2879.307 &  5.60866752 \\
$\nu_{59}$+$\nu_{51}$ & 1480.880 &  2.52875977 &
$\nu_{60}$+$\nu_{49}$ & 1480.791 &  8.09402614 \\
$\nu_{60}$+$\nu_{50}$ & 1475.855 &  3.33077431 &
$\nu_{67}$+$\nu_{36}$ & 1475.338 &  4.00969162 \\
$\nu_{72}$+$\nu_{29}$ & 1473.010 &  3.22996458 &
$\nu_{66}$+$\nu_{37}$ & 1471.116 &  2.20579940 \\
$\nu_{64}$+$\nu_{40}$ & 1468.954 &  6.24856503 &
$\nu_{62}$+$\nu_{43}$ & 1461.479 &  2.58867212 \\
$\nu_{61}$+$\nu_{45}$ & 1458.041 & 18.95606147 &
$\nu_{61}$+$\nu_{44}$ & 1453.966 &  2.07572570 \\
$\nu_{71}$+$\nu_{32}$ & 1447.689 &  2.49127779 &
$\nu_{68}$+$\nu_{35}$ & 1444.793 &  2.77107050 \\
$\nu_{64}$+$\nu_{45}$ & 1367.783 &  2.62009094 &
$\nu_{72}$+$\nu_{39}$ & 1294.021 & 10.73004112 \\
$\nu_{73}$+$\nu_{47}$ & 1125.730 &  2.53084322 &
$\nu_{67}$+$\nu_{56}$ & 1065.981 &  7.59556521 \\
$\nu_{63}$+$\nu_{62}$ &  857.117 &  2.89902719 &
                      &          &             \\
\hline
\end{tabular}
\end{table}

\clearpage

To summarize our analysis of the IR spectra based on a perturbative inclusion of cubic and semi-diagonal quartic force field contributions, we find, indeed, that an overall scaling of transition wavenumbers, which finds widespread application for harmonic force field studies, becomes to a large degree obsolete when applying our present combination of DFT flavor and atomic basis set to the class of chalcogenofenchones. Computed absorption coefficients in the anharmonic approximation match in some cases better to experiment and also transition wavenumbers fit sometimes better and sometimes worse to the measured band location than their scaled harmonic counterparts do. Having the possiblity to compare both harmonic and anharmonic treatments to experiment facilitates the assignment of the various IR bands observed.

\subsection{High-resolution spectra}

\begin{figure}[b]
\includegraphics[width=\linewidth]{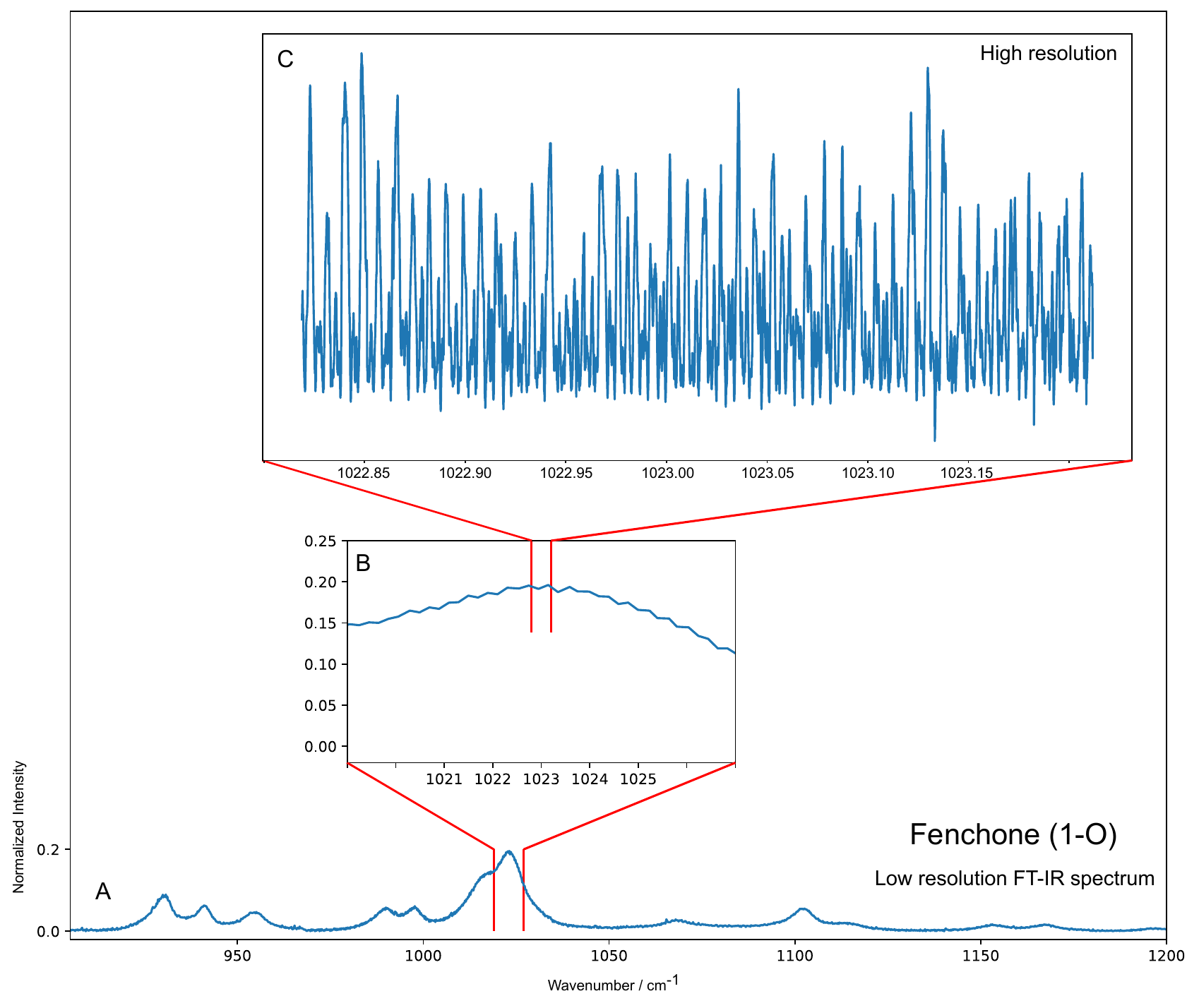}
\caption{High-resolution IR absorption spectrum of \textbf{1-O}. A) Part of the measured IR spectrum taken at room temperature using a low resolution Fourier transform (FT) spectrometer, B) Detail of spectrum shown in A, C) High-resolution rovibrational spectrum comprising lines of  \SI{38}{MHz} FWHM (\SI{0.001}{\per\centi\meter}) as measured using a quantum cascade laser setup in combination with a supersonic jet at temperatures around \SI{30}{\kelvin}. \label{fig:fenchone-highres}}
\end{figure}

The importance of precision spectroscopy is illuminated by the high-resolution gas-phase spectrum of fenchone shown in Figure~\ref{fig:fenchone-highres}, which demonstrates the capability to capture fine spectral details and achieve highly accurate measurements in this family of chiral compounds. A high level of precision potentially enables the detection of subtle phenomena such as parity violation (see subsequent section), which is crucial for understanding fundamental physical processes and interactions. To further improve high-resolution measurements, the combination of ultracold molecules and molecular beam techniques allows for enhanced experimental accuracy. Additionally, frequency comb-stabilized lasers provide a stable reference for spectral analysis, enabling even finer resolution. These advancements in precision spectroscopy offer new insights into molecular interactions and quantum mechanical effects, underscoring the necessity of such techniques for exploring complex molecular and physical phenomena.

\clearpage

\subsection{Parity-violating energy differences}

The detection of parity-violating splittings caused by electroweak interactions in the vibrational resonance wavenumbers of enantiomers serves as a prominent fundamental application for precision spectroscopy. To highlight the particular benefits of $Z$-dependent enhancements in chiral molecules, we present herein exploratory calculations of parity-violating potentials ($E_\mathrm{pv}$) computed at the equilibrium structure of the chalcogenofenchones. These energies are in general expected to scale approximately like $Z^5$, where $Z$ is the nuclear charge of the chalcogen in the fenchone derivatives. In a crude first approximation that neglects rovibrational effects, the absolute value of the parity violating energy difference between the two enantiomers is about two times the norm of $E_\mathrm{pv}$ at the equilibrium structure.

As we show in Table~\ref{tab:epv}, $E_\mathrm{pv}$ on the two-component Hartree--Fock (HF) level amounts to about \SI{8e-20}{\hartree} for the parent fenchone, whereas two different density functional theory (DFT) flavors, namely hybrid B3LYP and the local density approximation (LDA), predict slightly larger values for this system. This order of magnitude agrees well with that predicted for the constitutional isomer camphor \textbf{2} (\SI{6e-20}{\hartree} on the four-component Dirac--Hartree--Fock level) reported in Ref.~\onlinecite{schwerdtfeger:2004}. The steep $Z^5$ scaling is confirmed on the Hartree--Fock level, so that five orders of magnitude larger parity violating potentials are predicted for polonofenchone. On the DFT level, the increase of $E_\mathrm{pv}$ with growing $Z$ is smaller and the spread between predictions from different approaches becomes sizable for the heaviest compound in this series. The latter possibly emerges from the choice of one common reference structure for each \textbf{1-X} molecule, which was energy-minimized on the non-relativistic (\textbf{1-O} to \textbf{1-Se}) or on the scalar relativistic level with effective core potentials (\textbf{1-Te}, \textbf{1-Po}). If structures were energy-minimized on the respective two-component HF or DFT levels instead, one might expect a reduced spread in the predicted values, but we can not rule out at the current state that also electron correlation effects gain in importance for the heavier representatives \textbf{1-Te}, \textbf{1-Po} of the chalcogenofenchone family and thus contribute to considerably larger variations in predictions of the parity violating potentials. Despite these limitations of our present study, it clearly demonstrates the huge potential that is provided by a series of tunable compounds as the chalcogenofenchones introduced in this work. As pointed out by Letokhov, the relative parity-violating effects in electronic, vibrational and rotational transitions remain approximately constant. By this token, we can expect that also vibrational transition wavenumber differences in this series of compounds will scale with $Z^5$ and, thus, we are led to anticipate reaching from relative splittings of $\Delta \nu/\nu \approx \SI{e-19}{}$ in fenchone to sizable relative splittings of up to about $\Delta \nu/\nu \approx \SI{e-14}{}$ in polonofenchone. Specific calculations in the framework of electroweak quantum chemistry can in principle be performed for all vibrational modes of the present compounds to identify suitable transitions for such experiments. But these are beyond the scope of the present work, in which we primarily wanted to demonstrate the steep scaling of intriguing physical effects in a series of chiral molecules.

\begin{table}[htbp]
    \centering
    \caption{Parity violating energy \(E_\text{pv}\) in \(\SI{}{\hartree}\) for fenchone derivatives 
    calculated using HF and DFT with different functionals.}
    \begin{ruledtabular}
    \begin{tabular}{lS[table-format=1.1e2,round-mode=places,round-precision=1]S[table-format=1.1e2,round-mode=places,round-precision=1]S[table-format=1.1e2,round-mode=places,round-precision=1]S[table-format=1.1e2,round-mode=places,round-precision=1]S[table-format=1.1e2,round-mode=places,round-precision=1]}
         {Method} & {\textbf{1-O}} & {\textbf{1-S}} & {\textbf{1-Se}} & {\textbf{1-Te}} & {\textbf{1-Po}} \\\hline
         HF & 7.531012334133391e-20  & 3.2237615485589794e-19 &  6.922345312786151e-18 &  9.653431505374258e-17  &  5.73049304751787e-15 \\
         B3LYP &  1.0334977568383743e-19 & 3.9298431661153134e-19  & 4.715708612169703e-18   &  3.651851653343862e-17  & 9.112850807012725e-16  \\
         LDA &  1.1262381283172888e-19 & 4.293213417505908e-19  & 4.186386803134462e-18   &  1.875439048623183e-17  &  3.201166685979752e-17
    \end{tabular}
    \end{ruledtabular}
    \label{tab:epv}
\end{table}
\section{Conclusion}

\noindent In this work we have reported synthesis and characterization of thiofenchone \textbf{1-S} as well as selenofenchone \textbf{1-Se} and determined gas-phase spectral features in the microwave and infrared range of electromagnetic radiation for these compounds.  

Additionally, we investigated the gas-phase IR spectrum of the parent fenchone \textbf{1-O} in low resolution and presented a high-resolution spectrum measured with a quantum cascade laser at about \SI{1023}{\per\centi\meter} to observe rovibrational lines of full-width at half-maximum of \SI{0.0001}{\per\centi\meter} in a supersonic jet expansion. 

These experimental studies were accompanied by systematic theoretical investigations of the series of chalcogenofenchones from \textbf{1-O} to \textbf{1-Po} by computing equilibrium structures, IR spectra in the double-harmonic approximation and IR spectra that account for anharmonicities in a vibrational perturbation treatment based on quartic force fields computed on a hybrid density functional theory level.

Volatility of the series of relatively rigid fenchone derivatives allowed for a thorough, comparative analysis of molecular structures obtained from MW spectroscopy as well as fundamental transitions in the mid IR for \textbf{1-O}, \textbf{1-S}, \textbf{1-Se} and assignment of corresponding observed resonances by virtue of on overall good agreement between experiment and theory. We emphasize the importance of complete analysis, where we hold the view that this joint experimental and theoretical work provides a deep understanding about the molecular structure and opens a path towards similar heavy-elemental compounds.

Heavier chalcogenoketone derivatives are established herein as promising benchmark systems for gas-phase studies of molecular chirality in a versatile and tunable series of compounds that allow to explore systematically nuclear charge $Z$ dependent influences. As one of those properties that are particularly sensitive to $Z$-variations, we have estimated parity-violating energy differences between enantiomers of these compounds in the framework of electroweak quantum chemistry and highlighted the ongoing search for molecular parity-violation that benefits from increased spin-orbit coupling in heavy-elemental chiral compounds.

Another class of chirality-dependent properties favorably studied in volatile systems forms photoelectron circular dichroism (PECD) in chalcogenofenchones that we report in an accompanying paper \cite{vasudevan:2025}.

\subsection{Accession Code}
CCDC  2429055 contains the supplementary crystallographic data for this paper. These data can be obtained free of charge via www.ccdc.cam.ac.uk/data\_request/cif, or by emailing  ata\_request@ccdc.cam.ac.uk, or by contacting The
Cambridge Crystallographic Data Centre, 12 Union Road,
Cambridge CB2 1EZ, UK; fax: +44 1223 336033.

\begin{acknowledgments}
This work was supported by the Deutsche Forschungsgemeinschaft (DFG, German Research Foundation) --- Projektnummer 328961117 --- SFB 1319 ELCH and INST 159/146-1 FUGG. Computer time provided by the Center for Scientific Computing (CSC) Frankfurt is gratefully acknowledged.
\end{acknowledgments}

\clearpage

\appendix
\section{Appendix}

Herein we present in Appendix 1 additional information on the experiments related to synthesis and characterisation of the compounds \textbf{1-S} and \textbf{1-Se} combined with additional data from the MW study of \textbf{1-S} and \textbf{1-Se} and we report in Appendix 2 a compilation of selected force constant data as well as Cartesian coordinates of the Computed equilibrium structures of \textbf{1-O}, \textbf{1-S}, \textbf{1-Se}, \textbf{1-Te} and \textbf{1-Po},.

\subsection{Experimental}
All experiments were carried out under exclusion of moisture and air under an inert argon atmosphere. Work-up was performed under ambient, atmospheric conditions, where applicable, and products were stored under an inert argon atmosphere. Mesitylene was dried over sodium and distilled prior to use. Further solvents for reactions were dried over sodium potassium alloy and distilled prior to use where applicable. Column chromatography was performed using silica gel (puriFlash PF-25SIHC) without any pretreatment. Solvents used for column chromatography were used as purchased (\textit{n}-hexane (Chemsolute), \textit{n}-pentane (AnalaR Normapur)). The following chemicals were commercially available and used as received: D(+)-fenchone (97~\%) was purchased from Acros, L(–)-fenchone (98~\%, 84~\% \textit{ee}) was purchased from Merck, Lawesson’s reagent was purchased from Alfa Aesar, 9-borabicyclo[3.3.1]nonane-dimer (9-BBN) was purchased from Merck. Bis(1,5-cyclooctanediylboryl)monoselenide was synthesized according to literature procedures.\cite{koster:1992} \textsuperscript{1}H, \textsuperscript{13}C, and \textsuperscript{77}Se-NMR-data was recorded on Jeol JNM-ECZL500, Varian VNMRS-500 MHz or MR-400 MHz spectrometers at \SI{25}{\celsius}. Chemical shifts were referenced to residual protic impurities in the solvent ($^{1}$H) or the deuterio solvent itself (\textsuperscript{13}C) and reported relative to SiMe\textsubscript{4} = 0~ppm (\textsuperscript{1}H, \textsuperscript{13}C) and NMR spectra of hetero nuclei were referenced using the $\Xi$‐scale following IUPAC recommendations with SeMe\textsubscript{2} = 0~ppm as secondary reference for \textsuperscript{77}Se.\cite{Harris:2001} NMR-shifts were assigned according to 1D NOESY, 2D HSQC, HMBC and NOESY analysis. Because NMR-spectroscopy is an achiral analysis method, only one set of spectra corresponding to both enantiomers will be presented. IR-spectra were recorded on an Alpha Platinum ATR-spectrometer (diamond crystal, Bruker) with neat substances. Spectral assignment and depiction was carried out with Opus 6.5 (Bruker Optics) with following describtors: vs~=~very strong, s~=~strong, m~=~medium, mv~=~medium-weak, w~=~weak. UV/vis spectra were obtained using a Shimadzu UV-1900 spectrometer. CD spectra were obtained using a Jasco J-1500 CD spectrometer. APCI-HR mass spectra were recorded on a \textit{micrOTOF} (Bruker Daltonics) and Orbitrap-mass spectrometer \textit{Exactive} (Thermo Fisher Scientific, Bremen, D). Calibration was performed prior to data acquisition with a \textit{LTQ XL/Hybrid CalMix} ESI-standard mixture. Elemental analysis was performed with a \textit{HEKAtech Euro EA CHNS} elemental analyzer. The samples were prepared in a Sn cup and analyzed with V\textsubscript{2}O\textsubscript{5} added to ensure complete combustion.

\subsubsection{Crystallography}

For all data collections a single crystal was mounted on a \textit{micro mount} and all geometric and intensity data were taken from this sample by $\omega$-scans. Data collections were carried out on a Stoe IPDS2 diffractometer equipped with an area detector. The data sets were corrected for absorption (by multi scans), Lorentz and polarisation effects. The structures were solved by direct methods (SHELXT 2014/7) \cite{Sheldrick:2008} and refined using alternating cycles of least-squares refinements against F2 (SHELXL2014/7) \cite{Sheldrick:2008}. H atoms were included to the models in calculated positions with the 1.2 fold isotropic displacement parameter of their bonding partner. Experimental details for the diffraction experiment is given in table \ref{tab:xraytab1}. Data for all compounds have been deposited with the Cambridge Crystallographic Database (2429055).

\begin{table}[htbp]
        \caption{Crystal data and structure refinement for \textbf{1-Se}.}
        \begin{tabular}{ll}
        Empirical formula 	                   &C\textsubscript{10}H\textsubscript{16}Se \\
        Formula weight 	                       &215.19 \\
        \textit{T}/K 	                       &100 \\
        Crystal system 	                       &orthorhombic \\
        Space group 	                       &\textit{P}2\textsubscript{1}2\textsubscript{1}2\textsubscript{1} \\
        \textit{a}/Å 	                       &9.9995(8) \\
        \textit{b}/Å 	                       &11.746(2) \\
        \textit{c}/Å 	                       &25.632(3) \\
        $\alpha$/° 	                           &90 \\
        $\beta$/° 	                           &90 \\
        $\gamma$/° 	                           &90 \\
        \textit{V}/Å\textsuperscript{3}        &3010.5(7) \\
        \textit{Z} 	                           &12 \\
        $\rho$ calc g/cm\textsuperscript{3}    &1.424 \\
        $\mu$/mm\textsuperscript{‑1}           &3.681 \\
        \textit{F}(000) 	                   &1320.0 \\
        Crystal size/mm\textsuperscript{3}     &0.71 × 0.48 × 0.32 \\
        Radiation used 	                           &Mo \textit{K}\textsubscript{$\upalpha$} ($\lambda$ = 0.71073) \\
        2$\theta$ range for data collection/°  &3.178 to 53.86 \\
        Index ranges 	                       &-11 $\leq$ \textit{h} $\leq$ 12 \\
                                               &-13 $\leq$ \textit{k} $\leq$ 14 \\
                                               &-32 $\leq$ \textit{l} $\leq$ 28 \\
        Reflections collected 	               &10471 \\
        Independent reflections 	           &6323 [\textit{R}\textsubscript{int} = 0.1163] \\
        Data/restraints/parameters 	           &6323/0/309 \\
        Goodness-of-fit on F\textsuperscript{2}&1.044 \\
        Final \textit{R} indexes [\textit{I}$\geq$2$\upsigma$ (\textit{I})] 	       &\textit{R}1 = 0.1023, \textit{wR}2 = 0.2495 \\
        Final \textit{R} indexes [all data]    &\textit{R}1 = 0.1516, \textit{wR}2 = 0.2955 \\
        Largest diff. peak/hole / e Å\textsuperscript{-3} 	   &0.82/-0.83 \\
        Flack parameter	                       &0.02(6)     \\
        CCDC                                   &2429055
          \end{tabular}
        \label{tab:xraytab1}
\end{table}

\clearpage

\subsubsection{General procedure for the preparation of \textbf{1-S}}
In a \SI{100}{\milli\liter} Schlenk tube \SI{1.536}{\gram} (\SI{10.1}{\milli\mol}) of fenchone and \SI{8.209}{\gram} (\SI{20.3}{\milli\mol}) Lawesson’s reagent were suspended in \SI{20}{\milli\liter} of \textit{o}-xylene and heated to \SI{155}{\celsius} for \SI{3}{\hour} in an oil bath to give a bright orange solution. At \SI{50}{\celsius} oil bath temperature volatile compounds were collected in a cold trap at \SI{e-2}{\milli\bar} as a bright orange liquid. Collection was stopped when no liquid parts remained in the original \SI{100}{\milli\liter} flask. After this step, the substance handling and the work-up were performed under atmospheric conditions. The distillate was subjected to separation using column chromatography on silica gel with \textit{n}-hexane as eluent and an orange phase was collected. Removal of the solvent \textit{in vacuo} yielded \SI{1.513}{\gram} (\SI{9.0}{\milli\mol}, 90~\%) of orange, oily \textbf{1-S}, which was stored under argon atmosphere.\\
\textbf{$^{1}$H NMR} (500~MHz, CDCl\textsubscript{3}) $\updelta$ (ppm): 2.31 (dddd, \textit{J} ~ 3.6, 1.7, 1.7, 1.7~Hz, 1H, H4), 1.84 (dddd, \textit{J} = 10.3, 2.2, 2.2, 2.2~Hz, 1H, H7a), 1.79 (m, 1H, H5b), 1.75 (m, 1H, H5a), 1.66 (m, 1H, H7b), 1.65 (m, 1H, H6a), 1.32 (s, 3H, Me C10), 1.27 (dddd, \textit{J} = 14.8, 8.6, 5.6, 2.3~Hz, 1H, H6b), 1.18 (s, 3H, Me C9), 1.14 (s, 3H, Me C8); \textbf{\textsuperscript{13}C NMR} (126~MHz, CDCl\textsubscript{3}) $\updelta$ (ppm): 280.6 (C2, q), 66.4 (C1, q), 57.9 (C3, q), 47.0 (C4), 43.8 (C7), 35.6 (C6), 28.8 (C8), 26.6 (C9), 25.2 (C5), 19.3 (C10). \\
IR (\SI{}{\per\centi\meter}): 2963~s, 2923~m, 2864~mw, 1467~mw, 1447~m, 1376~mw, 1356~w, 1322~w, 1275~s, 1252~w, 1226~w, 1177~vs, 1141~w, 1111~w, 1091~s, 1059~mw, 1013~w, 1000~w, 962~w, 952~w , 938~w, 922~mw, 883~w, 817~w, 730~w, 667~w.\\
Elem. anal. calcd. (\%) for C\textsubscript{10}H\textsubscript{16}S: C~71.37, H~9.58, S~19.05 found: C~71.05, H~9.66, S~19.09.\\
APCI-HR: \textit{m/z} = 169.1044 (M+H, 100\%)\textsuperscript{+} calculated: 169.1045. 

\begin{figure}[h]
    \centering
    \includegraphics[width=0.8\linewidth]{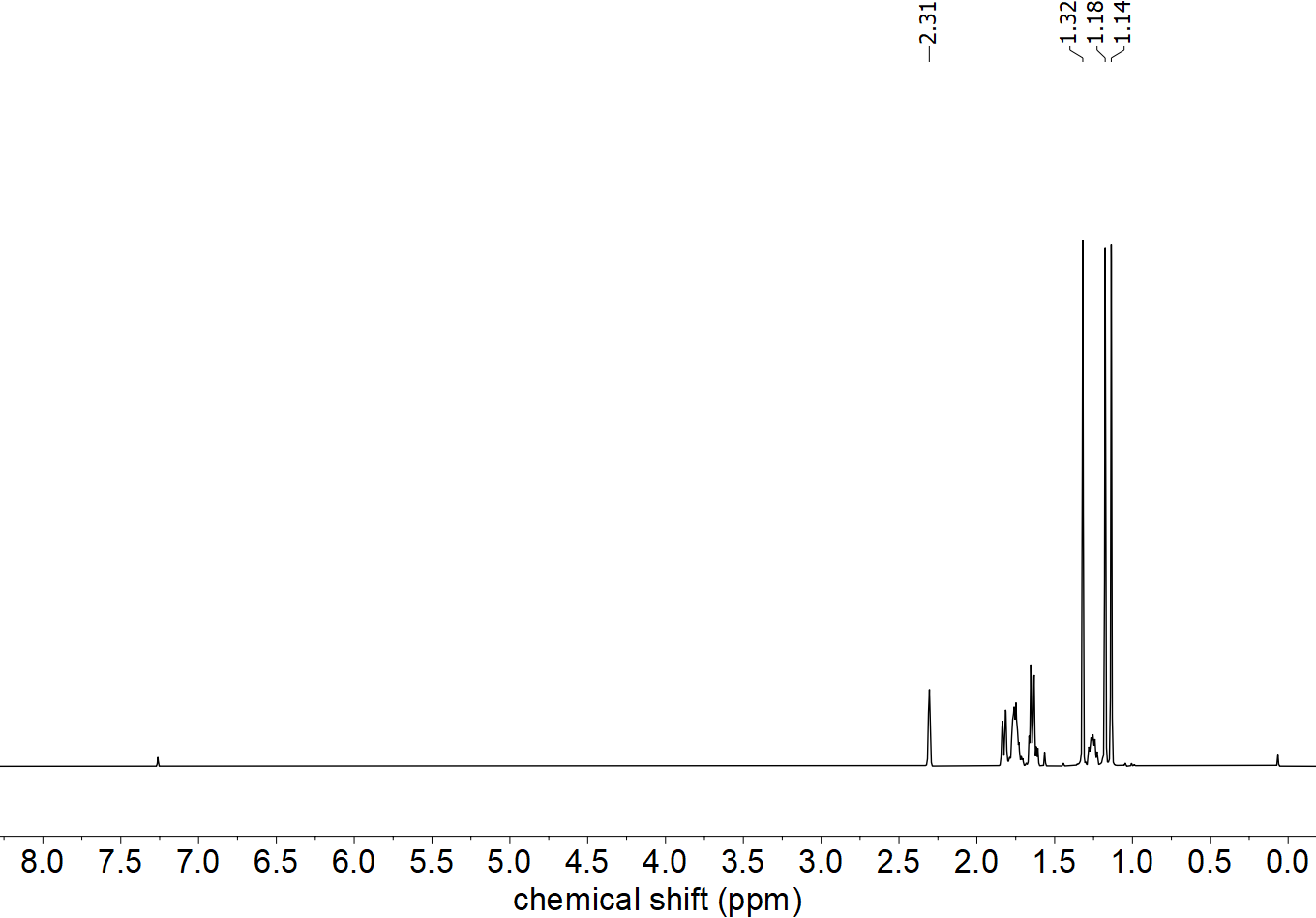}
    \caption{\textsuperscript{1}H NMR (CDCl\textsubscript{3}, 500~MHz) of \textbf{1-S}.}
    \label{fig:1-s hnmr}
\end{figure}

\begin{figure}[h]
    \centering
    \includegraphics[width=0.8\linewidth]{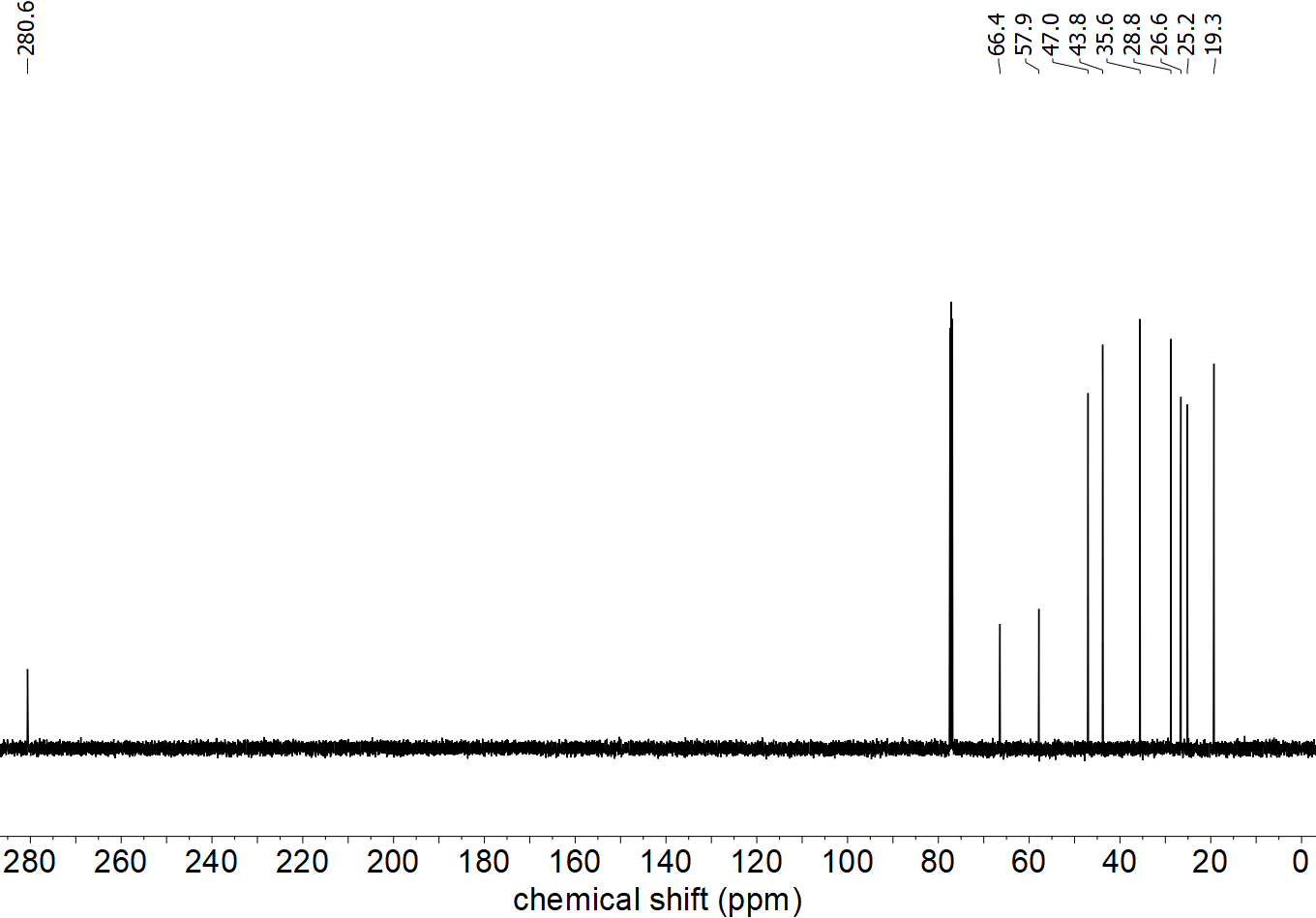}
    \caption{\textsuperscript{13}C NMR (CDCl\textsubscript{3}, 126~MHz) of \textbf{1-S}.}
    \label{fig:1-s cnmr}
\end{figure}

\begin{figure}[h]
    \centering
    \includegraphics[width=0.8\linewidth]{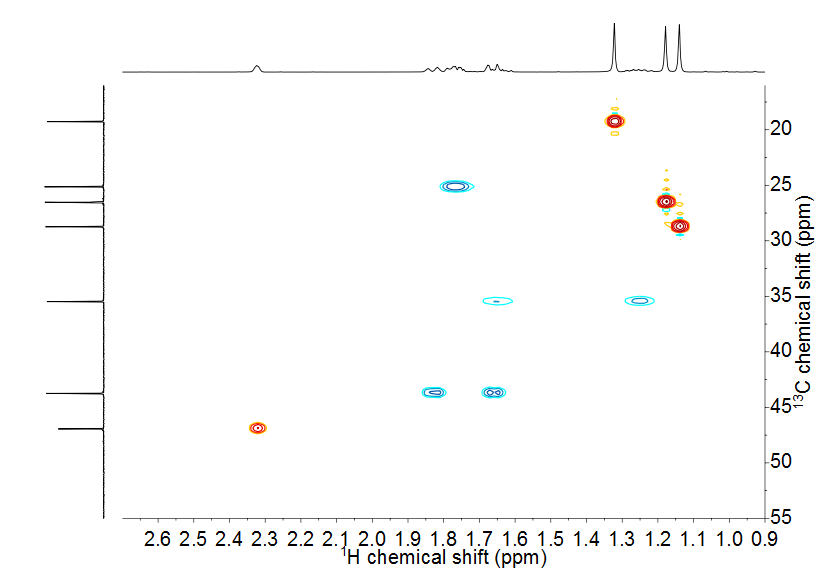}
    \caption{HSQC (CDCl\textsubscript{3}, 126~MHz) of \textbf{1-S}.}
    \label{fig:1-s hsqc}
\end{figure}

\begin{figure}[h]
    \centering
    \includegraphics[width=0.9\linewidth]{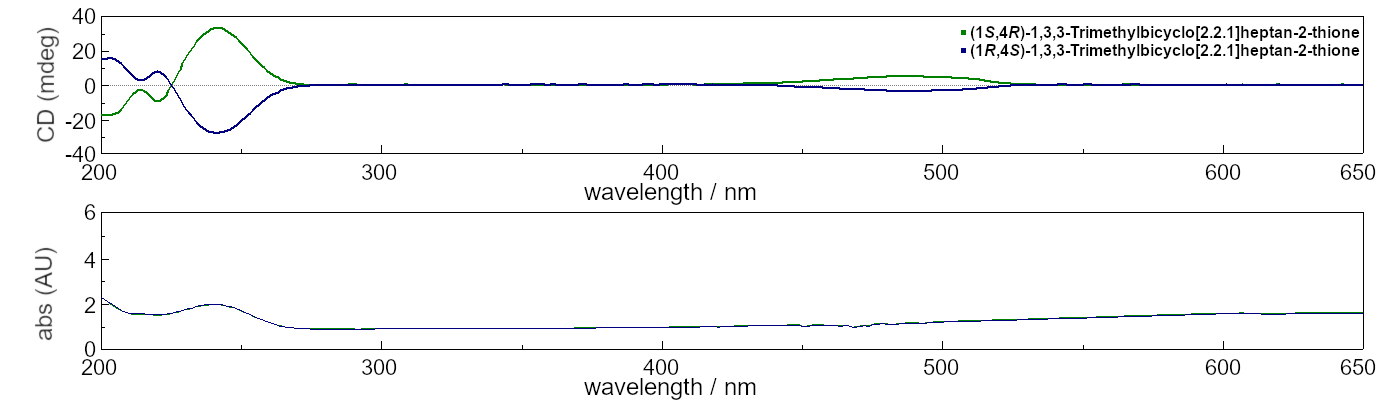}
    \caption{CD spectrum (top) and UV/vis absorption spectrum (bottom) of \textbf{1-S} measured in a hexane solution with concentration of \SI{1e-4}{\mol\per\liter}.}
    \label{fig:S-CD}
\end{figure}

\begin{figure}[h]
    \centering
    \includegraphics[width=0.9\linewidth]{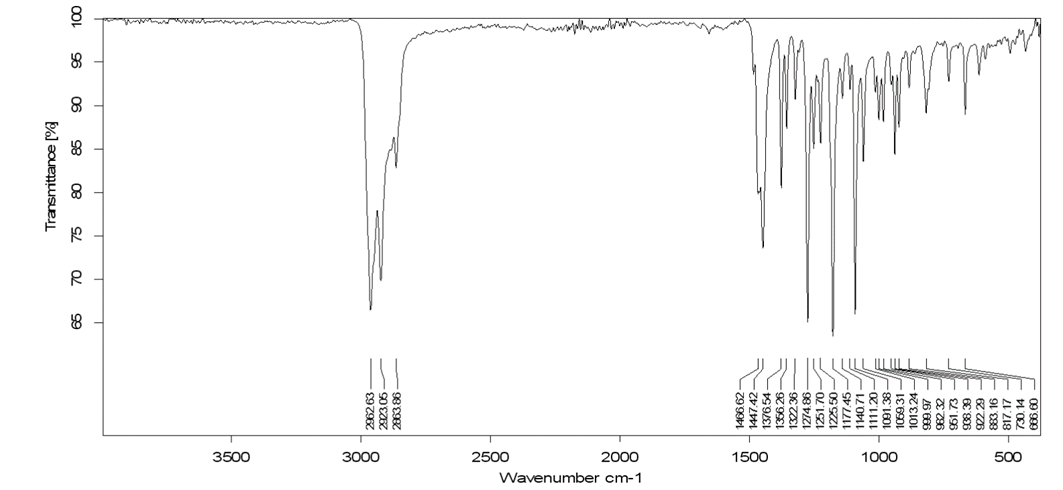}
    \caption{ATR-IR spectrum of neat \textbf{1-S}.}
    \label{fig:S-IR}
\end{figure}

\clearpage

\subsubsection{General procedure for the preparation of \textbf{1-Se}}
In a \SI{250}{\milli\liter} Schlenk flask \SI{11.118}{\gram} (\SI{45.6}{\milli\mol}) of 9-Borabicyclo[3.3.1]nonan-dimer (9-BBN) were suspended in \SI{40}{\milli\liter} of mesitylene and heated to \SI{150}{\celsius} in an oil bath to give a slightly yellow solution. To this solution \SI{3.598}{\gram} (\SI{45.6}{\milli\mol}) of grey selenium was added and the temperature raised to \SI{170}{\celsius}. Within the period of half an hour gas formation could be monitored over a connected oil bubbler. Once gas evolution ceased, the reaction was complete and bis(1,5-cyclooctanediylboryl)monoselenide was used \textit{in situ} without further purification. At this point a slightly yellow, greenish solution with a minor amount of black particles was obtained and temperature was lowered to \SI{120}{\celsius}. Thereafter \SI{6.935}{\gram} (\SI{45.6}{\milli\mol}) of fenchone was added. Upon addition, the color changed to blue and stirring at \SI{120}{\celsius} was continued for \SI{36}{\hour}. Thereafter, at \SI{160}{\celsius} oil bath temperature all volatile compounds were collected in a cold trap at \SI{e-2}{\milli\bar} as a deep blue liquid. Collection was stopped when only red and colorless substances remained in the original \SI{250}{\milli\liter} flask. From this point on, substance handling and the work-up were performed under atmospheric conditions. The majority of mesitylene was destilled off at \SI{10}{\milli\bar} employing a sufficiently long Vigreux-column setup while the \textbf{1-Se} remained within the Vigreux-column. The distillate, containing minor amounts of \textbf{1-Se}, which colored the collected mesitylene phase slightly blue, was discarded. The remaining \textbf{1-Se} within the Vigreux-column was dissolved with \textit{n}-hexane and afterwards, the solvent was removed in a rotary evaporater under reduced pressure. The remaining oil, containing traces of fenchone and mesitylene, was subjected to separation using column chromatography on silica gel with \textit{n}-hexane as eluent. Removal of the solvent \textit{in vacuo} yielded \SI{4.413}{\gram} (\SI{20.5}{\milli\mol}, 45\%) of blue, partly waxy, partly microcrystalline \textbf{1-Se}, which was stored under argon. Suitable crystals for crystal structure determination could be obtained by sublimation at room temperature after an extended period of time.\\

\textbf{\textsuperscript{1}H NMR} (500~MHz, CDCl\textsubscript{3}) $\updelta$ (ppm): 2.42 (dddd, \textit{J} = 3.3, 1.5, 1.5, 1.5~Hz, 1H, H4), 1.82 (dddd, \textit{J} = 10.3, 2.1, 2.1, 2.1~Hz, 1H, H7a), 1.76 (ddd, \textit{J} = 12.6, 2.4, 1.1~Hz, 1H, H5b) 1.72 (ddd, \textit{J} = 12.6, 5.9, 3.7~Hz, 1H, H5a), 1.53 (dd, \textit{J} = 10.3, 1.6~Hz, 1H, H7b), 1.46 (s, 3H, Me C10), 1.31 (dddd, \textit{J} = 12.6, 5.1, 2.3, 2.3~Hz, 1H, H6b), 1.26 (m, 1H, H6a), 1.24 (s, 3H, Me C9), 1.16 (s, 3H, Me C8); \textbf{\textsuperscript{13}C NMR} (101~MHz, CDCl\textsubscript{3}) $\updelta$ (ppm): 293.4 (C2, q, \textsuperscript{1}\textit{J}\textsubscript{CSe} = 221~Hz), 72.9 (C1, q), 63.7 (C3, q), 47.5 (C4), 42.8 (C7), 33.1 (C6), 27.8 (C8), 26.7 (C9), 25.2 (C5), 21.0 (C10); \textbf{\textsuperscript{77}Se NMR} (95~MHz, CDCl\textsubscript{3}) $\updelta$ (ppm): 1604.\\
IR (\SI{}{\per\centi\meter}): 2961~s, 2921~m, 2661~mw, 1465~mw, 1444~m, 1376~mw, 1355~w, 1319~w, 1245~w, 1161~w, 1142~w, 1129~mw, 1079~vs, 1046~vs, 1009~mw, 998~mw, 981~w, 950~w, 938~mw, 922~w, 882~w, 814~w, 716~w, 653~mw, 504~w.\\
Elem. anal. calcd. (\%) for C\textsubscript{10}H\textsubscript{16}Se: C~55.81, H~7.49, found: C~55.80, H~7.83.\\
APCI-HR: \textit{m/z} = 217.0490 (M+H, 100\%)\textsuperscript{+} calculated: 217.0490.

\begin{figure}[h]
    \centering
    \includegraphics[width=0.8\linewidth]{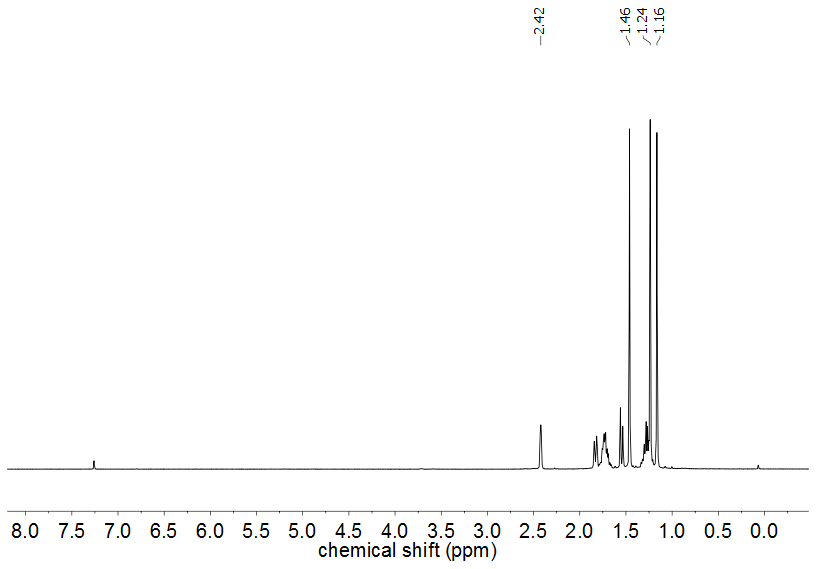}
    \caption{\textsuperscript{1}H NMR (CDCl\textsubscript{3}, 500~MHz) of \textbf{1-Se}.}
    \label{fig:1-se hnmr}
\end{figure}
\begin{figure}[h]
    \centering
    \includegraphics[width=0.8\linewidth]{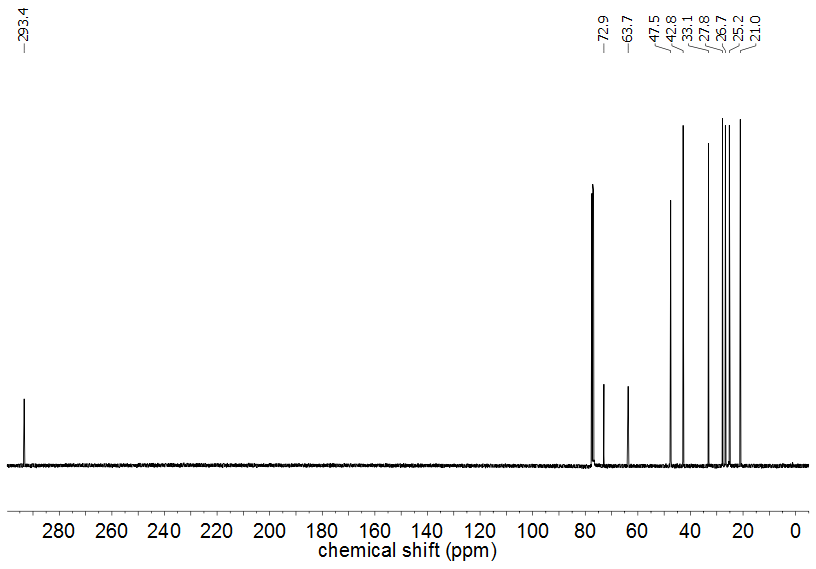}
    \caption{\textsuperscript{13}C NMR (CDCl\textsubscript{3}, 101~MHz) of \textbf{1-Se}.}
    \label{fig:1-se cnmr}
\end{figure}
\begin{figure}[h]
    \centering
    \includegraphics[width=0.8\linewidth]{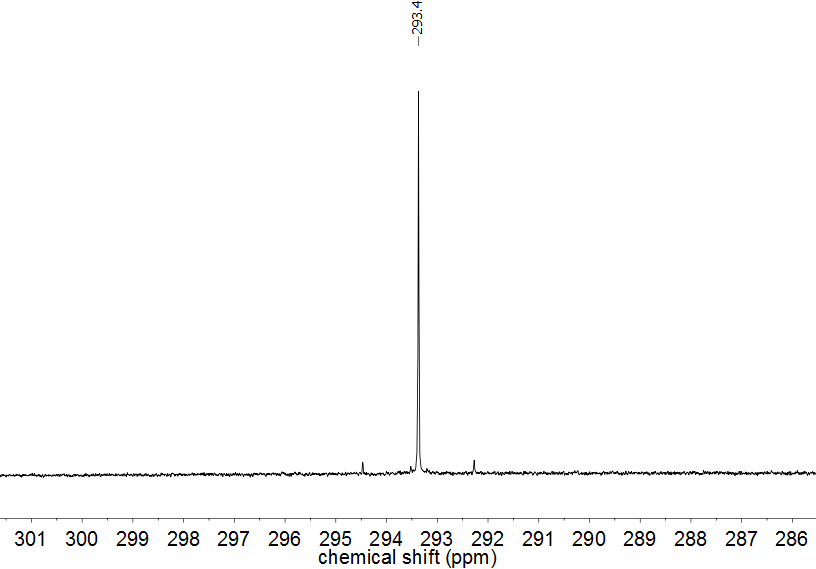}
    \caption{\textsuperscript{13}C NMR (CDCl\textsubscript{3}, 101~MHz) of \textbf{1-Se} displaying the selenium satellites.}
    \label{fig:1-se cnmrse}
\end{figure}

\begin{figure}[h]
    \centering
    \includegraphics[width=0.8\linewidth]{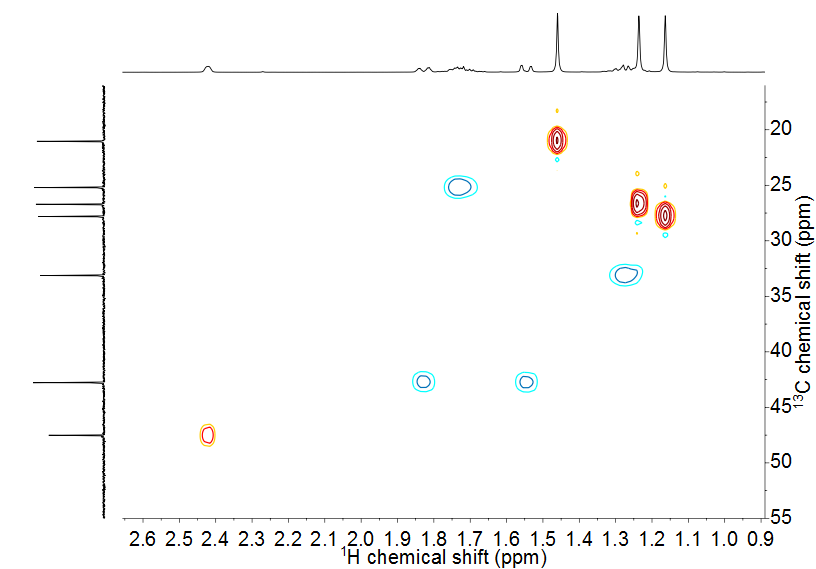}
    \caption{HSQC (CDCl\textsubscript{3}, 126~MHz) of \textbf{1-Se}.}
    \label{fig:1-se hsqc}
\end{figure}
\begin{figure}[h]
    \centering
    \includegraphics[width=0.8\linewidth]{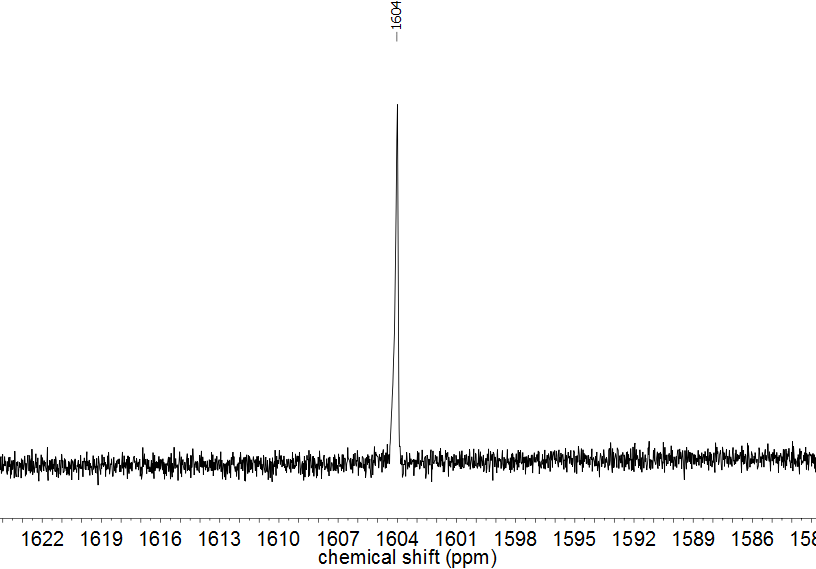}
    \caption{\textsuperscript{77}Se NMR (CDCl\textsubscript{3}, 95~MHz) of \textbf{1-Se}.}
    \label{fig:1-se senmr}
\end{figure}

\begin{figure}[h]
    \centering
    \includegraphics[width=0.9\linewidth]{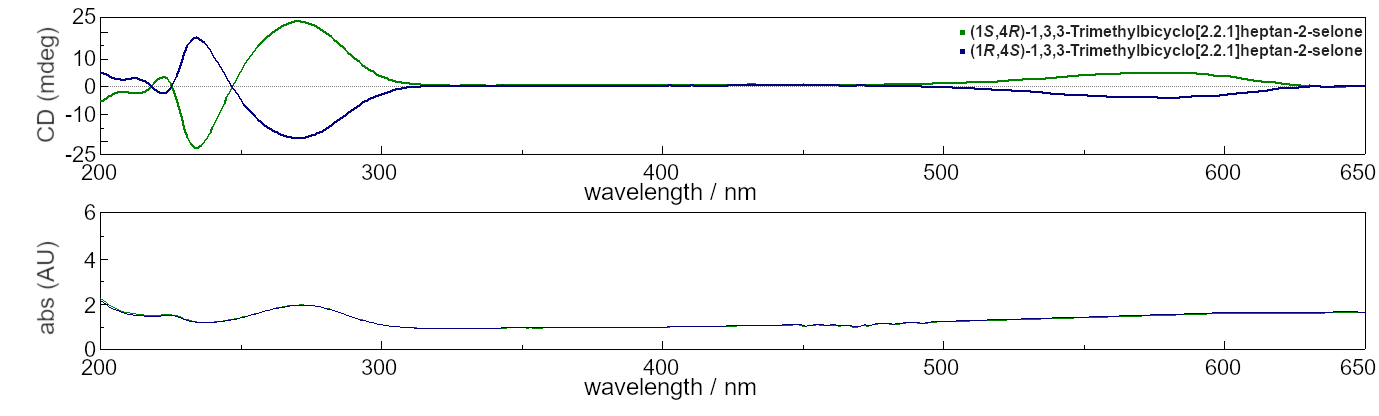}
    \caption{CD spectrum (top) and UV/vis absorption spectrum (bottom) of \textbf{1-Se} measured in hexane solution with concentration of \SI{1e-4}{\mol\per\liter}.}
    \label{fig:Se-CD}
\end{figure}

\begin{figure}[h]
    \centering
    \includegraphics[width=0.9\linewidth]{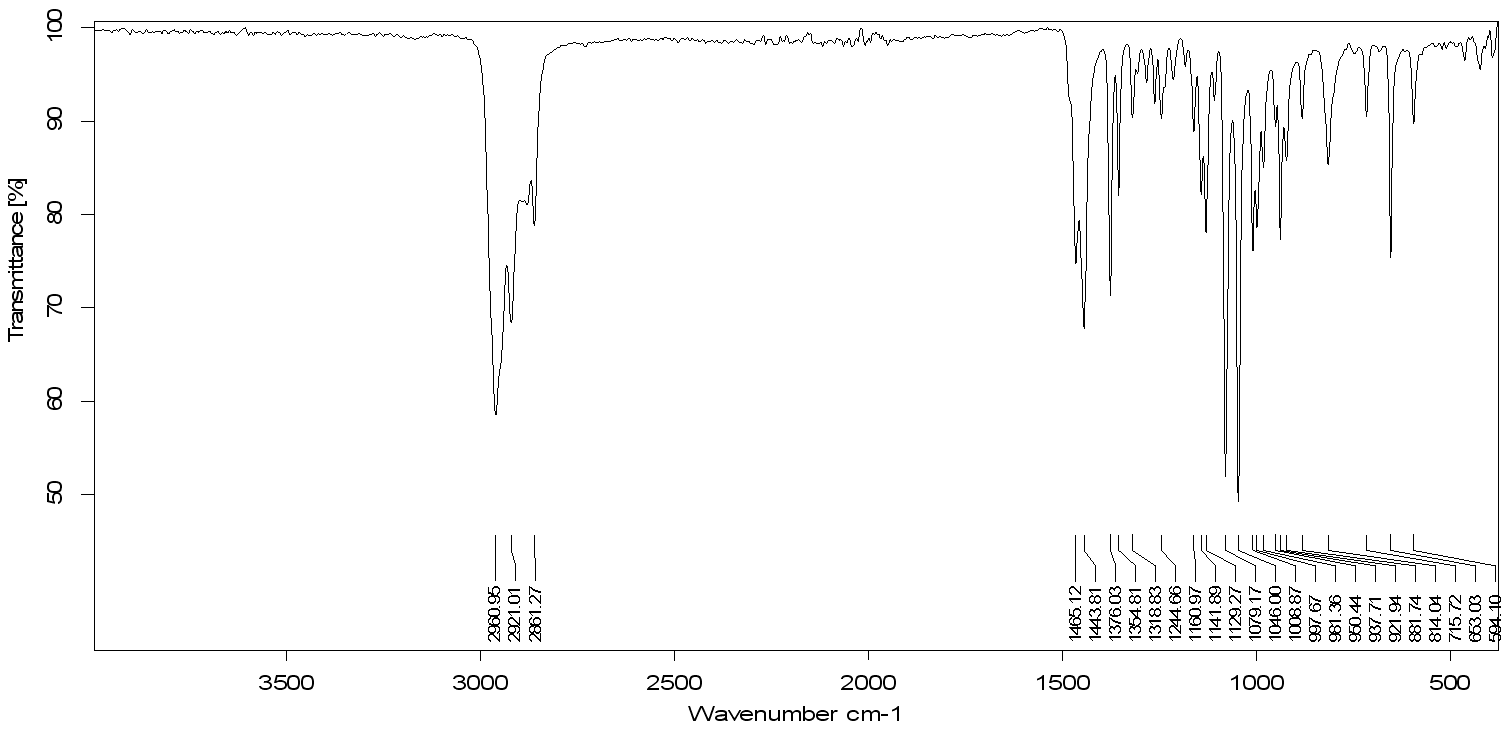}
    \caption{ATR-IR spectrum of neat \textbf{1-Se}.}
    \label{fig:Se-IR}
\end{figure}

\clearpage

\subsubsection{Microwave spectroscopy}

\begin{table}[ht]
  \centering
  \caption{Experimental rotational spectroscopic parameters (rotational constants $A$, $B$, $C$; quartic centrifugal distortion constant $\Delta_{JK}$) of thiofenchone (\textbf{1-S}) including its $^{13}$C and $^{34}$S singly substituted isotopologues. A centrifugal distortion constant given in square brackets was kept frozen at its value determined for the normal isotopologue that is composed only of the most abundant isotope of each element.\label{tab:1-S_RC}}
  \resizebox{\textwidth}{!}{%
    \begin{tabular}{cSSSSSS}
    \hline
               & normal     & $^{13}$C${1}$       & $^{13}$C${2}$       & $^{13}$C${3}$       & $^{13}$C${4}$       & $^{13}$C${5}$ \\
    \hline
    $A$/MHz      & 1156.59033(19) & 1154.06863(35) & 1156.32148(53) & 1153.07398(30) & 1152.43897(54) & 1152.79833(61) \\
    $B$/MHz      & 1121.03504(17) & 1119.34028(34) & 1120.25642(50) & 1120.59166(28) & 1117.11189(46) & 1109.71777(51) \\
    $C$/MHz      &  787.53933(16) &  785.61368(20) &  787.02678(30) &  785.72978(14) &  784.04328(28) &  782.09702(24) \\
    $\Delta_{JK}$/kHz    &   -0.0262(12)  & [-0.262]   & [-0.262]   & [-0.262]   & [-0.262]   & [-0.262] \\
    \# of transitions & 114        & 21         & 26         & 28         & 25         & 25 \\
    RMSD/kHz   & 5.2        & 5.3        & 5.0          & 3.2        & 5.2        & 5.5 \\
    \hline
               & $^{13}$C${6}$       & $^{13}$C${7}$       & $^{13}$C${8}$       & $^{13}$C${9}$       & $^{13}$C${10}$      & $^{34}$S \\
    \hline
    $A$/MHz      & 1152.71969(40) & 1152.71478(62) & 1145.24348(91) & 1144.64693(43) & 1141.18855(43) & 1152.04299(35) \\
    $B$/MHz      & 1111.77200(35) & 1112.49166(60) & 1112.24805(64) & 1115.51762(39) & 1117.75464(43) & 1099.56917(30) \\
    $C$/MHz      &  784.01489(18) &  785.11781(34) &  782.47865(40) &  782.00866(18) &  779.37535(29) &  775.34524(13) \\
    $\Delta_{JK}$/kHz    & [-0.262]   & [-0.262]   & [-0.262]   & [-0.262]   & [-0.262]   & [-0.262] \\
    \# of transitions & 27         & 21         & 25         & 32         & 25         & 51 \\
    RMSD/kHz   & 4.3        & 5.3        & 6.6        & 4.7        & 4.5        & 3.7 \\
    \hline
    \end{tabular}%
    }%
\end{table}%

\begin{table}[ht]
  \centering
  \caption{Experimental rotational spectroscopic parameters (rotational constants $A$, $B$, $C$; quartic centrifugal distortion constants $\Delta_{J}$, $\Delta_{JK}$) of selenofenchone (\textbf{1-Se}) including its $^{13}$C, $^{74}$Se, $^{76}$Se, $^{77}$Se, $^{78}$Se, and $^{82}$Se singly substituted isotopologues. Centrifugal distortion constants given in square brackets were kept frozen at their value determined for the normal isotopologue that is composed only of the most abundant isotope of each element. \label{tab:1-Se_RC}}
  \resizebox{\textwidth}{!}{%
    \begin{tabular}{cSSSSSS}
    \hline
               & normal     & $^{13}$C${1}$       & $^{13}$C${2}$       & $^{13}$C${3}$       & $^{13}$C${4}$       & $^{13}$C${5}$ \\
    \hline
    $A$/MHz      & 1139.63633(37) & 1135.7102(49)  & 1139.6574(38)  & 1135.7965(48)  & 1138.5635(41)  & 1136.9366(59)  \\
    $B$/MHz      &  738.68086(27) &  737.96137(36) &  738.70915(28) &  738.04221(37) &  733.49968(29) &  730.42393(44) \\
    $C$/MHz      &  576.74781(28) &  575.35617(32) &  576.78162(24) &  575.40076(31) &  573.41903(26) &  572.38097(48) \\
    $\Delta_{J}$/kHz     &    0.0175(49)  & [0.0175]   & [0.0175]   & [0.0175]   & [0.0175]   & [0.0175] \\
    $\Delta_{JK}$/kHz    &    0.0615(54)  & [0.0615]   & [0.0615]   & [0.0615]   & [0.0615]   & [0.0615] \\
    \# of transitions & 71         & 27         & 24         & 22         & 26         & 25 \\
    RMSD/kHz   & 5.5        & 5.1        & 3.8        & 4.8        & 4.1        & 5.4 \\    
    \hline
               & $^{13}$C${6}$       & $^{13}$C${7}$       & $^{13}$C${8}$       & $^{13}$C${9}$       & $^{13}$C${10}$      & $^{74}\mathrm{Se}$ \\
    \hline
    $A$/MHz      & 1132.3384(42)  & 1135.1765(42)  & 1123.9810(43)  & 1125.2847(39)  & 1122.3974(42)  & 1139.9489(62)  \\
    $B$/MHz      &  734.60763(30) &  733.41779(30) &  736.32125(32) &  737.18284(29) &  738.27870(33) &  762.68925(43) \\
    $C$/MHz      &  574.05060(27) &  574.12259(30) &  573.89650(30) &  573.53254(26) &  572.50610(30) &  591.21645(37) \\
    $\Delta_{J}$/kHz     & [0.0175]   & [0.0175]   & [0.0175]   & [0.0175]   & [0.0175]   & [0.0175] \\
    $\Delta_{JK}$/kHz    & [0.0615]   & [0.0615]   & [0.0615]   & [0.0615]   & [0.0615]   & [0.0615] \\
    \# of transitions & 26         & 29         & 26         & 27         & 25         & 25 \\
    RMSD/kHz   & 4.2        & 4.2        & 4.6        & 4.1        & 4.6        & 5.5 \\
    \hline
               & $^{76}\mathrm{Se}$       & $^{77}\mathrm{Se}$       & $^{78}\mathrm{Se}$       & $^{82}\mathrm{Se}$\\
    \hline
    $A$/MHz      & 1139.83335(20)  & 1139.78177(26) & 1139.73236(20) & 1139.54527(36)  &            &  \\
    $B$/MHz      &  754.368090(91) &  750.32298(14) & 746.370906(95) &  731.28036(10)  &            &  \\
    $C$/MHz      &  586.226297(78) &  583.79093(11) & 581.405901(90) &  572.243991(98) &            &  \\
    $\Delta_{J}$/kHz     & [0.0175]   & [0.0175]   & [0.0175]   & [0.0175]   &            &  \\
    $\Delta_{JK}$/kHz    & [0.0615]   & [0.0615]   & [0.0615]   & [0.0615]   &            &  \\
    \# of transitions & 38         & 32         & 51         & 42         &            &  \\
    RMSD/kHz   & 2.7        & 3.2        & 3.9        & 3.9        &            &  \\
    \hline
    \end{tabular}%
    }%
\end{table}%

\begin{table}
\caption{Structural parameters ($r_0$) of \textbf{1-S} and \textbf{1-Se} determined from microwave spectroscopy in this work, compared to data for \textbf{1-O} from Ref.~\onlinecite{loru:2016}. Bond lengths $r$(Z1-Z2) between atoms Z1 and Z2 are given in \SI{}{\angstrom}.\label{tab:1-S,Se_r}}
\begin{tabular}{llS[table-format=1.4(2),round-mode=uncertainty,round-precision=2]S[table-format=1.4(2),round-mode=uncertainty,round-precision=2]S[table-format=1.4(2),round-mode=uncertainty,round-precision=2]}
\hline
    &         &    {X=O$^\mathrm{a}$}  &    {X=S}               & {X=Se}                \\
Z1  & Z2      &    \multicolumn{3}{c}{$r$(Z1--Z2)}                                      \\
\hline
C2  & C1      &    1.526(29)           &   1.50589 ( 0.00429 ) &   1.50484 ( 0.01456 ) \\
C2  & X       &    1.214(5)            &   1.63412 ( 0.00300 ) &   1.77262 ( 0.02085 ) \\
C3  & C2      &    1.535(31)           &   1.52260 ( 0.00297 ) &   1.52392 ( 0.01304 ) \\
C4  & C3      &    1.549(30)           &   1.55682 ( 0.00314 ) &   1.55546 ( 0.00803 ) \\
C5  & C4      &    1.546(8)            &   1.54408 ( 0.00631 ) &   1.54949 ( 0.01806 ) \\
C6  & C5      &    1.526(9)            &   1.55955 ( 0.00604 ) &   1.55207 ( 0.01779 ) \\
C6  & C1      &    1.555(18)           &   1.56270 ( 0.00434 ) &   1.56599 ( 0.01193 ) \\
C7  & C1      &    1.541(25)           &   1.55501 ( 0.00565 ) &   1.55629 ( 0.01107 ) \\
C7  & C4      &    1.552(8)            &   1.54055 ( 0.00709 ) &   1.54289 ( 0.01537 ) \\
C8  & C3      &    1.545(16)           &   1.54496 ( 0.00785 ) &   1.55196 ( 0.00894 ) \\
C9  & C3      &    1.535(13)           &   1.53830 ( 0.00796 ) &   1.53271 ( 0.00865 ) \\
C10 & C1      &    1.521(11)           &   1.51768 ( 0.00313 ) &   1.51866 ( 0.01172 ) \\
\hline
\end{tabular}
\footnotetext[1]{From Ref.~\onlinecite{loru:2016}.}
\end{table}

\begin{table}
\caption{Structural parameters ($r_0$) of \textbf{1-S} and \textbf{1-Se} determined from microwave spectroscopy in this work, compared to data for \textbf{1-O} from Ref.~\onlinecite{loru:2016} where available. Bond angles $\alpha$(Z1-Z2-Z3) are given in degree.\label{tab:1-S,Se_alpha}}
\begin{tabular}{lllS[table-format=3.2(2),round-mode=uncertainty,round-precision=2]S[table-format=3.2(2),round-mode=uncertainty,round-precision=2]S[table-format=3.2(2),round-mode=uncertainty,round-precision=2]}
\hline
    &     &         &    {X=O$^\mathrm{a}$}  &  {X=S}                 & {X=Se}                 \\
Z1  & Z2  & Z3      &  \multicolumn{3}{c}{$\alpha$(Z1-Z2-Z3)}         \\
\hline
C1  & C2  & X       &                        &  126.60244 ( 0.15752 ) &  126.61351 ( 0.72805 ) \\
X   & C2  & C3      &  125.8(31)             &  125.52404 ( 0.19822 ) &  125.55072 ( 0.71040 ) \\
C1  & C2  & C3      &  107.5(12)             &  107.86757 ( 0.21773 ) &  107.82780 ( 1.32451 ) \\
C4  & C3  & C2      &  100.8(9)              &  100.98646 ( 0.18433 ) &  101.15508 ( 0.78309 ) \\
C4  & C7  & C1      &   95.2(6)              &   95.07480 ( 0.25755 ) &   95.14832 ( 0.61142 ) \\
C5  & C4  & C3      &  110.3(8)              &  110.16137 ( 0.49327 ) &  109.82120 ( 0.86961 ) \\
C6  & C5  & C4      &  102.8(3)              &  102.63841 ( 0.20575 ) &  102.78247 ( 0.55947 ) \\
C5  & C6  & C1      &  103.9(7)              &  104.15974 ( 0.24748 ) &  104.37621 ( 0.50236 ) \\
C7  & C4  & C3      &                        &  101.71286 ( 0.27506 ) &  101.87143 ( 0.49974 ) \\
C7  & C4  & C5      &                        &  100.43713 ( 0.23761 ) &  100.08108 ( 1.03916 ) \\
C7  & C1  & C2      &                        &  100.25557 ( 0.24526 ) &  100.39377 ( 0.71259 ) \\
C7  & C1  & C6      &  101.4(11)             &  100.53916 ( 0.24995 ) &  100.23735 ( 0.56377 ) \\
C8  & C3  & C2      &                        &  110.14934 ( 0.38658 ) &  109.99520 ( 0.70661 ) \\
C8  & C3  & C4      &  111.4(14)             &  110.71815 ( 0.53984 ) &  109.82295 ( 0.86923 ) \\
C9  & C3  & C2      &                        &  111.94385 ( 0.49271 ) &  112.40311 ( 0.65622 ) \\
C9  & C3  & C4      &  116.2(22)             &  115.11535 ( 0.42482 ) &  115.46330 ( 0.79418 ) \\
C10 & C1  & C2      &                        &  116.84628 ( 0.22951 ) &  117.24792 ( 1.13168 ) \\
C10 & C1  & C7      &                        &  117.79900 ( 0.32589 ) &  117.75605 ( 0.87901 ) \\
C10 & C1  & C6      &  115.3(11)             &  114.39370 ( 0.29136 ) &  114.16697 ( 0.64336 ) \\
\hline
\end{tabular}
\footnotetext[1]{From Ref.~\onlinecite{loru:2016}.}
\end{table}

\begin{table}
\caption{Structural parameters ($r_0$) of \textbf{1-S} and \textbf{1-Se} determined from microwave spectroscopy in this work, compared to data for \textbf{1-O} from Ref.~\onlinecite{loru:2016} where available. Dihedral angles $\tau$(Z1-Z2-Z3-Z4) are given in degree. \label{tab:1-S,Se_tau}}
\begin{tabular}{llllS[table-format=+3.2(2),round-mode=uncertainty,round-precision=2]S[table-format=+3.2(2),round-mode=uncertainty,round-precision=2]S[table-format=+3.2(2),round-mode=uncertainty,round-precision=2]}
\hline
   &      &     &         &    {X=O$^\mathrm{a}$}  &  {X=S}                 & {X=Se}                 \\
Z1 &  Z2  & Z3  & Z4      & \multicolumn{3}{c}{$\tau$(Z1-Z2-Z3-Z4)}         \\ 
\hline
C3 &  C2  & X   & C1      &                        &  179.00381 ( 0.62795 ) &  178.84635 ( 1.10698 ) \\
C4 &  C3  & C2  & C1      &                        &    1.58335 ( 0.62425 ) &    1.21230 ( 0.94460 ) \\
C4 &  C3  & C2  & X       & -177.7(10)             & -177.57640 ( 0.36138 ) & -177.81500 ( 0.71413 ) \\
C5 &  C4  & C3  & C2      &   70.2(13)             &   69.99178 ( 0.52675 ) &   70.06725 ( 1.09834 ) \\
C5 &  C6  & C1  & C2      &                        &   72.83148 ( 0.22479 ) &   72.77768 ( 0.96038 ) \\
C6 &  C5  & C4  & C3      &  -67.3(10)             &  -66.98958 ( 0.27365 ) &  -66.89420 ( 1.12692 ) \\
C6 &  C1  & C2  & C3      &                        &  -71.16633 ( 0.46263 ) &  -70.75175 ( 0.68976 ) \\
C7 &  C1  & C2  & C3      &                        &   32.68235 ( 0.47028 ) &   32.81400 ( 0.79870 ) \\
C7 &  C1  & C6  & C5      &  -30.3(7)              &  -30.80443 ( 0.28698 ) &  -30.90554 ( 0.76609 ) \\
C7 &  C4  & C3  & C2      &                        &  -35.86341 ( 0.60287 ) &  -35.35322 ( 0.94141 ) \\
C7 &  C4  & C5  & C6      &                        &   39.72053 ( 0.25863 ) &   39.73530 ( 0.74791 ) \\
C8 &  C3  & C2  & C1      &                        & -115.48818 ( 0.28493 ) & -114.86410 ( 0.78903 ) \\
C8 &  C3  & C4  & C5      &                        & -173.35509 ( 0.26280 ) & -173.72922 ( 0.93175 ) \\
C9 &  C3  & C2  & C1      &                        &  124.56376 ( 0.24868 ) &  124.94192 ( 0.69796 ) \\
C9 &  C3  & C4  & C5      &                        &  -50.76712 ( 0.38644 ) &  -51.54408 ( 1.09616 ) \\
C10&  C1  & C2  & C3      &                        &  161.20123 ( 0.44432 ) &  161.64572 ( 0.76270 ) \\
C10&  C1  & C7  & C4      &                        &  179.22221 ( 0.26344 ) &  178.84126 ( 0.93881 ) \\
C10&  C1  & C6  & C5      & -160.6(16)             & -158.05070 ( 0.23162 ) & -157.75706 ( 1.05184 ) \\
\hline
\end{tabular}
\footnotetext[1]{From Ref.~\onlinecite{loru:2016}.}
\end{table}

\clearpage

\subsection{Computational}

\begin{table}[h]
    \centering
    \caption{Calculated equilbrium C=X bond lengths, selected experimental IR spectral data and calculated scaled harmonic vibrational wavenumbers. \label{tab:Xstretch}}
    \renewcommand{\arraystretch}{1.3} 
    \resizebox{\textwidth}{!}{ 
    \begin{tabular}{c S[round-mode=figures,round-precision=3] c c c c}
        \hline
        \multirow{2}{*}{\textbf{Molecules}} & {\multirow{2}{*}{\textbf{C-X bond length (\AA)}}} & \multirow{2}{*}{\textbf{Experimental wavenumbers}} & \multicolumn{3}{c}{\textbf{Calculated (scaled harmonic) wavenumbers}} \\
        \cline{4-6}
        & & \textbf{C=X str (cm$^{-1}$)} & \textbf{C-X str (cm$^{-1}$)} & \textbf{C-X bend (in-plane) (cm$^{-1}$)} & \textbf{C-X bend (out-of-plane) (cm$^{-1}$)} \\
        \hline
        \textbf{1-O}  & 1.207 & 1756.46, 1741.83 & 1742 & 996 & 696 \\
        \textbf{1-S}  & 1.624 & 1178.57          & 1158 & 915 & 653 \\
        \textbf{1-Se} & 1.777 & 1053.15, 1047.39 & 1032 & 915 & 641 \\
        \textbf{1-Te} & 1.967 &                  & 1019 & 914 & 625 \\
        \textbf{1-Po} & 2.101 &                  &  965 & 913 & 619 \\
        \hline
    \end{tabular}
    }
\end{table}

\begin{table}[h]
    \centering
    \caption{Scaled internal force constants for C=X bond of chalcogenofenchones in gas phase obtained at the same level of theory.\label{tab:Xforce}}
    \renewcommand{\arraystretch}{1.3} 
    \begin{tabular}{c S[round-mode=places,round-precision=1]}
        \hline
        & \textbf{Force constant (gas phase)} \\
        & \textbf{(nN \AA$^{-1}$)} \\
        \hline
        \textbf{\textit{ƒ}(C=O)}  & 219.8 \\
        \textbf{\textit{ƒ}(C=S)}  & 18.3 \\
        \textbf{\textit{ƒ}(C=Se)} & 11.9 \\
        \textbf{\textit{ƒ}(C=Te)} & 12.4 \\
        \textbf{\textit{ƒ}(C=Po)} & 10.4 \\
        \hline
    \end{tabular}
\end{table}

\begin{table}[h]
    \centering
    \caption{Cartesian coordinates ($x,y,z$ in \AA) of the computed equilibrium structure of \textbf{1-O}.\label{tab:Ostruct}}
    \begin{tabular}{l S[round-mode=places,round-precision=8] 
                      S[round-mode=places,round-precision=8]
                      S[round-mode=places,round-precision=8]}
    \hline
    {Element symbol} & {$x$} & {$y$} & {$z$} \\  
    \hline
    C   &   2.0562232581625 & -0.6514866343295 &  1.2218239459896  \\
    C   &   1.2210522334510 &  0.0013593159563 &  0.1042962474049  \\
    C   &   2.1301297135477 &  0.2104466499587 & -1.1123798286501  \\
    O   &   0.1064905762696 & -2.0116019017455 & -0.7473814445800  \\
    C   &   0.0283564189488 & -0.9144691028568 & -0.2512331639664  \\
    C   &  -2.4777545810373 & -1.0477594868132 &  0.2729330151038  \\
    C   &  -1.2533930545524 & -0.1713786796272 &  0.1263507868548  \\
    C   &  -0.2559119591553 &  1.9771039587631 & -0.5437610752817  \\
    C   &  -1.3682531590036 &  0.9656074854058 & -0.9410239946880  \\
    C   &   0.4618000652774 &  1.2605438347183 &  0.6186866953549  \\
    C   &  -0.7496704131961 &  0.6366281620387 &  1.3437921198908  \\
    H   &   2.9136549109439 & -0.0187028882725 &  1.4587146036455  \\
    H   &   1.4901937304041 & -0.8054977957710 &  2.1392781390431  \\
    H   &   2.4275439926266 & -1.6215152099799 &  0.8922622900508  \\
    H   &   2.9225680595142 &  0.9243869000293 & -0.8784592626524  \\
    H   &   2.5952138410605 & -0.7349580930845 & -1.3900885039378  \\
    H   &   1.5917343380852 &  0.5758053328631 & -1.9851580667412  \\
    H   &  -2.3373698235935 & -1.7952319974228 &  1.0548011061464  \\
    H   &  -3.3552264694161 & -0.4498267732018 &  0.5253844988250  \\
    H   &  -2.6833859396850 & -1.5831753269966 & -0.6544530137624  \\
    H   &   0.4037903352746 &  2.2319490755147 & -1.3696964677547  \\
    H   &  -0.6915963817467 &  2.9104626512719 & -0.1863951508713  \\
    H   &  -2.3613803479883 &  1.4114737467954 & -0.8732854233567  \\
    H   &  -1.2570638777263 &  0.5820458726143 & -1.9553346188429  \\
    H   &   1.0899782310254 &  1.9163781779620 &  1.2203133361484  \\
    H   &  -1.4753515636104 &  1.3829925528932 &  1.6703134655851  \\
    H   &  -0.4936537662560 &  0.0110314036576 &  2.1976878841790  \\
    \hline
    \end{tabular}
\end{table}

\begin{table}[h]
    \centering
    \caption{Cartesian coordinates (in \AA) of the computed equilibrium structure of \textbf{1-S}.\label{tab:Sstruct}}
    \begin{tabular}{l S[round-mode=places,round-precision=8] 
                      S[round-mode=places,round-precision=8]
                      S[round-mode=places,round-precision=8]}
    \hline
    {Element symbol} & {$x$} & {$y$} & {$z$} \\  
    \hline
    C  &    1.4135888249303 & -0.0619516876299 &  1.2128279839609  \\
    C  &    0.8310642252847 &  0.9519485574420 &  0.1964902236689  \\
    C  &   -0.5613975072531 &  0.3831029596857 & -0.0069345533891  \\
    C  &   -0.4536363092012 & -1.1426821704218 &  0.1528502311834  \\
    C  &    1.0805936575897 & -1.3097527711163 &  0.3784093890947  \\
    C  &    1.8540715342138 & -0.9953379810763 & -0.9176474447012  \\
    C  &    1.6203047561879 &  0.5293358333443 & -1.0934782119578  \\
    C  &   -1.2419723860567 & -1.6006218562093 &  1.3986500702499  \\
    C  &   -1.0206097050668 & -1.8942466332534 & -1.0596833155148  \\
    S  &   -1.9149207335758 &  1.2129104065326 & -0.3463167537653  \\
    C  &    0.9243955717780 &  2.4256571841776 &  0.5290445249713  \\
    H  &   -1.1494425284264 & -2.6823991564854 &  1.5101055376418  \\
    H  &   -0.8892818530152 & -1.1342868032937 &  2.3167260260100  \\
    H  &   -2.2959175832238 & -1.3525953838720 &  1.2809452480963  \\
    H  &   -0.8199503469312 & -2.9633232004128 & -0.9650460502352  \\
    H  &   -2.0985919960147 & -1.7504792117858 & -1.1132461321607  \\
    H  &   -0.5986547260638 & -1.5489582321592 & -2.0014919782010  \\
    H  &    0.3726835002464 &  2.6602055162726 &  1.4399344570744  \\
    H  &    1.9667689020907 &  2.7149343289957 &  0.6733959384521  \\
    H  &    0.5056271993271 &  3.0365718038167 & -0.2703338356714  \\
    H  &    1.5226692042727 & -1.5797531478453 & -1.7724635549499  \\
    H  &    2.9131558309183 & -1.2120124978149 & -0.7759078628915  \\
    H  &    2.5599176152172 &  1.0825027311273 & -1.1169355829667  \\
    H  &    1.0710690817927 &  0.7818061942336 & -1.9993894281013  \\
    H  &    1.3356300522074 & -2.2699644623127 &  0.8246370152800  \\
    H  &    2.4851636748135 &  0.0820327425217 &  1.3581648735522  \\
    H  &    0.9252224345651 & -0.0207821969935 &  2.1847076140916  \\
    \hline
    \end{tabular}
\end{table}

\begin{table}[h]
    \centering
    \caption{Cartesian coordinates (in \AA) of the computed equilibrium structure of \textbf{1-Se}.\label{tab:Sestruct}}
    \begin{tabular}{l S[round-mode=places,round-precision=8] 
                      S[round-mode=places,round-precision=8]
                      S[round-mode=places,round-precision=8]}
    \hline
    {Element symbol} & {$x$} & {$y$} & {$z$} \\  
    \hline
    C &     1.9085519485081 &  0.6523352135084 &  1.1435447171163  \\
    C &     0.7995189384856 &  1.2269387706721 &  0.2237445575488  \\
    C &    -0.1115652641359 &  0.0267895446178 &  0.0932794156106  \\
    C &     0.7687191471649 & -1.2259087465076 &  0.1700920711726  \\
    C &     2.1878869552402 & -0.5844696035976 &  0.2756094178580  \\
    C &     2.5751723385432 &  0.0875483441815 & -1.0566428603899  \\
    C &     1.5813598757925 &  1.2761761252801 & -1.1403601428713  \\
    C &     0.4420880450232 & -2.0304771414138 &  1.4483942149703  \\
    C &     0.5693393859037 & -2.1632393084235 & -1.0302754575363  \\
   Se &    -1.8745939469570 &  0.0475569436457 & -0.1295612093224  \\
    C &     0.1664528840301 &  2.5438964861294 &  0.6199014648439  \\
    H &     1.0937697432305 & -2.9043473301512 &  1.5022931741595  \\
    H &     0.5769673014171 & -1.4494234659690 &  2.3585832869250  \\
    H &    -0.5926411720052 & -2.3684234970506 &  1.4172578991866  \\
    H &     1.3037999899077 & -2.9706058764356 & -0.9985010537437  \\
    H &    -0.4263920987559 & -2.6021057240440 & -0.9967782696554  \\
    H &     0.6667970567368 & -1.6482010516720 & -1.9837815378490  \\
    H &    -0.3509423691237 &  2.4608128597465 &  1.5760710914327  \\
    H &     0.9317744604564 &  3.3170678710984 &  0.7074474896674  \\
    H &    -0.5662908756023 &  2.8673652466993 & -0.1188346044414  \\
    H &     2.5148494174372 & -0.5807211423601 & -1.9120780350938  \\
    H &     3.6040464554409 &  0.4442289526806 & -1.0035224067529  \\
    H &     2.0995042641531 &  2.2340015032279 & -1.2001762381938  \\
    H &     0.9041620555279 &  1.2170614142576 & -1.9908342862987  \\
    H &     2.9338645786871 & -1.2817650544383 &  0.6541794030380  \\
    H &     2.7619792117486 &  1.3281238747472 &  1.2147308747374  \\
    H &     1.5567171860678 &  0.4330377679941 &  2.1498092710163  \\
    \hline
    \end{tabular}
\end{table}

\begin{table}[h]
    \centering
    \caption{Cartesian coordinates (in \AA) of the computed equilibrium structure of \textbf{1-Te}.\label{tab:Testruct}}
    \begin{tabular}{l S[round-mode=places,round-precision=8] 
                      S[round-mode=places,round-precision=8]
                      S[round-mode=places,round-precision=8]}
    \hline
    {Element symbol} & {$x$} & {$y$} & {$z$} \\  
    \hline
    C &     2.4061722524131 &  0.6594086488578 &  1.0891215529577  \\
    C &     1.2405916763955 &  1.2229029743903 &  0.2358366124248  \\
    C &     0.3354610256937 &  0.0206351560327 &  0.1348093978080  \\
    C &     1.2274607274558 & -1.2183254516603 &  0.1780315818222  \\
    C &     2.6458700425963 & -0.5721291612579 &  0.2051371617266  \\
    C &     2.9515493698226 &  0.1033999802531 & -1.1466061507417  \\
    C &     1.9563082167529 &  1.2933237092254 & -1.1651929663166  \\
    C &     0.9707895900853 & -2.0063378596591 &  1.4844012144740  \\
    C &     0.9816280203643 & -2.1824130665065 & -0.9915795125652  \\
   Te &    -1.6204199726380 &  0.0243204049404 & -0.0750514774561  \\
    C &     0.6268239158839 &  2.5348199250244 &  0.6767557723521  \\
    H &     1.6341663334826 & -2.8725126612348 &  1.5223740490748  \\
    H &     1.1388007451340 & -1.4089828197577 &  2.3781192389606  \\
    H &    -0.0608522119197 & -2.3542029019394 &  1.5049853297017  \\
    H &     1.7321690831030 & -2.9757558095922 & -0.9858795663718  \\
    H &    -0.0018497698793 & -2.6398675644678 & -0.8982415611075  \\
    H &     1.0163176325195 & -1.6828252325529 & -1.9575083196439  \\
    H &     0.1706652700898 &  2.4410432229459 &  1.6629422240392  \\
    H &     1.3917421702384 &  3.3117646030712 &  0.7224218728777  \\
    H &    -0.1518606312160 &  2.8597327096820 & -0.0126579762106  \\
    H &     2.8363833969781 & -0.5630607260223 & -1.9978311642468  \\
    H &     3.9823082283007 &  0.4583049748462 & -1.1571195463662  \\
    H &     2.4732402825546 &  2.2515495026448 & -1.2311804155807  \\
    H &     1.2390588352356 &  1.2501399581002 & -1.9826186417327  \\
    H &     3.4147684621193 & -1.2666506347821 &  0.5398136432272  \\
    H &     3.2553996678688 &  1.3433872231499 &  1.1095051466212  \\
    H &     2.1161842745703 &  0.4361813843632 &  2.1135767012235  \\
    \hline
    \end{tabular}
\end{table}

\begin{table}[h]
    \centering
    \caption{Cartesian coordinates (in \AA) of the computed equilibrium structure of \textbf{1-Po}.\label{tab:Postruct}}
    \begin{tabular}{l S[round-mode=places,round-precision=8] 
                      S[round-mode=places,round-precision=8]
                      S[round-mode=places,round-precision=8]}
    \hline
    {Element symbol} & {$x$} & {$y$} & {$z$} \\  
    \hline
    C &     2.4117102503063 &  1.2920809819647 & -1.1673173458454  \\
    C &     1.6830878662964 &  1.2285166411825 &  0.2297179430403  \\
    C &     2.8427030318041 &  0.6714436337388 &  1.1011282912762  \\
    C &     3.1032433248959 & -0.5583424016127 &  0.2203721827766  \\
    C &     3.4252879408947 &  0.1193827641333 & -1.1256913547105  \\
    C &     1.6871560844701 & -1.2178182787914 &  0.1719505684051  \\
    C &     0.7888492455488 &  0.0186583628057 &  0.1342406051124  \\
    C &     1.0660712939287 &  2.5449965552830 &  0.6541648359233  \\
   Po &    -1.3048540975750 &  0.0121317869668 & -0.0438898695627  \\
    C &     1.4342365360563 & -2.0321134250775 &  1.4633542304030  \\
    C &     1.4641346583611 & -2.1660562563977 & -1.0173317626048  \\
    H &     2.1161946884432 & -2.8841530618846 &  1.4919415723415  \\
    H &     1.5833648009503 & -1.4450655931184 &  2.3671162552092  \\
    H &     0.4104683228815 & -2.4033562031426 &  1.4708648262383  \\
    H &     2.2225721927985 & -2.9519689538689 & -1.0127516355039  \\
    H &     0.4838396099385 & -2.6338379157382 & -0.9422752662439  \\
    H &     1.5078413528862 & -1.6515395966504 & -1.9748708563113  \\
    H &     0.5879392281446 &  2.4593946718766 &  1.6303706564285  \\
    H &     1.8352685787957 &  3.3172949241954 &  0.7129655205835  \\
    H &     0.3048717742599 &  2.8729443249004 & -0.0534234327736  \\
    H &     3.3398443444442 & -0.5495619776604 & -1.9784341536704  \\
    H &     4.4499668802448 &  0.4916717854119 & -1.1142381530230  \\
    H &     2.9118537676071 &  2.2585799470364 & -1.2427090526509  \\
    H &     1.7015417779715 &  1.2244939636479 & -1.9893726452338  \\
    H &     3.8724610963627 & -1.2476769285387 &  0.5657269676491  \\
    H &     3.6859498970619 &  1.3629517029019 &  1.1360275654689  \\
    H &     2.5373896472316 &  0.4459524467771 &  2.1209231135794  \\
    \hline
    \end{tabular}
\end{table}

\clearpage

%\bibliography{Refence1.bib}

%aipnum4-2.bst 2019-01-14 (MD) hand-edited version of apsrev4-1.bst

\end{document}